\documentclass[10pt,preprint,5p,twocolumn,times]{elsarticle}

\usepackage{amsmath,amssymb}
\usepackage{dcolumn} %
\usepackage{graphicx}
\usepackage{hyperref}
\usepackage{color}
\usepackage{colortbl}
\usepackage[tabel]{xcolor}
\usepackage{braket}

\graphicspath{
  {figures/pdf/}
  {figures/eps/}
}

\renewcommand{\vec}[1]{\mathbf{#1}}

\begin{document}
\begin{frontmatter}

\journal{Computer Physics Communications}
\title{Lattice Dynamics Calculations\\ based on Density-functional Perturbation Theory in Real Space}

\author[FHI]{Honghui Shang}
\author[FHI]{Christian Carbogno}
\author[FHI,Aalto]{Patrick Rinke}
\author[FHI]{Matthias Scheffler}

\address[FHI]{Fritz-Haber-Institut der Max-Planck-Gesellschaft, Faradayweg 4--6, D-14195 Berlin, Germany}
\address[Aalto]{COMP/Department of Applied Physics, Aalto University, P.O. Box 11100, Aalto FI-00076, Finland}

\date{\today}

\begin{abstract}
A real-space formalism for density-functional perturbation theory~(DFPT) is derived and applied 
for the computation of harmonic vibrational properties in molecules and solids. The practical implementation
using numeric atom-centered orbitals as basis functions is demonstrated exemplarily for the all-electron Fritz Haber Institute \textit{ab initio} molecular simulations~({\it FHI-aims}) package. The convergence of the calculations 
with respect to numerical parameters is carefully investigated and a systematic comparison with finite-difference approaches is performed both for finite~(molecules) and extended~(periodic) systems. 
Finally, the scaling tests and scalability tests on massively parallel computer systems demonstrate the computational efficiency.
\end{abstract}

\end{frontmatter}
\section{Introduction}
Density-functional theory (DFT)~\cite{Hohenberg1964,Kohn1965} is to date the most widely applied method to compute the ground-state electronic structure and total energy for polyatomic systems in chemistry, physics, and material science. Via the Hellmann-Feynman~\cite{Hellmann1937,Feynman1939} theorem the DFT ground state density also provides access to the first derivatives of the total energy,~i.e.,~the forces acting on the nuclei and the stresses acting on the lattice degrees of freedom. The forces and stress in  turn can be used to determine equilibrium geometries with optimization algorithms~\cite{BFGS}, to traverse thermodynamic phase space with {\it ab initio} molecular dynamics \cite{Car-1985}, and even to search for transition states of chemical reactions or structural transitions \cite{Henkelman2000}. Second and higher order derivatives, however, cannot be calculated on the basis of the ground state density alone, but also require knowledge of its response to the corresponding perturbation: The $2n+1$~theorem~\cite{Gonze-1989} proves that the $n$-th order derivative of the density/wavefunction is required to determine the $2n+1$-th derivative of the total energy. For example, for the calculation of vibrational frequencies and phonon band-structures~(second order derivative) the response of the electronic structure to a nuclear displacement~(first order derivative) is needed. These derivatives can be calculated in the framework of density-functional perturbation theory~(DFPT)~\cite{Gonze1997-1,Gonze1997-2,Baroni-2001} viz.~the coupled perturbed self-consistent field (CPSCF) method~\cite{Gerratt-1967,Pople-1979,Dykstra-1984,Frisch-1990,Ochsenfeld-1997, Liang-2005} \footnote{Formally,
DFPT and CPSCF are essentially equivalent, but the term DFPT is more widely used in the physics community, whereas CPSCF is better known in quantum chemistry.}. DFPT and CPSCF then provide access to many fundamental physical phenomena, such as superconductivity~\cite{J.R.Schrieffer1964,Cardona2006}, phonon-limited carrier lifetimes~\cite{Bostwick2006,Lazzeri2005,Park2009} in electron transport and hot electron relaxation~\cite{Bernardi2014,Bernardi2015},  Peierls instabilities~\cite{Peierls1964}, the renormalization of the electronic structure due to nuclear motion~\cite{Eiguren2008,Giustino2010,Cannuccia2011,Kawai2014,Ponce2014a,Ponce2014b,Antonius2014,Ponce2015,Sezen2015,Gonze2016},  Born effective charges~\cite{Giannozzi1991}, phonon-assisted transitions in spectroscopy \cite{Kioupakis_PRB_2010,Kioupakis/etal:2011,Noffsinger/etal:2012}, infrared~\cite{Maschio2012} as well as Raman spectra~\cite{Maschio2013}, and much more~\cite{Giustino2016RevModPhys}.

In the literature, implementations of DFPT using a \emph{reciprocal-space} formalism have been mainly reported for plane-wave (PW) basis sets for norm-conserving pseudopotentials~\cite{Giannozzi1991,Gonze1997-1,Gonze1997-2}, for ultrasoft pseudopotentials~\cite{DalCorso2001}, and for the projector augmented wave method~\cite{DalCorso2010}. 
These techniques were also used for all-electron, full-potential implementations with linear muffin tin orbitals~\cite{Savrasov1996} and 
linearized augmented plane-waves~\cite{Yu1994,Kouba2001}. 
For codes using localized atomic orbitals, DFPT has been mainly implemented to treat finite, isolated systems~\cite{Gerratt-1967,Pople-1979,Dykstra-1984,Frisch-1990,Ochsenfeld-1997, Liang-2005}, but only a few literature
reports exist for the treatment of periodic boundary conditions with such basis sets~\cite{Hirata1998,Izmaylov2007,Pruneda2002}.
In all these cases, which only considered perturbations commensurate with the unit cell~($\Gamma$-point perturbations), the exact same \emph{reciprocal-space} formalism has been used as in the case of plane-waves. Sun and Bartlett~\cite{Sun1998} have analytically generalized the formalism to account for non-commensurate perturbations (corresponding to non-$\Gamma$ periodicity in reciprocal-space), but no practical implementation has been reported.    

\begin{figure}
\centering
\includegraphics[width=0.95\columnwidth]{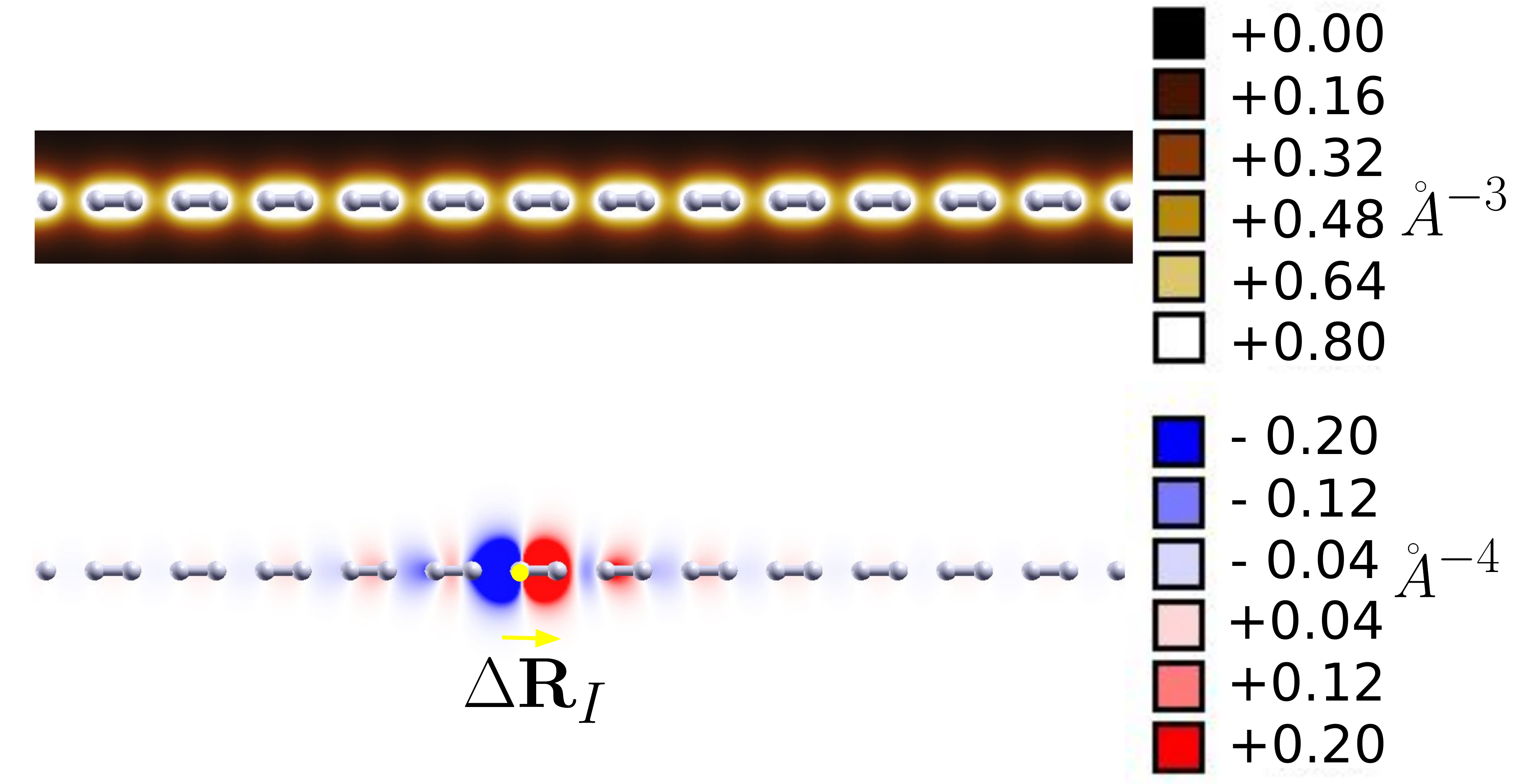}
\caption{Periodic Electronic density~$n(\vec{r})$ and spatially localized response of the electron density~$dn(\vec{R})/d\vec{R}_I$ to a perturbation viz.~displacement of atom~$\Delta \vec{R}_I$ shown exemplarily for an infinite line of H$_2$~molecules.}
\label{fig:H2_line_rho1}
\end{figure}

In the aforementioned reciprocal-space implementations, each perturbation characterized by its reciprocal-space vector~$\vec{q}$ requires an individual DFPT calculation. Accordingly, this formalism can become computationally expensive
quite rapidly, whenever the response to the perturbations is required to be known on a very tight $\vec{q}$~grid.
To overcome this computational bottleneck, various interpolation techniques have been proposed in literature: For instance,
Giustino {\it et al.}~\cite{Giustino2007} suggested to Fourier-transform the reciprocal-space electron-phonon coupling elements to real-space. The spatial localization of the perturbation in real-space~(see Fig.~\ref{fig:H2_line_rho1}) allows an accurate interpolation by using Wannier functions as a compact, intermediate representation. In turn, this then enables a back-transformation onto a dense $\vec{q}$~grid in reciprocal-space.

To our knowledge, however, no {\it real-space} DFPT formalism that {\bf directly} exploits the spatial localization of the perturbations under periodic boundary conditions has been reported in the literature, yet. This is particularly surprising,
since real-space formalisms have attracted considerable interest for standard ground-state DFT  calculations~\cite{Delley1990,Soler2002,Blum2009,GAPW,Bowler2010,Bowler2012,Mohr2014} in the last decades due to their favorable scaling with respect to the number of atoms and their potential for massively parallel implementations. Formally, one would expect a real-space DFPT formalism to exhibit similar beneficial features and thus to facilitate calculations of larger systems with less computational expense on modern multi-core architectures. 

We here derive, implement, and validate a real-space formalism for DFPT. The inspiration for this approach comes from the work of Giustino {\it et al.}~\cite{Giustino2007},
who demonstrated that Wannierization~\cite{Marzari1997} can be used to map reciprocal-space DFPT results to real-space, which in turn enables numerically efficient interpolation strategies~\cite{Giustino2007_PRL}. In contrast to these previous approaches, however, our DFPT implementation is formulated directly in real space and utilizes the exact same localized, atom-centered basis set as the underlying ground-state DFT calculations.  This allows us to exploit the  inherent locality of the basis set to  describe the spatially localized perturbations and thus to take advantage of the numerically favorable scaling of such a localized basis set.
In addition, all parts of the calculation consistently
rely on the same real-space basis set. Accordingly, all computed response 
properties are known in an accurate real-space representation from the start
and no potentially error-prone interpolation~(re-expansion) is required.
 However, this reformulation
of DFPT also gives rise to many non-trivial terms that are discussed in this paper. For instance, the fact that we utilize
atom-centered orbitals require accounting for various Pulay-type terms~\cite{Pulay1969}. Furthermore, the treatment of spatially localized perturbations that are not translationally invariant with respect to the lattice vectors requires specific adaptions of the algorithms used in ground-state DFT to compute electrostatic interactions, electronic densities, etc. We also note that the proposed approach facilitates the treatment of isolated molecules, clusters, and periodic systems on the same footing. Accordingly, we demonstrate the validity and reliability of our approach by using the proposed real-space DFPT formalism to compute the electronic response to a 
displacement of nuclei and harmonic vibrations in molecules and phonons in solids.

The remainder of the paper is organized as follows. In Sec.~\ref{sec:theory} we succinctly summarize the fundamental theoretical framework used in DFT, in DFPT, and in the evaluation of harmonic force constants. Starting from the established real-space formalism for ground-state DFT calculations, we derive the fundamental relations required to perform DFPT and lattice dynamics calculations in section ~\ref{sec:real-space DFPT}.
The practical and computational implications of these equations are then discussed in Sec.~\ref{sec:Implementation} using 
our own implementation in the all-electron, full-potential, numerical atomic orbitals based code {\it FHI-aims}~\cite{Blum2009,Havu/etal:2009,Ren/etal:2012} as an example. In Section~\ref{sec:results} we validate our method and implementation for both molecules and extended systems by comparing  vibrational and phonon frequencies computed with
DFPT to the ones computed via finite-differences. Furthermore,  we exhaustively investigate the convergence behavior with respect to the numerical parameters of the implementation~(basis set, system sizes, integration grids, etc.) and we discuss the performance and scaling with system size. Eventually, Sec.~\ref{Conclusions} summarizes the main ideas and findings of this work and highlights
possible future research directions, for which the developed formalism seems particularly promising.

\section{Fundamental Theoretical Framework}
\label{sec:theory}

\subsection{Density-functional theory}
\label{sec:DFT}
In DFT, the total energy is uniquely determined by the electron density $n(\mathbf{r})$  
\begin{equation}
E_{KS}= T_{s}[n]+E_{ext}[n]+E_{H}[n]+E_{xc}[n] + E_{ion-ion}\;,
\label{eq:KSTOT}
\end{equation}
in which $T_{s}$ is the kinetic energy of non-interacting electrons, $E_{ext}$ the electron-nuclear, $E_H$ the Hartree, $E_{xc}$ the exchange-correlation, and $E_{ion-ion}$ the ion-ion repulsion energy. All energies are functionals of the electron density. Here we avoid an explicitly spin-polarized notation, a formal generalization to collinear (scalar) spin-DFT is straightforward.

The ground state electron density~$n_0(\vec{r})$ (and the associated ground state total energy) 
is obtained by variationally minimizing Eq.~(\ref{eq:KSTOT})
\begin{equation}
\dfrac{\delta}{\delta n}\left[E_{KS}- \mu\left(\int \! \! n(\vec{r}) \: d\vec{r} - N_e\right) \right] 
 =  0 \;,
\label{eq:ks-variational}
\end{equation}
whereby the chemical potential $\mu=\delta E_{KS}/\delta n$ ensures that the number of electrons $N_e$ is conserved. This yields the Kohn-Sham single particle equations
\begin{equation}
\hat{h}_{KS}\psi_i = \left[ \hat{t}_s + \hat{v}_{ext}(r)+\hat{v}_{H}+\hat{v}_{xc}\right] \psi_i = \epsilon_{i} \psi_i \;,
\label{eq:ks-equation}
\end{equation}
for the Kohn-Sham Hamiltonian~$\hat{h}_{KS}$. In Eq.~(\ref{eq:ks-equation}) $\hat{t}_s$ is the single particle kinetic operator, $\hat{v}_{ext}$ the
(external) electron-nuclear potential, $\hat{v}_{H}$ the Hartree potential, and $\hat{v}_{xc}$ the exchange-correlation potential. Solving Eq.~(\ref{eq:ks-equation}) yields the Kohn-Sham single particle states~$\psi_i$ and their eigenenergies~$\epsilon_{i}$. The single particle states determine the electron density via
\begin{equation}
n(\mathbf{r})=\sum_i f(\epsilon_{i}) |\psi_i(r)|^2\;,
\end{equation}
in which $f(\epsilon_{i})$ denotes the Fermi-Dirac distribution function.

To solve Eq.~(\ref{eq:ks-equation}) in numerical implementations,
the Kohn-Sham states are expanded in a finite basis set~$\chi_\mu(\vec{r})$
\begin{equation}
\psi_i(\mathbf{r})=\sum_{\mu}C_{\mu i} \: \chi_{\mu}(\mathbf{r})\;,
\label{eq:expansion}
\end{equation}
using the expansion coefficients $C_{\mu i}$. In this expansion, Eq.~(\ref{eq:ks-equation}) becomes a generalized algebraic 
eigenvalue problem 
\begin{equation}
\sum_{\nu} H_{\mu\nu} C_{\nu i} = \epsilon_{i}  \sum_{\nu} S_{\mu\nu} C_{\nu i}\;.
\end{equation}
Using the bra-ket notation~$<.|.>$ for the inner product in Hilbert space,
$H_{\mu\nu}$ denotes the elements~$\braket{\chi_{\mu}|\hat{h}_{KS}|\chi_{\nu}}$ of 
the Hamiltonian matrix and $S_{\mu\nu}$ the elements~$\braket{\chi_{\mu}|\chi_{\nu}}$ of the overlap matrix.

Accordingly, the variation with respect to the density in Eq.~(\ref{eq:ks-variational})
becomes a minimization with respect to the expansion coefficients~$C_{\nu i}$
\begin{equation}
E_{tot} = E_{KS}[n_0(\vec{r})] = \min_{C_{\nu i}}\left[E_{KS}-\sum_{i}f(\epsilon_{i})\epsilon_{i}(\braket{\psi_i|\psi_i} -1)\right] \;,
\label{eq:ks-variationalC}
\end{equation}
in which the eigenstates~$\psi_i$ are constrained to be orthonormal. Typically, the ground state density~$n_0(\vec{r})$ and the associated total energy~$E_{tot}$ are determined numerically by solving Eq.~(\ref{eq:ks-variationalC}) iteratively, until self-consistency is achieved.

To determine the force~$\vec{F}_I$ acting on nucleus~$I$ at position~$\vec{R}_I$ in the electronic ground state, 
it is necessary to compute the respective gradient of the total energy,~i.e.,~its \emph{total} derivative~\cite{Matthias1985,Kohler1995,Kohler1996}
\begin{eqnarray}
\vec{F}_{I} & = & - \dfrac{d E_{tot}}{d\mathbf{R}_{I} } \nonumber \\
& = & -\dfrac{\partial{E_{tot}} }{\partial{\mathbf{R}_{I}}} - \sum_\mu \dfrac{\partial{E_{tot}} }{\partial \chi_\mu}\dfrac{\partial \chi_\mu}{\partial{\mathbf{R}_{I}}}-\sum_{\mu i}\underbrace{\dfrac{\partial E_{tot}}{\partial C_{\mu i}}}_{=0} \dfrac{\partial C_{\mu i}} {\partial{\mathbf{R}_{I}}} \;. 
\label{eq:force_derivative}
\end{eqnarray}
In Eq.~(\ref{eq:force_derivative}) we have used the notation $\partial/\partial {\mathbf{R}_{I}}$ to highlight \emph{partial} derivatives.  The
first term in Eq.~(\ref{eq:force_derivative}) describes the direct dependence of the total energy on the nuclear degrees of freedom. The second term,
the so-called {\it Pulay} term~\cite{Pulay1969},
captures the dependence of the total energy on the basis set chosen
for the expansion in Eq.~(\ref{eq:expansion}). It vanishes for a complete basis set or if the chosen basis set does not depend on the nuclear coordinates,~e.g.,~in the case of plane-waves. The last term vanishes, if Eq.~(\ref{eq:ks-variationalC}) has been variationally minimized with respect to the expansion coefficients~$C_{\mu i}$ to obtain the ground state total energy and density.
That this holds true also in practical numerical implementations is demonstrated in Sec.~\ref{sec:eq9}.

However, for higher order derivatives of the total energy,~e.g.,~the Hessian,  
\begin{eqnarray}
\dfrac{d^2 E_{tot}}{d\mathbf{R}_{I}  d\mathbf{R}_{J} } & = & -\dfrac{d }{d\mathbf{R}_{J} } \vec{F}_{I} \label{Hessian}\\
& = & -\dfrac{\partial{\vec{F}_{I}} }{\partial{\mathbf{R}_{J}}} - \sum_\mu \dfrac{\partial{\vec{F}_{I}} }{\partial \chi_\mu}\dfrac{\partial \chi_\mu}{\partial{\mathbf{R}_{J}}}-\sum_{\mu i}\underbrace{\dfrac{\partial \vec{F}_I}{\partial C_{\mu i}}}_{\neq 0} \dfrac{\partial C_{\mu i}} {\partial{\mathbf{R}_{J}}} \;,\nonumber
\end{eqnarray}
the last term no longer vanishes since the forces are \emph{not} variational with respect to the expansion coefficients~$C_{\mu i}$. Accordingly,
a calculation of the Hessian does not only require the analytical derivatives appearing in the first two terms, but also the \emph{response} of the expansion coefficients and the basis functions to a nuclear displacement ($\partial C_{\mu i} / \partial{\mathbf{R}_{J}}$ and  $\partial \chi_\mu  / \partial{\mathbf{R}_{J}}$, respectively). More generally, according to the $(2n+1)$~theorem, knowledge of the $n$-th order response~(i.e.~the $n$-th order total derivative) of the electronic structure 
with respect to a perturbation is required to determine the respective $(2n+1)$-th total derivatives of
the total energy~\cite{Gonze-1989}. These response quantities are, however, not directly accessible within DFT,
but require the application of first order perturbation theory.

\subsection{Density-functional perturbation theory}
To determine the $\partial C_{\mu i} / \partial{\mathbf{R}_{J}}$ and $\partial \chi_\mu  / \partial{\mathbf{R}_{J}}$ needed for the computation of the Hessian~(Eq.~\ref{Hessian}), we assume that the displacement from equilibrium~$\Delta \vec{R}_J$ only results in a minor perturbation~(linear response)
\begin{equation}
\hat{h}_{KS}(\Delta\vec{R}_J) = \hat{h}_{KS}^{(0)} + \underbrace{\frac{d\hat{h}_{KS}}{d\vec{R}_J}}_{\hat{h}_{KS}^{(1)}}\Delta \vec{R}_J \;,
\end{equation}
of the original Hamiltonian~$\hat{h}_{KS}^{(0)}$. We then expand the wave functions $\psi_i(\Delta\vec{R}_J)=\psi_i^{(0)} + \psi_i^{(1)}(\Delta\vec{R}_J)$ and eigenvalues $\epsilon_i(\Delta\vec{R}_J)=\epsilon_i^{(0)}+\epsilon_i^{(1)}(\Delta\vec{R}_J)$ linearly and apply the normalization condition $\braket{\psi_i(\Delta\vec{R}_J)|\psi_i(\Delta\vec{R}_J)}=1$. From the perturbed Kohn-Sham equations
\begin{equation}
\hat{h}_{KS}(\Delta\vec{R}_J) \ket{\psi_i(\Delta\vec{R}_J)} = 
\epsilon_i(\Delta\vec{R}_J) \ket{\psi_i(\Delta\vec{R}_J)} \;,
\end{equation}
we then immediately obtain the Sternheimer equation ~\cite{Sternheimer1951}
\begin{equation}
 (\hat{h}_{KS}^{(0)}-\epsilon_i^{(0)})\ket{\psi_{i}^{(1)}}
 =-(\hat{h}_{KS}^{(1)}-\epsilon_i^{(1)})\ket{\psi_{i}^{(0)}}\;.
 \label{eq:dfpt_1}
\end{equation}
The corresponding first order density is given by
\begin{equation}
n(\mathbf{r})^{(1)}=\sum_{i}f(\epsilon_{i})\left[
\psi_i^{*(0)}(\mathbf{r})\psi_i^{(1)}(\mathbf{r}) + 
\psi_i^{*(1)}(\mathbf{r})\psi_i^{(0)}(\mathbf{r})
\right]\;.
\label{eq:dfpt_3}
\end{equation}

To solve the Sternheimer equation~(Eq.~\ref{eq:dfpt_1}),
we use the DFPT formalism~\cite{Gonze1997-1,Baroni-2001} and thus the same expansion for~$\psi_i^{(1)}$ as used in Eq.~(\ref{eq:expansion}) for $\psi_i^{(0)}$, which gives
\begin{equation}
\psi_i^{(1)}(\mathbf{r})=\sum_{\mu}\left[ C^{(1)}_{\mu i} \: \chi^{(0)}_{\mu}(\mathbf{r}) + C^{(0)}_{\mu i} \: \chi^{(1)}_{\mu}(\mathbf{r}) \right ] \;.
\label{expansionDFPT}
\end{equation}
To determine the unknown coefficients~$C^{(1)}_{\mu i}$, it 
is necessary to iteratively solve Eq.~(\ref{eq:dfpt_1}) until
self-consistency is achieved. This is best done in matrix form:  
\begin{align}
 \sum_{\nu}(H_{\mu\nu}^{(0)}-\epsilon_i^{(0)} S_{ \mu\nu}^{(0)})C_{\nu i}^{(1)}    
-\sum_{\nu}{\epsilon_i^{(0)} S_{\mu\nu }^{(1)}C_{\nu i}^{(0)}}\label{eq:matrix-form}\\ \nonumber
= -\sum_{\nu}\left( H_{\mu\nu}^{(1)} - \epsilon_i^{(1)}S_{\mu\nu}^{(0)}\right) C_{\nu i}^{(0)} \;.
\end{align}
Formally, DFPT and CPSCF are equivalent and only differ in the way the first order wave function coefficients $C^{(1)}$ are obtained. In the DFPT formalism,  $C^{(1)}$ is calculated directly by solving Eq.~(\ref{eq:matrix-form}) self-consistently. In the CPSCF formalism, the coefficients~$C^{(1)}$ are further expanded in terms of the coefficients of the unperturbed system~\cite{Gerratt-1967,Pople-1979}
\begin{equation}
C^{(1)}_{\mu i}= \sum_{p}C^{(0)}_{\mu p} U^{(1)}_{pi} \;,
\label{expansionCPSCF}
\end{equation}
whereby the respective expansion~$U_{pi}$ coefficients are given by
\begin{equation}
U_{pi}=\dfrac{(C^{(0)\dagger}S^{(1)}C^{(0)}E^{(0)} - C^{(0)\dagger}H^{(1)}C^{(0)} )_{pi}}{\epsilon_p^{(0)}- \epsilon^{(0)}_i} \;.
\end{equation}
Here, the $^\dagger$ is used to denote the respective Hermitian conjugate of the matrices, and $E^{(0)}$ denotes the diagonal matrices containing the eigenvalues~$\epsilon_i$.

\begin{figure}
\includegraphics[width=0.9\columnwidth]{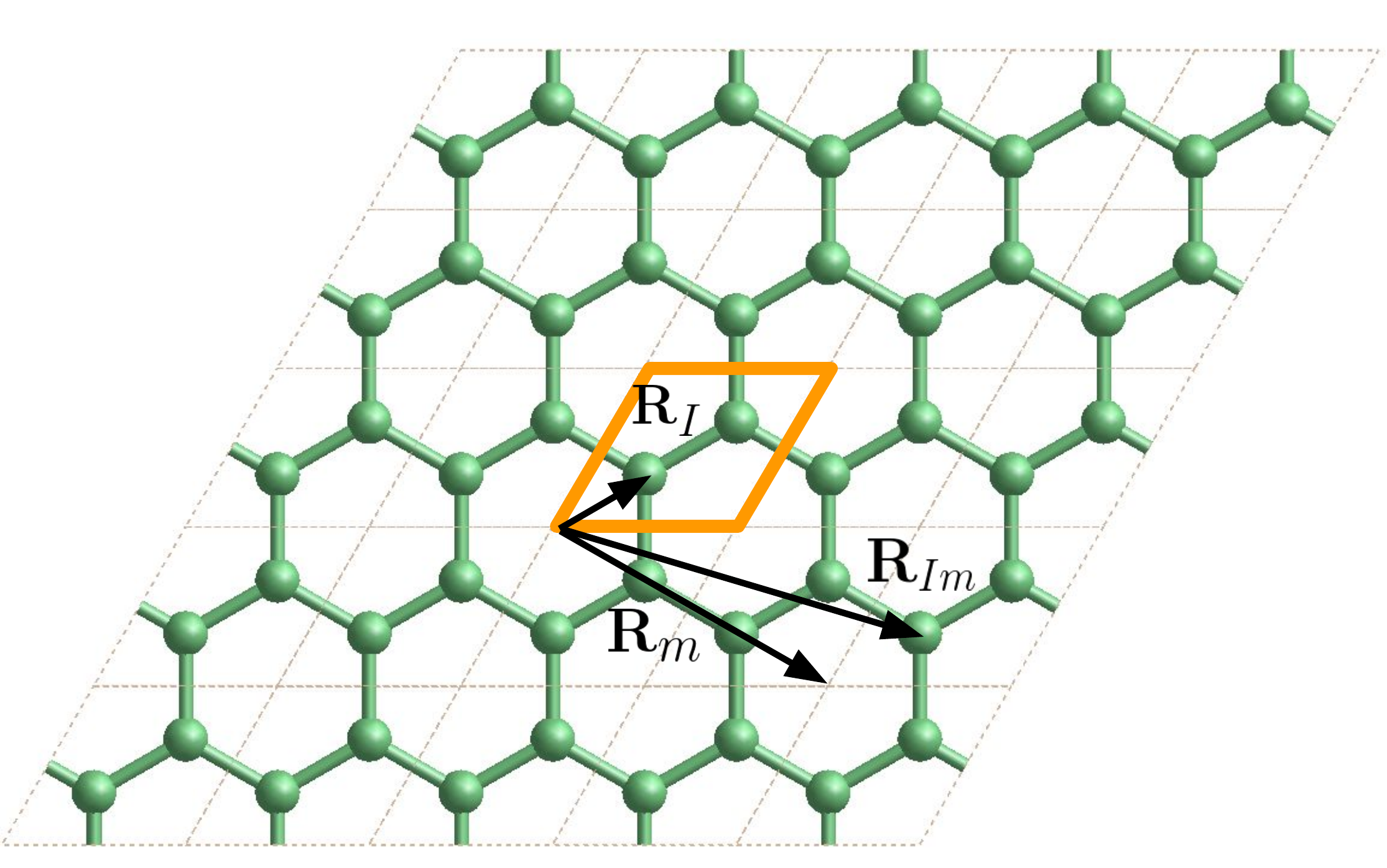}
\caption{Illustration of the atomic coordinates in the unit cell $\mathbf{R}_{I}$, its lattice vectors $\mathbf{R}_{m}$, and the atomic coordinates in a 
supercell~$\mathbf{R}_{Im}=\mathbf{R}_{m}+\mathbf{R}_{I}$.
}
\label{fig:centers}
\end{figure}

\subsection{The harmonic approximation: Molecular vibrations and phonons in solids}
\label{HARMPHON}
DFPT is probably most commonly applied to calculate molecular vibrations or phonon dispersions in solids in the harmonic approximation, although its capabilities extend much beyond this~\cite{Giustino2016RevModPhys}. Since we will later use vibrational and phonon frequencies to validate our implementation, we will now briefly present the harmonic approximation to nuclear dynamics.

To approximately describe the dynamics for a set of nuclei~$\{\vec{R}_I\}$, 
the total energy~Eq.~(\ref{eq:ks-variationalC}) is Taylor-expanded up to second order around the nuclei's equilibrium positions~$\{\vec{R}_I^0\}$ ({\it harmonic approximation})
\begin{align}
E_{tot} & \approx  E^{harm}_{tot}(\{\vec{R}_I\}) \nonumber \\
         &= E_{tot}(\{\vec{R}_I^0\}) + \frac{1}{2} \sum_{I,J} \dfrac{d^{2}E_{tot}}{d{\mathbf{R}_I}d{\mathbf{R}_J}}(\mathbf{R}_I - \mathbf{R}_I^0)(\mathbf{R}_J - \mathbf{R}_J^0) \;. 
\label{eq:Taylor}
\end{align}

The linear term in this expansion is not noted because it vanishes at the equilibrium positions. The Hessian in the second term (often referred to as {\it force constants}) can be determined with DFPT as described in the previous section. The equations of motions for the nuclei in this potential $E_{tot}^{harm}(\{\vec{R}_I\})$ are analytically solvable and yield a superposition of independent 
{\it harmonic oscillators} for the displacements from equilibrium~$\Delta \vec{R}_I(t) = \mathbf{R}_I(t) - \mathbf{R}_I^0$. 
In the complex plane, these displacements correspond to the real part of 
\begin{equation}
\Delta \vec{R}_I(t)  =  \operatorname{Re} \left( \frac{1}{\sqrt{M_I}}\sum_\lambda A_\lambda  \exp(i\omega_\lambda t) \left[\mathbf{e}_\lambda\right]_I \right) \;,
\label{eq:dynamics}
\end{equation}
in which the complex amplitudes (and phases)~$A_\lambda$ are dictated by the initial conditions;
the eigenfrequencies~$\omega_\lambda$ and the individual components~$\left[\mathbf{e}_\lambda\right]_I$ of the eigenvectors~$\mathbf{e}_\lambda$ are given by the solution of the eigenvalue problem:
\begin{equation}
\vec{D}\mathbf{e} = \omega^{2} \mathbf{e}\;,
\end{equation}
for the dynamical matrix
\begin{equation}
D_{IJ} = \dfrac{\Phi_{IJ}^{harm}}{\sqrt{M_I M_J}} =\dfrac{1}{\sqrt{M_I M_J}} \dfrac{d^{2}E_{tot}} {d{\mathbf{R}_I}d{\mathbf{R}_J}} \;.
\end{equation}

A technical complication arises for periodic solids, which are characterized by a translationally invariant unit cell defined by the lattice vectors~$\vec{a}_1$, $\vec{a}_2$, and $\vec{a}_3$. Each of the $N$ atoms~$\vec{R}_I$ in the primitive unit cell thus has an infinite number of periodic replicas
\begin{equation}
\mathbf{R}_{Im} = \mathbf{R}_I+\mathbf{R}_m,
\label{PeriodicImage}
\end{equation}
whereby $\vec{R}_m$ denotes an arbitrary linear combination of~$\vec{a}_1$, $\vec{a}_2$, and $\vec{a}_3$~(see Fig.~\ref{fig:centers}). Accordingly, also the size of the Hessian becomes in principle infinite, since also vibrations that break the perfect translational symmetry need to be accounted for.
This problem can be circumvented by transforming the harmonic force constants~$\Phi_{Im,J}^{harm}$ into reciprocal space.
Formally, this transforms this problem of infinite size into an infinite number of problems of finite size~\cite{Ashcroft1976}
\begin{align}
D_{IJ}(\mathbf{q}) &=\dfrac{1}{\sqrt{M_I M_J}} \sum_{m} \Phi_{Im,J}^{harm} \exp{\left(i\mathbf{q}\cdot\mathbf{R_m}\right)}  \nonumber \\
&=\dfrac{1}{\sqrt{M_I M_J}} \sum_{m}\dfrac{d^2 E_{tot}}{d \mathbf{R}_{Im} d \mathbf{R}_{J}  }\exp{\left(i\mathbf{q}\cdot\mathbf{R_m}\right)} \;,
\label{dynmat_FT}
\end{align}
since the finite~($3N\times 3N$) dynamical matrix~$\vec{D}(\vec{q})$ would in principle have to be determined for an infinite number of $\vec{q}$-points in the Brillouin zone. Its diagonalization would produce a set of $3N$ $\vec{q}$-dependent eigenfrequencies~$\omega_\lambda(\vec{q})$ and -vectors~$\vec{e}_\lambda(\vec{q})$. Furthermore, the displacements defined in Eq.~(\ref{eq:dynamics}) acquire an additional phase
factor:
\begin{equation}
\Delta \vec{R}_{Im}(t)  =  \operatorname{Re} \left( \frac{1}{\sqrt{M_I}} \sum_{\lambda,\vec{q}} A_\lambda(\vec{q}) e^{i\left[ \omega_\lambda(\vec{q}) t +   \mathbf{q}\cdot\mathbf{R_m}\right]}  {\left[\mathbf{e}_\lambda(\vec{q})\right]_I} \right) \; .
\end{equation}

In reciprocal-space DFPT implementations~\cite{Giannozzi1991,Gonze1997-1,Gonze1997-2,Kouba1999,Kouba2001}, perturbations that are incommensurate with the unit cell~($\vec{q}\neq 0)$ are typically directly incorporated into the DFPT formalism itself. For instance, a perturbation vector
\begin{equation}
\left[\vec{u}_{\lambda}(\mathbf{q})\right]_{Im}  =  \frac{\left[\mathbf{e}_\lambda(\vec{q})\right]_I}{\sqrt{M_I}}  \exp{\left(i\mathbf{q}\cdot\mathbf{R_m}\right)} \; .
\end{equation}
leads to a density response
\begin{equation}
n^{(1)}(\vec{r}+\vec{R}_m) = \dfrac{d n (\mathbf{r}+\vec{R}_m)}{d\vec{u}_{\lambda}(\mathbf{q})} = \dfrac{d n (\mathbf{r})}{d\vec{u}_{\lambda}(\mathbf{q})}\exp(i\mathbf{q}\mathbf{R}_m) \;,
\end{equation}
that is not commensurate with the primitive unit cell.
By adding an additional phase factor to the perturbation
\begin{equation}
\overline{\vec{u}}_{\lambda}(\mathbf{q},\vec{r})=  \vec{u}_{\lambda}(\mathbf{q}) \exp\left(-i\mathbf{q}\mathbf{r}\right)\;,
\end{equation}
the translational periodicity of the unperturbed system can be restored
\begin{equation}
\overline{n}^{(1)}(\vec{r}+\vec{R}_m) = \dfrac{d n (\mathbf{r}+\vec{R}_m)}{d\overline{\vec{u}}_{\lambda}(\mathbf{q},\vec{r})} = \dfrac{d n (\mathbf{r})}{d\overline{\vec{u}}_{\lambda}(\mathbf{q},\vec{r})} \;,
\end{equation}
so that also $\vec{q}\neq 0$ perturbations become tractable within 
the original, primitive unit cell, which is computationally advantageous. However, one DFPT calculation for each $\vec{q}$ point is required in such cases. In our implementation, we take a different route by choosing a real-space representation, as discussed in detail in the next section.

\section{DFT, DFPT, and Harmonic Lattice Dynamics in Real-space}
\label{sec:real-space DFPT}

\subsection{Total energies and forces in a real-space formalism}
 In practice, {\it FHI-aims} uses the Harris-Foulkes total energy functional~\cite{Harris1985,Foulkes1989} 
\begin{eqnarray}
\label{eq:E_KS_sc_new-functional}
E_{KS}& = & \sum_{i}{f_i \varepsilon_{i}}-\int{[n(\mathbf{r}) v_{xc}(\mathbf{r})] d\mathbf{r}} +E_{xc}(n) \\
&& -\int\left( n(\mathbf{r}) - \frac{1}{2} n^{MP}(\mathbf{r}) \right) [\sum_{I} V^{es,tot}_I(|\mathbf{r}-\mathbf{R}_{I}|)]   d\mathbf{r} \nonumber  \\
&& -\dfrac{1}{2}\sum_{I}Z_{I} [V^{es,tot}_I(0)+\sum_{J\ne I} V^{es,tot}_J(|\mathbf{R}_{J}-\mathbf{R}_{I}|) ] \nonumber \;.
\end{eqnarray}
to determine  the Kohn-Sham energy~$E_{KS}$ entering Eq.~(\ref{eq:ks-variationalC}) during the self-consistency cycles. 
Here, $v_{xc}=\frac{\delta E_{xc}}{\delta n}$ is the exchange-correlation potential and $E_{xc}[n]$ is the exchange-correlation energy.
For a fully converged density, the Harris-Foulkes formalism is equivalent to~\cite{Blum2009}
 \begin{eqnarray}
E_{KS} & = &
\sum_i \bra{\psi_i}\hat{t}_s\ket{\psi_i} + E_{xc}[n] \\
&&+\int \left( n(\mathbf{r}) - \frac{1}{2} n^{MP}(\mathbf{r}) \right) 
\left(\sum_{I} V^{es,tot}_I(|\mathbf{r}-\mathbf{R}_{I}|)\right) d\mathbf{r} \nonumber \\
&&-\dfrac{1}{2}\sum_{I}Z_{I} \left(V^{es}_I(0)+\sum_{J\ne I} V^{es,tot}_J(|\mathbf{R}_{J}-\mathbf{R}_{I}|)\right) \;. \nonumber 
\end{eqnarray}
In both Eq.~(\ref{eq:E_KS_sc_new-functional}) and here, $Z_I$~is the nuclear charge, and $n^{MP}(\mathbf{r})$ the multipole 
density obtained from partitioning the density~$n(\vec{r})$ into individual atomic multipoles to treat the electrostatic interactions in 
a computationally efficient manner. Accordingly, 
\begin{equation}
V^{es,tot}_I(\vec{r} - \vec{R}_I) = V^{es}_I(\vec{r} - \vec{R}_I) - \frac{Z_I}{|\vec{r}-\vec{R}_I|}\;,
\label{VestotI}
\end{equation}
is the full electrostatic potential stemming from atom~$I$, which includes the electronic
\begin{equation}
V^{es}(\vec{r}) = \sum_I V^{es}_I(\vec{r} - \vec{R}_I) = \int \frac{n(\vec{r}')}{|\vec{r}-\vec{r}'|}d\vec{r}' \;,
\end{equation}
 and nuclear contributions.

The respective forces
\begin{equation}
\vec{F}_I = -\dfrac{d E_{tot}}{d \mathbf{R}_{I}} = \vec{F}_I^{HF} + \vec{F}_I^{P} + \vec{F}_I^{MP} \;,
\label{forces}
\end{equation}
can be split into three individual terms. The Hellmann-Feynman force is
\begin{equation}
\vec{F}_I^{HF}= Z_{I}\left(\dfrac{\partial V^{es}_{I}(0)} {\partial \mathbf{R}_{I} } +  \sum_{J\neq I}
\dfrac{\partial V^{es,tot}_{J}(|\mathbf{R}_{I}-\mathbf{R}_{J}) |} {\partial \mathbf{R}_{I} }  \right) \;.
\label{eq:F_HF} 
\end{equation}
The Pulay force can be written as
\begin{equation}
\vec{F}_I^{P} = - 2\sum_{i,\mu,\nu} f_{i} C^{*}_{\mu i}C_{\nu i} \int  \dfrac{\partial \chi_{\mu}(\mathbf{r})}{\partial  \mathbf{R}_{I} }(\hat{h}_{ks}-\epsilon_{i}) \chi_{\nu}(\mathbf{r}) \, d\mathbf{r}    \;,
\label{eq:F_pulay} 
\end{equation}
and the force arising from the multipole correction is
\begin{equation}
\vec{F}_I^{MP}=-\int{\left(n(\mathbf{r})-n^{MP}(\mathbf{r})\right) \dfrac{\partial{ V_I^{es,tot}(\mathbf{r}-\vec{R}_I)} }
{\partial{\mathbf{R}_{I}} } d\mathbf{r} 
} \;.
\label{eq:F_multipole}
\end{equation}

\subsection{Periodic boundary condition}
\label{PBCSec}
\begin{figure}
\includegraphics[width=\columnwidth]{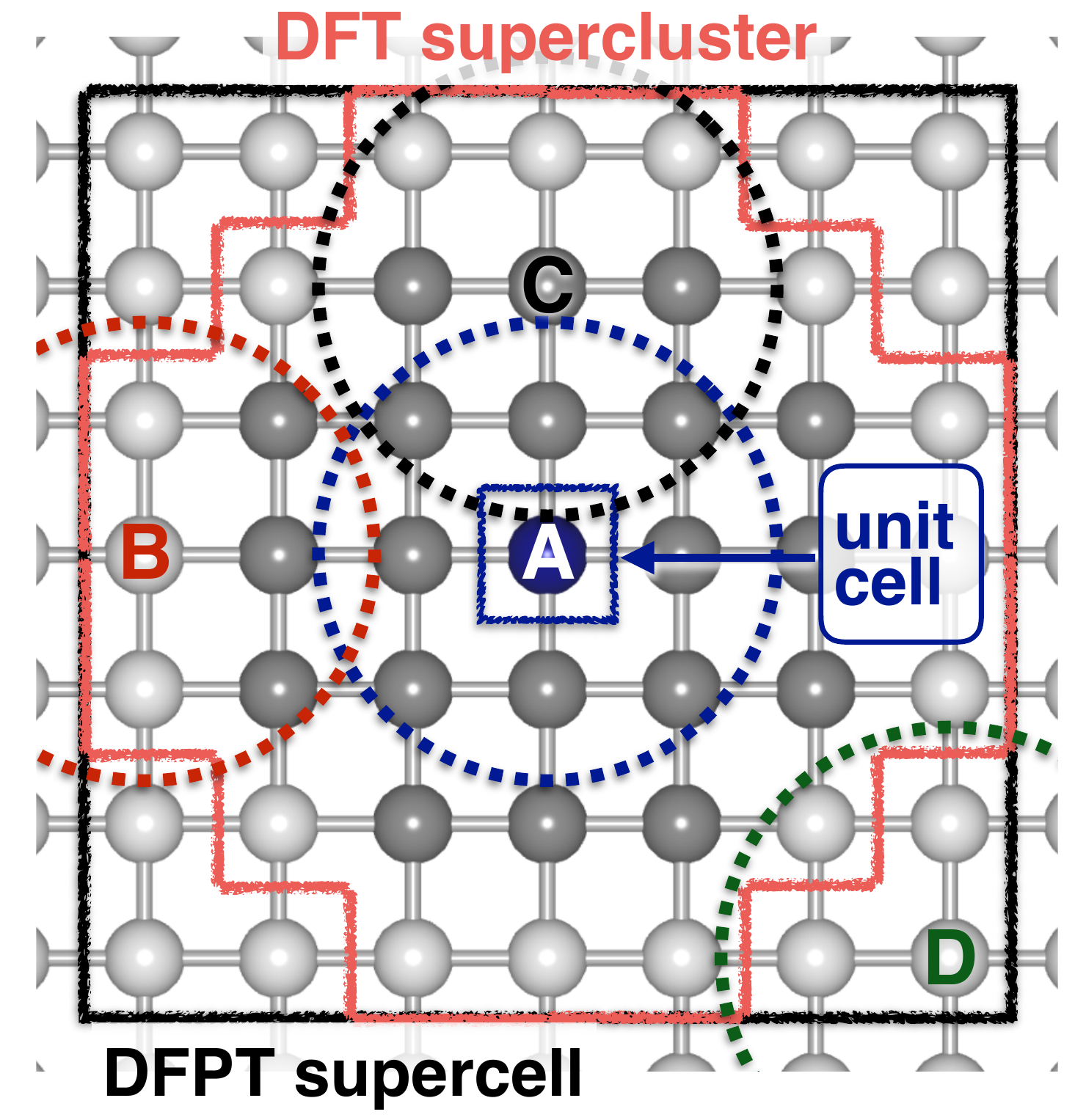}
\caption{Sketch of the real space approach for the treatment of periodic boundary conditions: The blue square indicates the unit cell, which contains one blue atom (label A). The blue dashed line shows the maximum extent of its orbitals. To treat periodic boundary conditions in DFT in real space, it is 
necessary to construct a supercluster (red solid line) which includes 
all periodic images that have non-vanishing overlap with the orbitals of the atoms in the original unit cell, as exemplarily shown here for atom A and B. In practice, it is
sufficient to carry out the integration in the unit cell alone, since translational symmetry then allows to reconstruct the full information, as discussed in more detail in Sec.~\ref{PBCSec} and~\ref{sec:Implementation}. In turn, only the dark gray atoms that have non-vanishing overlap with the unit cell need to be
accounted for in the integration, as exemplarily shown here for atom C. The DFPT supercell highlighted in black is the smallest possible supercell that encompasses the DFT supercluster and exhibits the same translational Born-von K\'arm\'an periodicity as the original unit cell. Accordingly, it contains slightly more atoms than the DFT supercluster,~e.g.,~atom D.} 
\label{fig:uc_and_sc}
\end{figure}
To treat extended systems with periodic boundary conditions in a real-space formalism, the equations for the total-energy and the forces given in the previous section need to be slightly adapted. The general idea follows this line of thought: A periodic solid is characterized by a (not-necessarily primitive) unit cell that contains atoms at the positions~$\vec{R}_I$, whereby the lattice vectors~$\vec{a}_1,\vec{a}_2,\vec{a}_3$ characterize the extent of this unit cell and impose translational invariance. To compute the properties of such a unit cell, it is not sufficient to only consider the mutual interactions between the electronic density~$n(\vec{r})$ and atoms~$\vec{R}_I$ in the unit cell, but it is also necessary to account for the interactions of the $N_{uc}$~atoms in the unit cell with the respective periodic images of the atoms~$\vec{R}_{Im}$ and of the density~$n(\vec{r}+\vec{R}_m)=n(\vec{r})$, as introduced and discussed in Eq.~(\ref{PeriodicImage}). 
 Accordingly, the double sum in Eq.~(\ref{eq:E_KS_sc_new-functional})
and the single sum in Eq.~(\ref{eq:F_HF}) become
\begin{equation}
\sum_{I}\sum_{J\neq I} \rightarrow \sum_{I}\sum_{Jm\neq I0} \qquad\text{and}\qquad \sum_{J\neq I} \rightarrow \sum_{Jm\neq I0} \;.
\end{equation}
Given that the extent of our atom-centered basis set is confined~\cite{Blum2009}, only a finite number of periodic images needs to be accounted for in these sums, since only
a finite number of periodic images feature atomic orbitals that have non-zero overlap with the orbitals of the atoms in the unit cell, as sketched in Fig.~\ref{fig:uc_and_sc}. In practical calculations, 
these periodic images are accounted for explicitly by the construction of superclusters that encompass all $N_{sc}$ atoms with non-vanishing overlap with the orbitals of the $N_{uc}$ atoms in the original unit cell~(see Fig.~\ref{fig:uc_and_sc}). 
As discussed in detail in Ref.~\cite{Blum2009,Knuth:2015kc}, also the basis set needs to be adapted to reflect the translational symmetry.  Since each local atomic orbital~$\chi_\mu(\vec{r})$ in Eq.~(\ref{eq:expansion}) is associated with an atom~$I(\mu)$, we first introduce periodic images $\chi_{\mu m}(\vec{r}) = \chi_{\mu}(\vec{r}-\vec{R}_{I(\mu)}+\vec{R}_m)$ for them as well.
Following the exact same reasoning as in Sec.~\ref{HARMPHON}, the
atomic orbitals used for the expansion of the eigenstates~(\ref{eq:expansion}) are then replaced by Bloch-like generalized basis functions 
\begin{equation}
\varphi_{\mu,\vec{k}}(\mathbf{r}) = \sum_m \chi_{\mu m}(\vec{r}) \exp{\left(- i \vec{k} \vec{R}_m\right)} \;,
\label{eq:Bloch1}
\end{equation}
so that all matrix elements~$\braket{.|.}$ become $\vec{k}$-dependent,~e.g.,
\begin{eqnarray}
H_{\mu\nu}(\vec{k}) & = & \bra{\varphi_{\mu,\vec{k}}}\hat{h}_{ks}\ket{\varphi_{\nu,\vec{k}}} \label{eq:Bloch2}\\
& = & \sum_{m,n}e^{\left(-i\vec{k} \cdot [\vec{R}_n-\vec{R}_m] \right)}\int_{uc} \chi_{\mu m}(\vec{r})\, \hat{h}_{ks} \, \chi_{\nu n}(\vec{r}) d\vec{r} \; .\nonumber
\end{eqnarray}
Please note that for practical reason the integration has been restricted  to the unit cell~(uc) in this case. To reconstruct the full information, e.g., of the $N_{uc}\times N_{sc}$ overlap matrix, the double sum and the associated phase factors run over all periodic images~$N_{sc}\times N_{sc}$, whereby only atoms with non-vanishing overlap in the unit cell contribute (see Fig.~\ref{fig:uc_and_sc} and Ref.~\cite{Knuth:2015kc}). These sums are finite, since all basis functions are bounded by a confinement potential~\cite{Blum2009}. In the expression for the Kohn-Sham energy~(Eq.~(\ref{eq:E_KS_sc_new-functional})) and the Pulay force~(Eq.~(\ref{eq:F_pulay})), the sum over electronic states now also runs over the $N_k$~$\vec{k}$-points  
\begin{equation}
\sum_i \rightarrow \frac{1}{N_k}\sum_{i,\vec{k}} \;.
\end{equation}

\begin{table}
\centering
\begin{tabular}{lc|cc}\hline\hline
System &        atoms in &  atoms in &   atoms in    \\ 
       &      unit cell   & DFT supercluster & DFPT supercell \\
 \hline
Polyethylene &  6 &   66 & 66   \\
Graphene &    2    & 200 & 242 \\
Si~(diamond)       &    2    & 368 & 686 \\
\hline\hline
\end{tabular}
\caption{Number of atoms in the unit cell and the corresponding number of atoms in the supercluster used in the ground-state DFT calculations~(atoms in the red box in Fig.~\ref{fig:uc_and_sc}) and in the DFPT supercell~(black box in Fig.~\ref{fig:uc_and_sc}). Please note that in the case of Si the increased number of atoms in the DFPT supercell  originates from the fact that in this case the circle-like DFT supercluster is encompassed by an oblique DFPT supercell with the same shape as the primitive unit cell of the diamond structure.}
\label{tab:Numberofcells}
\end{table}

Formally, the infinite periodic solid is thus treated in real-space within a finite DFT ``supercluster''~(see Fig.~\ref{fig:uc_and_sc}), which explicitly includes all periodic images~$\vec{R}_{Im}$ that have non-vanishing orbital overlap with the unit cell.  Thereby, Eq.~(\ref{eq:Bloch1}) enforces the translational symmetries to be retained. Accordingly, this real-space formalism for periodic solids leads to a notable, but reasonably tractable computational overhead for DFT calculations,~e.g.,~when comparing calculations with $N$ primitive atoms in a unit cell to calculations with $N$ atoms in an isolated molecule. This becomes immediately evident from Tab.~\ref{tab:Numberofcells}, which lists some typical supercell sizes that are used in the ground state total energy calculations at the DFT level for representative 1D, 2D, and 3D systems. 
However, the fact that the underlying DFT formalism explicitly accounts for all periodic images~$\vec{R}_{Im}$ turns out to  even be advantageous in DFPT calculations. For instance, the computation of the dynamical matrix in Eq.~(\ref{dynmat_FT}) explicitly requires the derivatives with respect to all periodic replicas~$\vec{R}_{Im}$. As discussed in detail in the Sec.~\ref{sec:real-space-FC}, the real-space formalism allows to reconstruct \emph{all} the necessary, non-vanishing elements of the Hessian that enter Eq.~(\ref{dynmat_FT}) within \emph{one} DFPT run. In turn, this allows us to \emph{exactly} compute the dynamical matrix~(Eq.~(\ref{dynmat_FT})) --~ and thus all eigenvalues~$\omega_\lambda^2(\vec{q})$ and -vectors~$\vec{e}_\lambda(\vec{q})$ ~--  at \emph{arbitrary} $\vec{q}$-points by simple Fourier transforms. 
In practice, we achieve this goal by computing the Hessian in a slightly larger Born-von K\'arm\'an~\cite{Ashcroft1976} DFPT supercell that encompasses the supercluster used for DFT ground state calculations~(cf. Fig.~\ref{fig:uc_and_sc}). By these means, the minimum image convention associated with translational symmetry can be straightforwardly exploited also in the case of perturbations 
that break the original symmetry of the crystal.

It should be noted that, for semiconductors and insulators, the size of the DFPT supercell is typically determined by the extent of the orbitals. However, for metals, this may not be enough since a large number of $\mathbf{k}$-points is required for convergence. To be consistent with this finer $\mathbf{k}$-mesh, the DFPT supercell would have to be extended to a much larger size for metals. The traditional reciprocal space approach~\cite{Gonze1997-1,Gonze1997-2,Baroni-2001} might therefore be computationally advantageous for metal. For this reason, we only apply our real-space formalism to semiconductors and insulators in the following sections.

\subsection{Real-Space force constants calculations}
\label{sec:real-space-FC}

To derive the expressions for the force constants in real-space, we will directly use the general case of periodic boundary conditions, as introduced in the previous section.  
Analogously to Eq.~(\ref{forces}) we can split the contributions to the Hessian (or to the force constants) defined in Eq.~(\ref{Hessian}) into the respective derivatives of the contributions to the force 
\begin{equation}
\Phi^{harm}_{Is,J}= \frac{d^2\vec{E}^{tot}}{d\vec{R}_{Is}d\vec{R}_{J}} 
=- \frac{d\vec{F}_{J}}{d\vec{R}_{Is}}
=- \frac{d\vec{F}_{Is}}{d\vec{R}_{J}} 
=\Phi_{Is,J}^{HF} + \Phi_{Is,J}^{P}  \;.
\label{eq:force_constants}
\end{equation}
Please note that we have omitted the multipole term here, since its contribution is already three orders of magnitude smaller at the level of the forces.

Due to the permutation symmetry~($\Phi_{Is,J}=\Phi_{J,Is}$) of the 
force constants, the order in which the derivatives are taken is irrelevant. The formulas given above for the forces~$\vec{F}_I$ acting on the atoms in the unit cell are equally valid for the forces~$\vec{F}_{Is}$ acting on its periodic images~$\vec{R}_{Is}$, 
as long as the sums and integrals in the supercell~(see Fig.~\ref{fig:uc_and_sc}) are performed using the minimum image convention. In the following, we will exploit this fact so that
only total derivatives with respect to the atoms in the primitive
unit cell need to be taken.
Consequently, the total derivative of the Hellmann-Feynman force yields 
\begin{eqnarray}
\Phi_{Is,J}^{HF} & = & -Z_{I}\left(\frac{d}{d\vec{R}_J}\dfrac{\partial V^{es}_{J}(0)} {\partial \mathbf{R}_{J} }\right) \delta_{Is,J0}\label{PhiHF}\\ && - Z_I\left(\dfrac{d}{d\vec{R}_{J}} \dfrac{\partial V^{es,tot}_{J}(|\mathbf{R}_{Is}-\mathbf{R}_{J}) |} {\partial \mathbf{R}_{Is} }  \right) \left(1-\delta_{Is,J0}\right)\;,\nonumber
\end{eqnarray}
in which $\delta_{Is,J0} = \delta_{IJ}\delta_{s0}$ denotes
a multi-index Kronecker delta. 

To determine the total derivative of the Pulay force, we
first split Eq.~(\ref{eq:F_pulay}) into two terms
\begin{eqnarray}
\vec{F}_{Is}^{P} & = & - 2 \left(\sum_{\mu m,\nu n} P_{\mu m,\nu n} \int  \dfrac{\partial \chi_{\mu m}(\mathbf{r})}{\partial  \mathbf{R}_{Is} }\, \hat{h}_{ks} \,\chi_{\nu n}(\mathbf{r}) \, d\mathbf{r} \right. \\
&& \left. - \sum_{\mu m,\nu n} W_{\mu m,\nu n} \int  \dfrac{\partial \chi_{\mu m}(\mathbf{r})}{\partial  \mathbf{R}_{Is} }\chi_{\nu n}(\mathbf{r}) \, d\mathbf{r}
\right)\;,
\end{eqnarray}
using the density matrix
\begin{equation}
P_{\mu m,\nu n}=\dfrac{1}{N_k}\sum_{i,\vec{k}}{f(\epsilon_{i}) C^{*}_{\mu i}(\mathbf{k})C_{\nu i} (\mathbf{k}) \exp{\left(i\vec{k}\cdot\left[\vec{R}_m-\vec{R}_n\right]\right)}}\;,
\label{eq:DM0-k}
\end{equation}
and the energy weighted density matrix
\begin{equation}
W_{\mu m,\nu n}=\dfrac{1}{N_k}\sum_{i,\vec{k}}{f(\epsilon_{i}) \epsilon_i(\vec{k}) C^{*}_{\mu i}(\mathbf{k})C_{\nu i} (\mathbf{k}) \exp{\left(i\vec{k}\cdot\left[\vec{R}_m-\vec{R}_n\right]\right)}} \;,
\label{eq:EDM0-k}
\end{equation}
which also incorporate the phase factors arising due to periodic boundary conditions. 
Using this notation, the total derivative of the Pulay term can be split into four terms for
the sake of readability:
\begin{equation}
\Phi_{Is,J}^{P} = \Phi_{Is,J}^{P-P} + \Phi_{Is,J}^{P-H} + \Phi_{Is,J}^{P-W} + \Phi_{Is,J}^{P-S}\;.
\end{equation}

The first term
\begin{equation}
\Phi_{Is,J}^{P-P} =  {2} \sum_{\mu m,\nu n} 
\left( \frac{d P_{\mu m,\nu n}}{d \vec{R}_{J}}\right)
\int  \dfrac{\partial \chi_{\mu m}(\mathbf{r})}{\partial  \mathbf{R}_{Is} }\hat{h}_{ks} \chi_{\nu n}(\mathbf{r}) \, d\mathbf{r}\;,
\end{equation}
account for the response of the density matrix~$P_{\mu m,\nu n}$. The second term 
\begin{eqnarray}
\Phi_{Is,J}^{P-H} & = &  
2\sum_{\mu m,\nu n}  
P_{\mu m,\nu n} \cdot\\ 
&&\left(
\int  \dfrac{\partial^2 \chi_{\mu m}(\mathbf{r})}{\partial  \mathbf{R}_{Is}\partial  \mathbf{R}_{J} }\,\hat{h}_{ks}\, \chi_{\nu n}(\mathbf{r}) \, d\mathbf{r}\right.\\
&&+\int  \dfrac{\partial \chi_{\mu m}(\mathbf{r})}{\partial  \mathbf{R}_{Is} }\frac{d \hat{h}_{ks}}{d\vec{R}_{J}}\chi_{\nu n}(\mathbf{r}) \, d\mathbf{r}\\
&&\left.+\int  \dfrac{\partial \chi_{\mu m}(\mathbf{r})}{\partial  \mathbf{R}_{Is} }\hat{h}_{ks}\frac{\partial \chi_{\nu n}(\mathbf{r})}{\partial \vec{R}_{J}} \, d\mathbf{r}
\right)\;,
\end{eqnarray}
account for the response of the Hamiltonian~$\hat{h}_{ks}(\vec{k})$, while the third  and fourth term
\begin{eqnarray}
\Phi_{Is,J}^{P-W} & = &  - 2\sum_{\mu m,\nu n }
\frac{d W_{\mu m,\nu n}}{d\vec{R}_{J}}
\int  \dfrac{\partial \chi_{\mu m}(\mathbf{r})}{\partial  \mathbf{R}_{Is} }\chi_{\nu n}(\mathbf{r}) \, d\mathbf{r}\;,
\label{PhiPW}\\
\Phi_{Is,J}^{P-S} & = &  - 2\sum_{\mu m,\nu n } 
W_{\mu m,\nu n} \dfrac{\partial}{\partial\vec{R}_{J} }
\int  \dfrac{\partial \chi_{\mu m}(\mathbf{r})}{\partial  \mathbf{R}_{Is} }\chi_{\nu n}(\mathbf{r}) \, d\mathbf{r} \;,\label{PhiPS}
\end{eqnarray}
account for the response of the energy weighted density matrix~$W_{\mu m,\nu n}$ and the overlap matrix~$S_{\mu m,\nu n}$, 
respectively~(cf. Sec.~\ref{RespS})
Please note that in all four contributions many terms vanish due to the fact that the localized atomic orbitals~$\chi_{\mu m}(\vec{r})$ are associated with one specific atom/periodic image~$\vec{R}_{J(\mu) m}$, which implies,~e.g.,
\begin{equation}
\frac{\partial \chi_{\mu m}(\vec{r})}{\partial \vec{R}_{Is}} = 
\frac{\partial \chi_{\mu m}(\vec{r})}{\partial \vec{R}_{Is}} \delta_{J(\mu)m,Is} \; .
\end{equation}

This allows us to re-index the sums over~$(\mu m,\nu n)$ in a computationally efficient, sparse matrix formalism~(cf. Ref.~\cite{sparse_matrix}). Similarly, it is important to
realize that all partial derivatives that appear in the force constants can be readily computed numerically, since the $\chi_{\mu m}$ are numeric atomic orbitals, which are defined using a splined radial function and spherical harmonics for the angular dependence~\cite{Blum2009}.

\section{Details of the Implementation}
\label{sec:Implementation}

\begin{figure}
\centering
 \includegraphics[width=0.8\columnwidth]{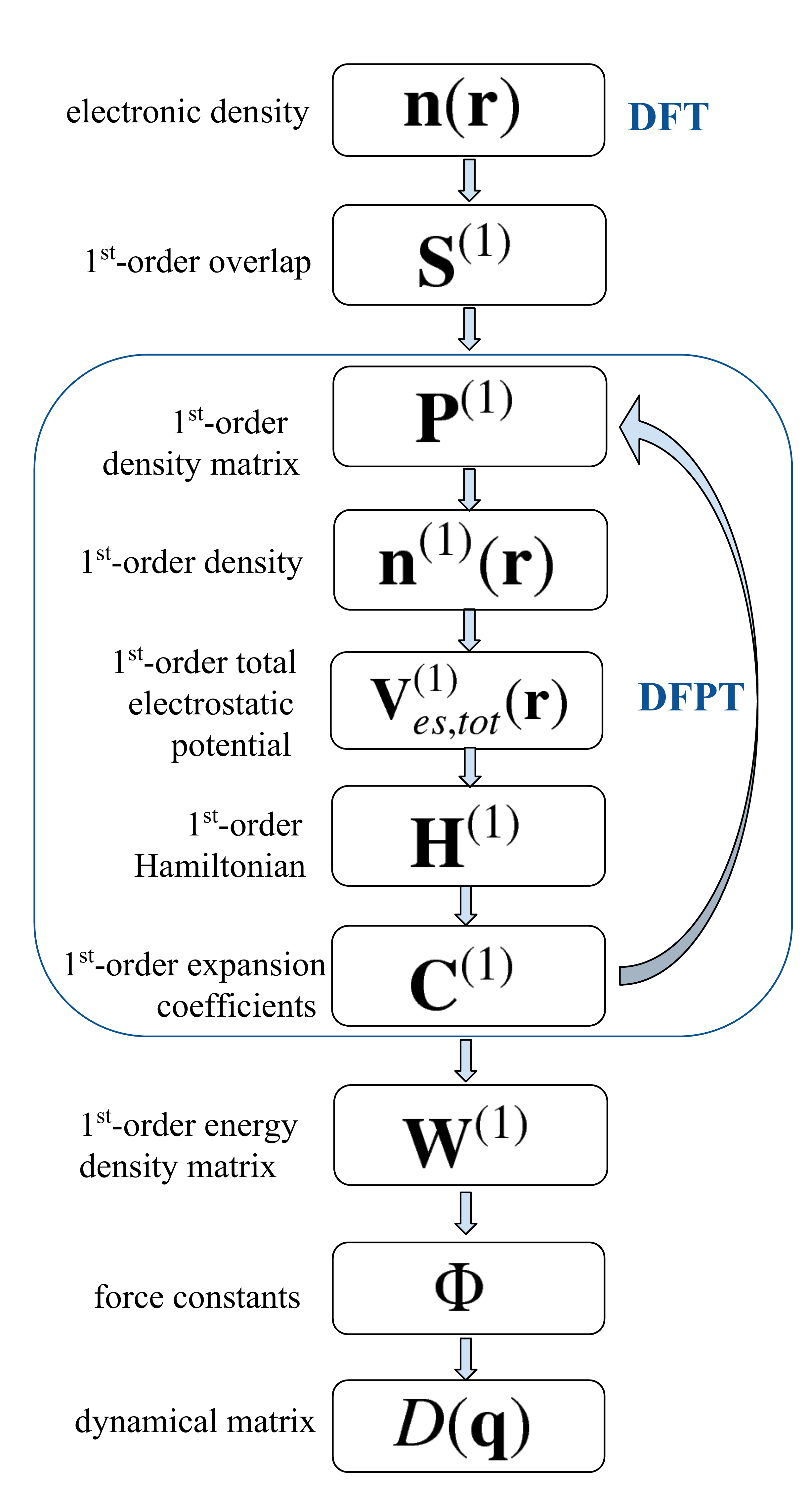}
 \caption{Flowchart of the lattice dynamics implementation using a real-space DFPT formalism.}
 \label{fig:DFPT_flowchart}
\end{figure}

The practical implementation of the described formalism closely follows the flowchart shown in
Fig.~\ref{fig:DFPT_flowchart}. For the sake of readability we use the notation
\begin{equation}
M^{(1)}=\dfrac{d {M^{(0)}}}{d {\mathbf{R}_{Is}}}\;,
\end{equation}
to highlight that in each step of the flowchart a loop over all 
atoms in the unit cell~$\vec{R}_I$ viz.~all periodic replicas~$\vec{R}_{Is}$ is performed to compute all associated derivatives. In the following chapters, we will use subscripts $i,j$ for occupied KS orbitals in the DFPT supercell, and $a$ for the corresponding unoccupied (virtual) KS orbitals, and $p,q$ for the entire set of KS orbitals in the DFPT supercell.

After the ground state calculation (see Sec.~\ref{sec:DFT} and Ref.~\cite{Blum2009}) is completed, 
the first step is to compute the response of the overlap matrix~$S^{(1)}$. We then use 
$U_{ai}^{(1)}=0$ (\ref{sec:DM1_derivation}) as the initial guess for the response of the expansion coefficients and determine the
response of the density matrix~$P^{(1)}$, which then allows to construct the respective density~$n^{(1)}(\vec{r})$. 
Using that, we compute the associated response of the electrostatic potential 
and of the Hamiltonian~$\hat{h}_{KS}^{(1)}$. In turn, all these ingredients then allow
to set up the Sternheimer equation, the solution of which  
allows to update the response of the expansion coefficients~$C^{(1)}$. Using a linear mixing
scheme, we iteratively restart the DFPT loop until self-consistency is reached,~i.e.,~until
the changes in~$C^{(1)}$ become smaller than a user-given threshold. In the last steps,
the response of the energy weighted density matrix~$W^{(1)}$, the force-constants~$\Phi_{Im,J}$, and
the dynamical matrix~$\vec{D}(\vec{q})$ are computed and diagonalized on user-specified paths and 
grids in reciprocal space.

\subsection{Response and Hessian of the Overlap Matrix}
\label{RespS}
\begin{figure}
\centering
\includegraphics[width=0.9\columnwidth]{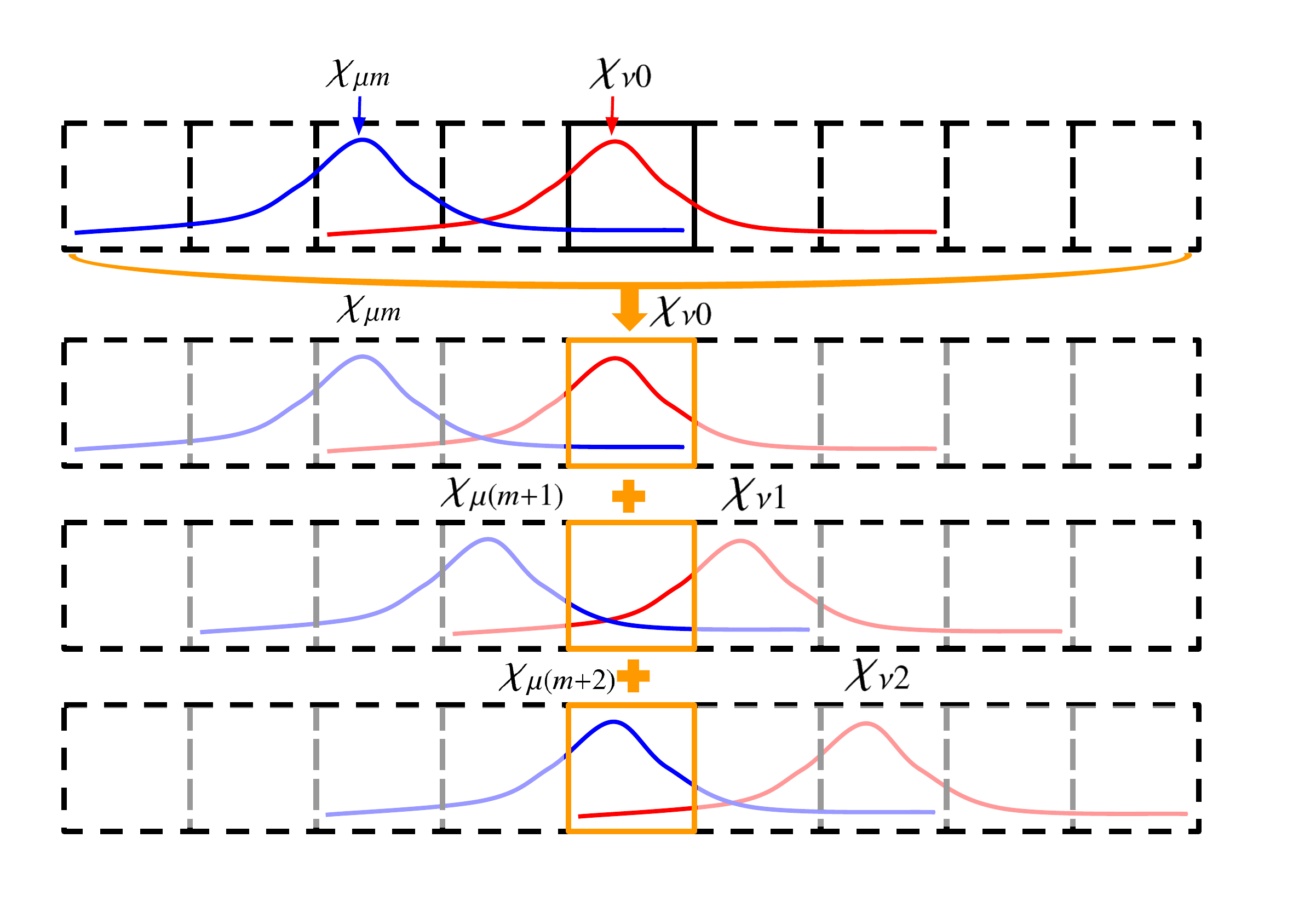}
\caption{Integration strategy for the computation of matrix elements, here shown exemplarily for the overlap matrix elements, see Eq.~(\ref{eq:S0}). Instead of integrating over the whole space, the integration is restricted to the unit cell and the individual contributions arising from translated basis function pairs are summed up. }
\label{fig:PBC_S0}
\end{figure}

The first step after completing the ground state DFT calculation is to compute
the first order response of the overlap matrix, a quantity that is not required in plane-wave 
implementations, but that needs to be accounted for when using localized atomic orbitals~\cite{Pulay1969}.
Using the definition of the overlap matrix $S$ given in Eq.~(\ref{eq:S0}), it becomes clear
that the individual elements are related by translational symmetry
\begin{eqnarray}
S_{\mu m, \nu n}^{(0)} & = & \int \chi_{\mu m}(\mathbf{r}) \chi_{\nu n}(\mathbf{r}) d\mathbf{r} =  S_{\mu (m-n), \nu 0}^{(0)} \; .
\end{eqnarray}
Therefore, it is possible to restrict the integration to the unit cell~(uc)
\begin{eqnarray}
S_{\mu m, \nu 0}^{(0)} & = & \sum_{n}\int_{uc} \chi_{\mu (m+n)}(\mathbf{r}) \chi_{\nu n}(\mathbf{r})  d\mathbf{r} \;,
\label{eq:S0}
\end{eqnarray}
and to reconstruct the whole integral by summing over all periodic replicas~$n$, as illustrated in Fig.~ \ref{fig:PBC_S0}.

\begin{figure}
\centering
\includegraphics[width=0.9\columnwidth]{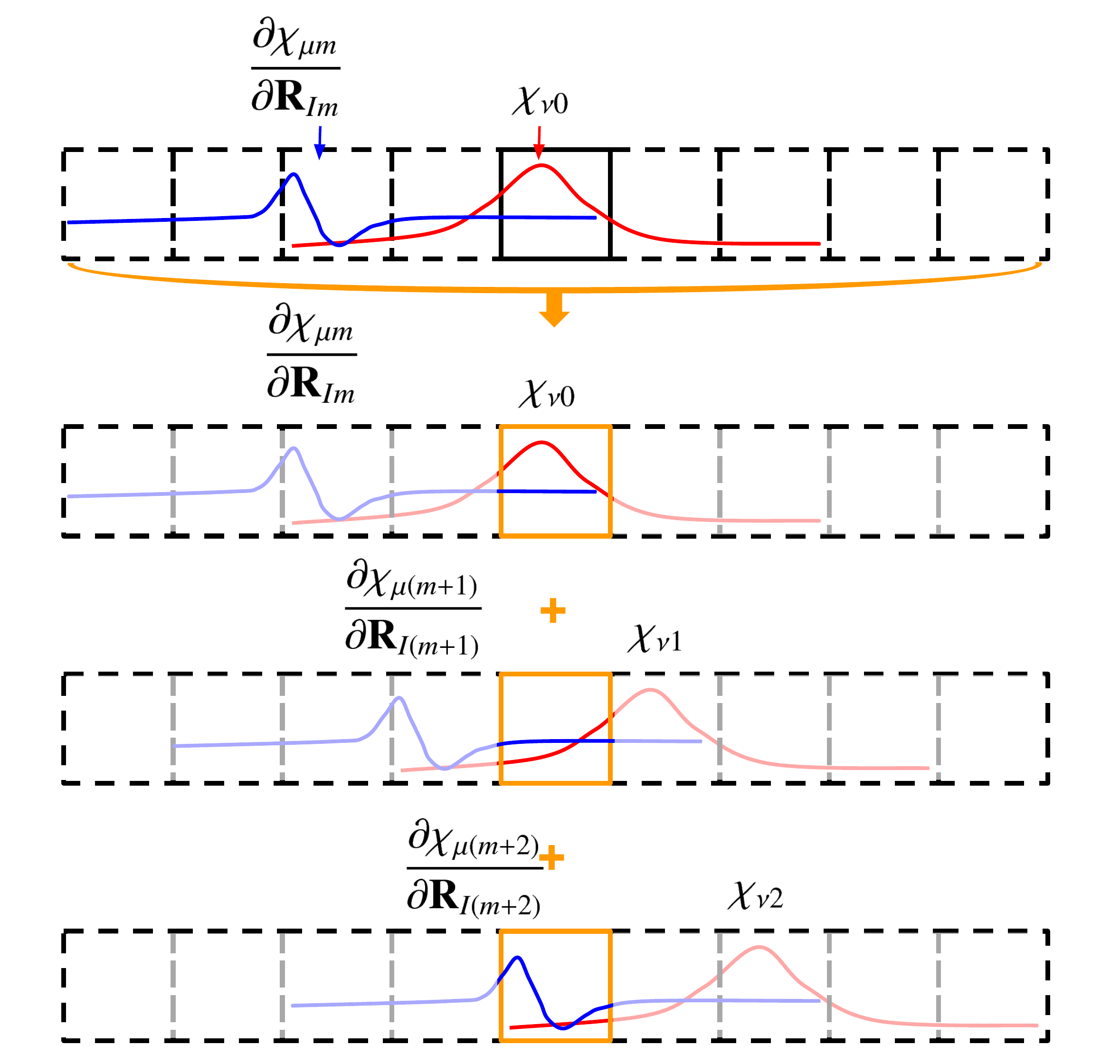}
\caption{Integration strategy for the computation of the response matrix elements, here shown for the first order overlap matrix~$S^{(1)}$~in Eq.~(\ref{eq:S1}). Please note that to be able to restrict the integration to the unit cell, the derivative has to be translated together with the orbital as shown in Eq.~(\ref{TransS1}).}
\label{fig:PBC_S1}
\end{figure}

For the response of the overlap matrix, translational symmetry 
\begin{equation}
S_{\mu m, \nu n }^{(1)} =  \dfrac{\partial{S_{\mu m, \nu n}^{(0)}}}{\partial{\mathbf{R}_{Is}}} = \dfrac{\partial{S_{\mu (m-n), \nu 0 }^{(0)}}}{\partial{\mathbf{R}_{I(s-n)}}}\;,
\label{TransS1}
\end{equation}
enables us again to restrict the integration to the unit cell
\begin{eqnarray}
\label{eq:S1}
S_{\mu m, \nu 0}^{(1)} &=&  \sum_{n} \left( \int_{uc} \dfrac{\partial \chi_{\mu (m+n)}(\mathbf{r})}{\partial {\mathbf{R}_{I(s+n)}}} \chi_{\nu n}(\mathbf{r}) d\mathbf{r} \right. \\
                       && + \left. \int_{uc} \chi_{\mu (m+n)}(\mathbf{r})  \dfrac{\partial \chi_{\nu n}(\mathbf{r})}{\partial {\mathbf{R}_{I(s+n)}}} d\mathbf{r} \right ) \nonumber \;,
\end{eqnarray}
as illustrated in Fig. \ref{fig:PBC_S1}. Please note that 
only very few non-vanishing contributions exist, since every orbital only depends on the position of one specific atom or replica
\begin{equation}
\int_{uc} \dfrac{\partial \chi_{\mu (m+n)}(\mathbf{r})}{\partial {\mathbf{R}_{I(s+n)}}} \chi_{\nu n}(\mathbf{r}) d\mathbf{r} = 
\delta_{J(\mu)m,Is}\int_{uc} \dfrac{\partial \chi_{\mu (m+n)}(\mathbf{r})}{\partial {\mathbf{R}_{I(s+n)}}} \chi_{\nu n}(\mathbf{r}) d\mathbf{r} \nonumber \;.
\end{equation}

Following the same strategy, also the second order derivatives of the overlap matrix required in Eq.~(\ref{PhiPS}) can be computed using:
\begin{align}
   &\dfrac{\partial}{\partial\vec{R}_{J} }
\int  \dfrac{\partial \chi_{\mu m}(\mathbf{r})}{\partial  \mathbf{R}_{Is} }\chi_{\nu n}(\mathbf{r}) \, d\mathbf{r} \\
  =&  \sum_{t} \left( \int_{uc} \dfrac{\partial^2 \chi_{\mu (m+t)}(\mathbf{r})}{\partial {\mathbf{R}_{I(s+t)}} \partial {\mathbf{R}_{J t}}} \chi_{\nu (n+t)}(\mathbf{r}) d\mathbf{r} \right. \nonumber  \\
   & + \left. \int_{uc}  \dfrac{\partial \chi_{\mu (m+t)}(\mathbf{r})}{\partial {\mathbf{R}_{I(s+t)}}} \dfrac{\partial \chi_{\nu (n+t)}(\mathbf{r})}{\partial {\mathbf{R}_{J t}}} d\mathbf{r} \right )  \nonumber \;.
\end{align}
Again, only a few contributions exist for the first term 
\begin{align}
\int_{uc} \dfrac{\partial^2 \chi_{\mu (m+t)}(\mathbf{r})}{\partial {\mathbf{R}_{I(s+t)}} \partial {\mathbf{R}_{J t}}} \chi_{\nu (n+t)} d\mathbf{r} =  \\
\delta_{K(\mu)m,Is}\delta_{K(\mu)m,J0} \int_{uc} \dfrac{\partial^2 \chi_{\mu (m+t)}(\mathbf{r})}{\partial {\mathbf{R}_{I(s+t)}} \partial {\mathbf{R}_{J t}}} \chi_{\nu (n+t)}  \nonumber  \;.
\end{align}
and for the second term
\begin{align}
 \int_{uc}  \dfrac{\partial \chi_{\mu (m+t)}(\mathbf{r})}{\partial {\mathbf{R}_{I(s+t)}}} \dfrac{\partial \chi_{\nu (n+t)}(\mathbf{r})}{\partial {\mathbf{R}_{J t}}} d\mathbf{r}   =  \\
\delta_{K(\mu)m,Is}\delta_{K(\nu)n,J0}   \int_{uc}  \dfrac{\partial \chi_{\mu (m+t)}(\mathbf{r})}{\partial {\mathbf{R}_{I(s+t)}}} \dfrac{\partial \chi_{\nu (n+t)}(\mathbf{r})}{\partial {\mathbf{R}_{J t}}} d\mathbf{r}  \nonumber  \;.
\end{align}

\subsection{Response of the Density Matrix}
\label{sec:DM1}
The first step in the DFPT self-consistency cycle is to calculate of the response of the density matrix 
using the given expansion coefficients~$C^{(0)}$ and~$C^{(1)}$. Using the discrete Fourier transform
\begin{equation}
C_{\mu m,i}^{(0)}=\sum_{\mathbf{k}} C_{\mu,i}^{(0)}(\mathbf{k})\exp{\left(-i\mathbf{k}\cdot\mathbf{R}_m\right)} \;,
\end{equation}
to get real-valued coefficients~$C_{\mu m,i}^{(0)}$, the  
density matrix defined in Eq.~(\ref{eq:DM0-k}) becomes:
\begin{equation}
P_{\mu m,\nu n}^{(0)}=\sum_{i}{f(\epsilon_{i}) C_{\mu m,i}^{(0)}C_{\nu n,i}^{(0)}} \;.
\end{equation}
Accordingly, its response is 
\begin{equation}
P_{\mu m,\nu n}^{(1)}=\sum_{i}{ f(\epsilon_{i})
\left(C_{\mu m,i}^{(1)} C_{\nu n,i}^{(0)} + C_{\mu m,i}^{(0)} C_{\nu n,i}^{(1)}    \right)  }  \;.
\label{eq:DM1}
\end{equation}
In the practical solution of the Sternheimer equation~(cf.~Sec.~\ref{sec:C1}), we use the CPSCF approach~(Eq.~\ref{expansionCPSCF}) and use matrix~$U^{(1)}$ 
to expand the response of the expansion coefficients~$C^{(1)}$ 
\begin{equation}
C^{(1)}=C^{(0)}U^{(1)} \;
\end{equation}
We have also solved the Sternheimer equation use DFPT approach~(Eq.~\ref{eq:matrix-form}) directly, and obtained exactly the same results as with Eq.~(\ref{expansionCPSCF}) for the systems~(e.g. molecules) discussed in this paper.  
In praxis, the density matrix can then be directly evaluated
in terms of $U^{(1)}$, as shown in~\ref{sec:DM1_derivation}.

\subsection{Response of the Electronic Density}
\label{sec:n1}
To determine the electronic density~$n(\vec{r})$, we use a
density matrix based formalism
\begin{equation}
n^{(0)}(\mathbf{r})= \sum_{\mu m,\nu n}{P_{\mu m,\nu n}^{(0)}\chi_{\mu m}^{(0)}(\mathbf{r})\chi_{\nu n}^{(0)}}(\mathbf{r}) \;.
\end{equation}
Similarly, the response of the electronic density
can thus be expressed as 
\begin{eqnarray}
n^{(1)}(\mathbf{r}) & = & %
\sum_{\mu m,\nu n}{P_{\mu m,\nu n}^{(1)}\chi_{\mu m}^{(0)}(\mathbf{r})\chi_{\nu n}^{(0)}}(\mathbf{r}) \\
&& +\sum_{\mu m,\nu n}{P_{\mu m,\nu n}^{(0)}\left(\chi_{\mu m}^{(1)}(\mathbf{r})\chi_{\nu n}^{(0)}(\mathbf{r})+\chi_{\mu m}^{(0)}(\mathbf{r})\chi_{\nu n}^{(1)}(\mathbf{r})\right)} \nonumber\;.
\end{eqnarray}
Please note that the ground state density is periodic~(translationally invariant)
\begin{equation}
n^{(0)}(\mathbf{r+R_m})=n^{(0)}(\mathbf{r}) \;,
\end{equation}
but its response is not
\begin{equation}
n^{(1)}(\mathbf{r+R_m})\neq n^{(1)}(\mathbf{r}) \;.
\end{equation}
As already discussed for the response of the overlap matrix in Sec.~\ref{RespS},
the individual contributions to the response are however related to each other
via their translation property
\begin{equation}
\dfrac{d n^{(0)}(\mathbf{r+R_m})}{d \mathbf{R}_{Is}} =
\dfrac{d n^{(0)}(\mathbf{r})}{d \mathbf{R}_{I,s-m}} \;.
\label{eq:translation}
\end{equation}

\subsection{Response of the Total Electrostatic Potential}
\label{sec:Ves1} 
In a real-space formalism~\cite{Blum2009,Delley1990} such as {\it FHI-aims} 
it is necessary to treat the electrostatic interactions (electronic Hartree potential~$v_{es}$
and nuclear external potential~$v_{ext}$ in a unified formalism~\cite{Blum2009,Knuth:2015kc}. Using 
Eq.~(\ref{VestotI}), the electrostatic potential entering the zero-order Kohn-Sham Hamiltonian~$\hat{h}_{KS}^{(0)}(\vec{k})$ is thus defined as
\begin{equation}
V_{es,tot}(\vec{r}) = \sum_{Jn} V_{Jn}^{es,tot}(\vec{r}-\vec{R}_{Jn}) \; .
\end{equation}
The contribution of each atom~$\vec{R}_{Jn}$ consists of two contributions
\begin{equation}
V_{Jn}^{es,tot} = V_{Jn}^{\text{free}}(\mathbf{r}-\vec{R}_{Jn}) + \delta V_{Jn}(\mathbf{r}-\vec{R}_{Jn}) \;.
\end{equation}
In this expression 
\begin{equation}
V_{Jn}^{\text{free}}(\mathbf{r}-\vec{R}_{Jn}) = -\dfrac{Z_I}{\mathbf{r}-\vec{R}_{Jn}}+
\int{\dfrac{n_{Jn}^{\text{free}}(\mathbf{r'}-\vec{R}_{Jn})}{|\mathbf{r}-\vec{r'}|}d \mathbf{r'}} \;.
\end{equation}
denotes the electrostatic potential associated with an isolated (``free'') atom of the same species with the electron density~$n^{\text{free}}(\mathbf{r}-\vec{R}_{Jn})$. Both the free-atom electron densities~$n^{\text{free}}(\mathbf{r}-\vec{R}_{Jn})$ and the electrostatic potential~$V_{Jn}^{\text{free}}(\mathbf{r}-\vec{R}_{Jn})$ are accurately known 
as cubic spline functions on dense grids. 
The second term in the total electrostatic potential~$V_{Jn}^{es,tot}$ is computed by partitioning~\cite{Knuth:2015kc} the difference density~$\delta n(\vec{r}) = n(\vec{r}) -\sum_{J,n} n^{\text{free} }(\mathbf{r}-\vec{R}_{Jn})$ into individual contributions $\delta_I n(\vec{r})$.
Their contribution~$\delta V_{Jn}(\mathbf{r}-\vec{R}_{Jn})$ to the translationally invariant and periodic electrostatic potential is computed using a combined multipole expansion and Ewald summation formalism proposed by Delley~\cite{Delley1990}.

As the perturbations break the local periodicity of the crystal, also, their response is localized in non-polar materials~\cite{Giustino2007}. Accordingly, no Ewald summation 
is needed for the response potential. Instead, we use a real-space multipole expansion for the computation of the first order potential~$V_{es,tot}^{(1)}(\vec{r})$.
From the given first-order density~$n^{(1)}(\vec{r})$, we first construct
\begin{eqnarray}
\delta n^{(1)}(\vec{r}) & = & n^{(1)}(\vec{r}) - \frac{d}{d\vec{R}_{Is}} \sum_{Jn} n^{\text{free}}(\mathbf{r}-\vec{R}_{Jn}) \\
                        & = & n^{(1)}(\vec{r}) - \frac{\partial }{\partial\vec{R}_{Is}}n^{\text{free}}(\mathbf{r}-\vec{R}_{Is}) \;,
\end{eqnarray}
whereby $n^{\text{free}}(\mathbf{r}-\vec{R}_{Is})$ and its first derivative is available by splines~\cite{Blum2009}. The respective
first order potential thus becomes
\begin{equation}
V_{es,tot}^{(1)}(\vec{r}) = \left( \frac{\partial }{\partial\vec{R}_{Is}}V^{\text{free}}(\mathbf{r}-\vec{R}_{Is}) \right) + \sum_{Jn} \delta V_{Jn}^{(1)}(\mathbf{r}-\vec{R}_{Jn})\;.
\end{equation}
The first term is readily accessible, given that~$V^{\text{free}}(\mathbf{r}-\vec{R}_{Is})$ is accurately known as a cubic spline. For the second term, we
first partition~$\delta n^{(1)}$ into individual contributions stemming from
the different atoms and periodic replicas~$\vec{R}_{Is}$, so we have the radial part of density:
\begin{equation}
\delta \widetilde{n}_{Jn}^{(1)lm}(r) = \int { d^2 \Omega_{J}  p_J(\mathbf{r}) \dfrac{d \delta n(\mathbf{r})}{d \mathbf{R}_{I(s+n)}}  Y^{lm}(\Omega_{J})}\;.
\end{equation}
Here the upper index~$(lm)$ refers to the quantum numbers of the spherical harmonics. The $p_J(\mathbf{r})$ are the atom-centered partition functions~\cite{Blum2009}. 
From that, we get the radial part of the electrostatic potential:
\begin{align}
\delta \widetilde{V}_{Jn}^{(1)lm}(r)=\int_{0}^{r}{
dr_{<} r_{<}^2}g_l(r_{<},r)\delta \widetilde{n}_{Jn}^{(1)lm}(r_{<})\\
+\int_{r}^{\infty}{
dr_{>} r_{>}^2}g_l(r,r_{>})\delta \widetilde{n}_{Jn}^{(1)lm}(r_{>}) \nonumber \;.
\end{align}
Here, $g_l(r_{<},r_{>})=r^{l}_{<}/r^{l+1}_{>}$ is the Green function for the unscreened Hartree potential~\cite{Blum2009}.
Then the full electrostatic potential is reassembled using
\begin{equation}
\delta V_{Jn}^{(1)}(\vec{r})=\sum_{lm}\delta \widetilde{V}_{Jn}^{(1)lm}(r)  Y^{lm}(\Omega_{J})\;,
\end{equation}
and 
\begin{equation}
\delta V_{es}^{(1)}(\vec{r}) =  \sum_{Jn} \delta V_{Jn}^{(1)}(\mathbf{r})\;.
\end{equation}

\begin{figure}
\includegraphics[width=0.95\columnwidth]{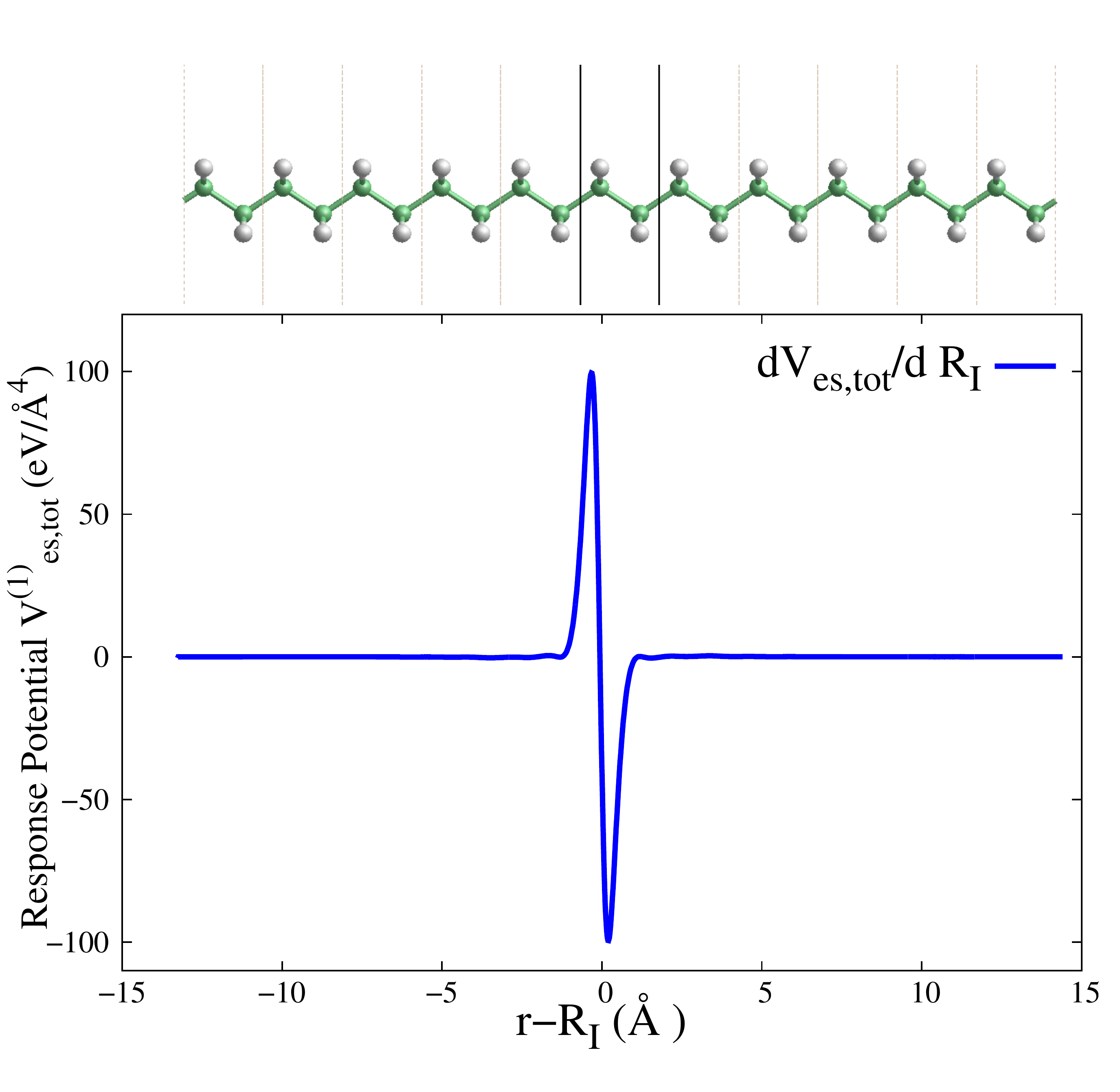}
 \caption{Response of the total electrostatic potential~$d V_{es,tot}/d\vec{R}_i$ as function of the
 distance from the perturbed nucleus~$\vec{R}_I$ in a linear polyethylene~(C$_2$H$_4$) chain. The calculation
 was performed at the LDA level of theory using fully converged numerical parameters~(cf.~Sec.~\ref{section:vibration}).
 In this non-polar system, the response of the electrostatic potential is strongly localized at the perturbation
 and thus contained in the DFPT supercell used in the calculation~(cf.~Fig.~\ref{fig:uc_and_sc} and Tab.~\ref{tab:Numberofcells}).}
 \label{fig:first_order_pot_c2h4}
\end{figure}

Please note that the chosen approach is valid to describe the electrostatics
in non-polar materials, in which the perturbation of the electrostatic  potential
is indeed spatially localized~\cite{Giustino2007}. Accordingly, it can be treated
accurately within the finite supercells used in our real-space DFPT approach~(see Sec.~\ref{sec:real-space DFPT}). 
Exemplarily, this is demonstrated in Fig.~\ref{fig:first_order_pot_c2h4} for the response of the electrostatic potential 
computed in a one-dimensional, infinite chain of polyethylene~(C$_2$H$_4$).
In polar materials, long-ranged dipole interactions can arise, which would extend beyond
the boundaries of the DFPT supercells used in the real-space formalisms. In that case, additional 
correction terms to the electrostatic perturbation potential~\cite{Verdi2015} need to be accounted for.

\subsection{Response of the Kohn-Sham Hamiltonian}
\begin{figure}
\centering
\includegraphics[width=0.8\columnwidth]{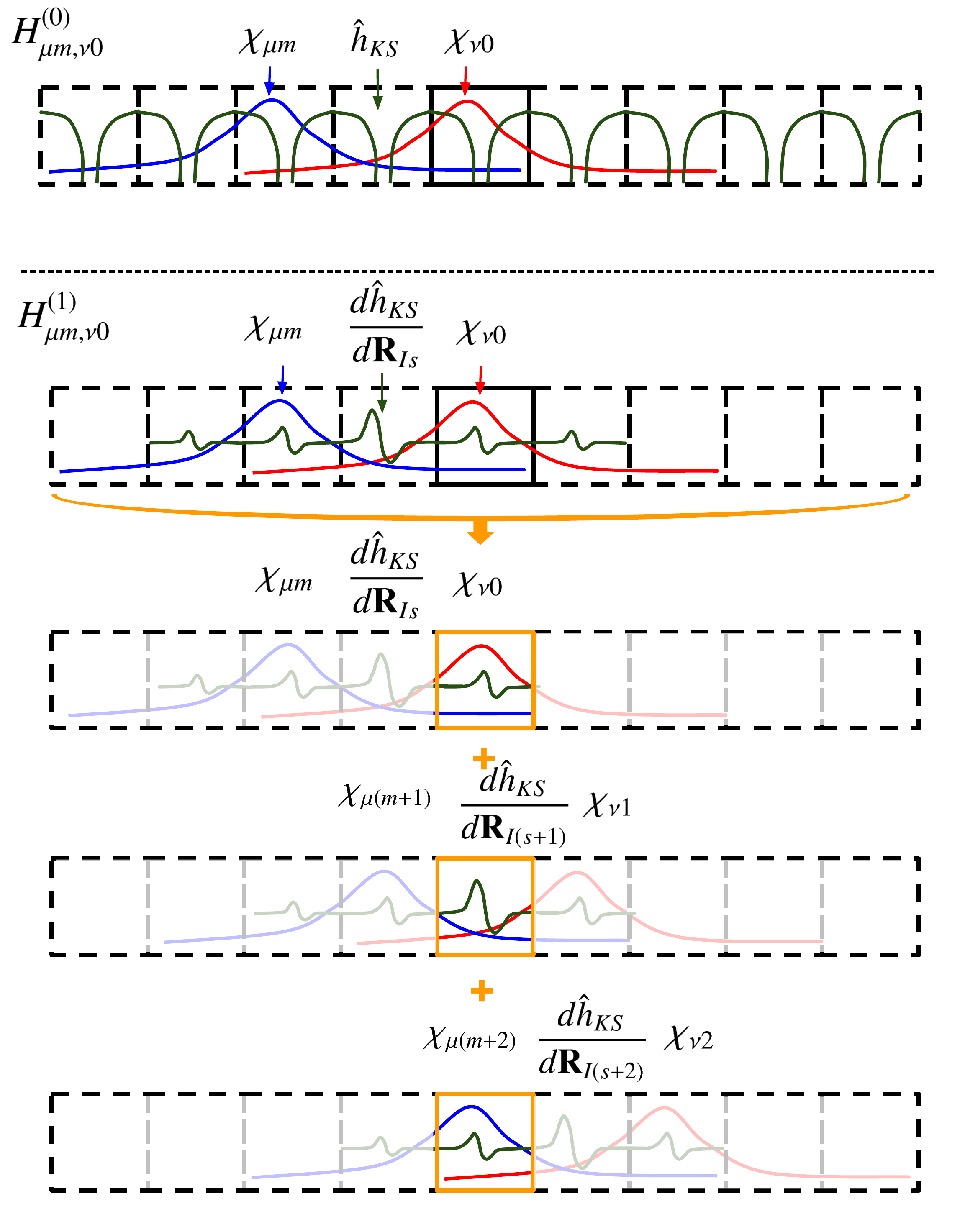}
\caption{Integration strategy for the computation of the Hamiltonian matrix elements~$H_{\mu m,\nu 0}^{(0)}$ and the response elements~$H_{\mu m,\nu 0}^{(1)}$. The first row~(a) shows the ground-state Kohn-Sham Hamiltonian, which --due to its periodicity-- can be integrated
using the exact same strategy used for the overlap matrix~$S^{(0)}$~(see Fig.~\ref{fig:PBC_S0}). The remaining rows~(b) highlight that  the response~$H_{\mu m,\nu 0}^{(1)}$ requires to account for derivatives of the Kohn-Sham Hamiltonian~$d\hat{h}_{KS}/d\vec{R}_{Is}$, which is not periodic. To restrict the integration to the unit cell, it is thus necessary to translate also this perturbation
accordingly. For this exact reason, a Born-von K\'arm\'an supercell~\cite{Ashcroft1976} supercell is needed in DFPT, but not in the case of a periodic Hamiltonian as in DFT.}
\label{fig:PBC_H1}
\end{figure}

To determine the Hamiltonian matrix and its response, we again exploit their properties
under translations already discussed for the overlap matrix in Sec.~\ref{RespS}: 
\begin{eqnarray}
\label{eq:H0}
H_{\mu m, \nu n}^{(0)} & = & \int \chi_{\mu m}(\mathbf{r}) \hat{h}_{KS}\chi_{\nu n}(\mathbf{r}) d\mathbf{r} =  H_{\mu (m-n), \nu 0}^{(0)} \;,\\
H_{\mu m, \nu 0}^{(0)} & = & \sum_{n}\int_{uc} \chi_{\mu (m+n)}(\mathbf{r}) \hat{h}_{KS}  \chi_{\nu n}(\mathbf{r}) d\mathbf{r} \;,\\
H_{\mu m, \nu n }^{(1)} & = & \dfrac{d {H_{\mu m, \nu n}^{(0)}}}{d{\mathbf{R}_{Is}}} = \dfrac{\partial{H_{\mu (m-n), \nu 0 }^{(0)}}}{\partial{\mathbf{R}_{I(s-n)}}} \;.
\end{eqnarray}
Accordingly, the response of the Hamiltonian matrix can be calculated using
\begin{eqnarray}
\label{eq:H1}
H_{\mu m, \nu 0}^{(1)} & = & \sum_{n} \left( \int_{uc} \dfrac{\partial \chi_{\mu (m+n)}(\mathbf{r})}{\partial {\mathbf{R}_{I(s+n)}}}\hat{h}_{KS} \chi_{\nu n}(\mathbf{r}) d\mathbf{r} \right.  \\
                        && + \int_{uc} \chi_{\mu (m+n)} \dfrac{d \hat{h}_{KS}}{d {\mathbf{R}_{I(s+n)}}} \chi_{\nu n}(\mathbf{r}) d\mathbf{r}                 \nonumber   \\
                       && + \left. \int_{uc} \chi_{\mu (m+n)}(\mathbf{r})\hat{h}_{KS}  \dfrac{\partial \chi_{\nu n}(\mathbf{r})}{\partial {\mathbf{R}_{I(s+n)}}} d\mathbf{r} \right ) \nonumber \;.
\end{eqnarray}
The response of the Hamiltonian operator 
\begin{equation}
\hat{h}_{KS}^{(1)}=\frac{d \hat{h}_{KS}}{d {\mathbf{R}_{Is}}}= V_{es,tot}^{(1)} + V_{xc}^{(1)} \;,
\end{equation}
includes the response of the total electrostatic potential~$V_{es,tot}^{(1)}$ discussed in the previous section and
the response of the exchange-correlation potential~$V_{xc}^{(1)}$. In the case of the LDA~\cite{Perdew/Zunger:1981,Ceperley/Alder:1980} functional considered in this work, 
evaluating the functional derivative in the latter term yields:
\begin{equation}
V_{xc}^{(1)}[n(\vec{r})] = \dfrac{\delta V_{xc}[n(\vec{r})]}{\delta n(\mathbf{r})} n^{(1)}(\mathbf{r}) \;.
\end{equation}
A sketch of the employed integration strategy to compute the response of the Hamiltonian is shown in
Fig.~\ref{fig:PBC_H1}.

\subsection{Solution of the Sternheimer Equation}
\label{sec:C1}
Using the notations introduced in this section, 
the Sternheimer equation given defined in Eq.~(\ref{eq:matrix-form})
becomes
\begin{align}
\sum_{\nu n}(H_{\mu m,\nu n}^{(0)}-\epsilon_i^{(0)} S_{\mu m,\nu n}^{(0)})C_{\nu n, i}^{(1)}-\sum_{\mu m}{\epsilon_i^{(0)} S_{\mu m,\nu n}^{(1)}C_{\nu n, i}^{(0)}}\label{eq:DFPT-LCAO}
\\
=-\sum_{\nu n}{ \left( H_{\mu m,\nu n}^{(1)} -\epsilon_i^{(1)}S_{\mu m, \nu n}^{(0)}\right)  C_{\nu n, i}^{(0)}} \;, \nonumber
\end{align}
More conveniently, it can be written in matrix form as
\begin{align}
H^{(0)}C^{(1)}-S^{(0)}C^{(1)}E^{(0)}-{S^{(1)}C^{(0)}}E^{(0)} & \\ =-  H^{(1)}C^{(0)} + S^{(0)}C^{(0)}E^{(1)}  \;, \nonumber
\end{align}
whereby $E^{(0)}$ and $E^{(1)}$ denote the diagonal matrices containing the eigenvalues~$\epsilon_i$ and their
responses respectively. By multiplying with the Hermitian conjugate~$C^{(0)\dagger}$ and by expanding 
the response~$C^{(1)}$ in terms of the zero-order expansion coefficients~$C^{(0)}$ using
\begin{equation}
C^{(1)}=C^{(0)}U^{(1)} \quad \text{i.e.} \quad C^{(1)}_{\nu n, p}=\sum_{q}{C_{\nu n, q}^{(0)}U_{qp}^{(1)}}\;, 
\end{equation}
we get
\begin{align}
\label{eq:CPSCF-U}
E^{(0)}U^{(1)}-U^{(1)}E^{(0)}- C^{(0)\dagger}S^{(1)}C^{(0)}E^{(0)} & \\ = -C^{(0)\dagger}H^{(1)}C^{(0)}+E^{(1)} \;. \nonumber
\end{align}
Thereby, we have used the orthonormality relation:
\begin{equation}
C^{(0)\dagger}S^{(0)}C^{(0)}=1 \;.
\label{eq:orthonormal}
\end{equation}
Due to the diagonal character of~$E^{(0)}$ and~$E^{(1)}$,
this matrix equation contains the response of the eiqenvalues on
its diagonal
\begin{equation}
\epsilon_p^{(1)}= \left[ C^{(0)\dagger}H^{(1)}C^{(0)}-C^{(0)\dagger}S^{(1)}C^{(0)}E^{(0)} \right]_{pp}\;.
\label{eq:epsilon1}
\end{equation}
Conversely, the off-diagonal elements determine the response of
the expansion coefficients for $p\neq q$
\begin{equation}
U_{pq}^{(1)}=\dfrac{(C^{(0)\dagger}S^{(1)}C^{(0)}E^{(0)}-C^{(0)\dagger}H^{(1)}C^{(0)})_{pq}}{(\varepsilon_{p}-\epsilon_{q})} \; . 
\label{eq:cpscf-LCAO}
\end{equation}
The orthogonality relation
\begin{equation}
\braket{\Psi_p^{(0)}|\Psi_p^{(1)}}+ \braket{\Psi_p^{(1)}|\Psi_p^{(0)}} = 0 \;,
\end{equation}
then also yields the missing diagonal elements
\begin{equation}
U^{(1)}_{pp}=-\dfrac{1}{2}\left(C^{(0)\dagger}S^{(1)}C^{(0)}\right)_{pp} \;.
\end{equation}

\subsection{Response of the Energy Weighted Density Matrix}
\label{sec:EDM1}
After achieving self-consistency in the DFPT loop, the last task is to determine the response
of the energy weighted density matrix
\begin{equation}
W_{\mu m,\nu n}^{(0)}=\sum_{i}{f(\epsilon_{i}) \epsilon_{i} C_{\mu m, i}^{(0)}C_{\nu n, i}^{(0)}} \;,
\end{equation}
that is required for the evaluation of Eq.~(\ref{PhiPW}). In close analogy to the density matrix formalism
discussed in Sec.~\ref{sec:DM1}, the response of the energy weighted density matrix can be expressed as:
\begin{equation}
W_{\mu m,\nu n}^{(1)} =  \sum_{i}f(\epsilon_{i}) \left( \epsilon_{i}^{(1)} C_{\mu m, i}^{(0)}C_{\nu n, i}^{(0)} \right. 
+ \epsilon_{i} C_{\mu m, i}^{(1)}C_{\nu n, i}^{(0)} 
\left. +\epsilon_{i} C_{\mu m, i}^{(0)}C_{\nu n, i}^{(1)} \right)\,.
\label{eq:EDM1}
\end{equation}
In close analogy to our discussion of the density matrix, 
the energy weighted density matrix is also evaluated in
practice directly in terms of~$U^{(1)}$, as detailed in~\ref{sec:EDM1_derivation}.

\subsection{Symmetry of the Force Constants}
\label{sec:ASR}
As mentioned above, the individual force constant elements are related to each other
by translational symmetry
\begin{equation}
\Phi_{Is,J0} = \Phi_{I(s+m),Jm}\;,
\end{equation}
and permutation symmetry
\begin{equation}
\Phi_{Is,Jm} = \Phi_{Jm,Is} \;.
\end{equation}
Due to these symmetries, only a subset~$N_{uc} \times N_{sc}$ of the complete $N_{sc} \times N_{sc}$~force constant matrix 
needs to be computed for a supercell containing $N_{sc}$ atoms~(see Fig.~\ref{fig:uc_and_sc} and Tab.~\ref{tab:Numberofcells}). Similarly, invariance under a complete translation of the
system implies the so called ``acoustic sum rule''
\begin{equation}
\Phi_{J0s,J0} = -\sum_{(Is)\neq(J0)}\Phi_{Is,J0} \;,
\end{equation}
which enables us to determine the entries on the diagonal~$\Phi_{J0,J0}$ from the off-diagonal elements.
For our implementation, this is computationally favorable, since no special treatment of ``on-site'' terms,~i.e.,~contributions stemming from one individual atom, is required,~e.g.,~in Eq.~(\ref{PhiHF})
or for the integration of ``on-site'' matrix elements~\cite{Knuth:2015kc}.

Please note that space and point group symmetries~\cite{Ashcroft1976}, which would allow to further reduce 
the amount of force constants that need to be computed, are not exploited in the implementation, yet.

\begin{figure}
 \centering
 \includegraphics[width=0.9\columnwidth]{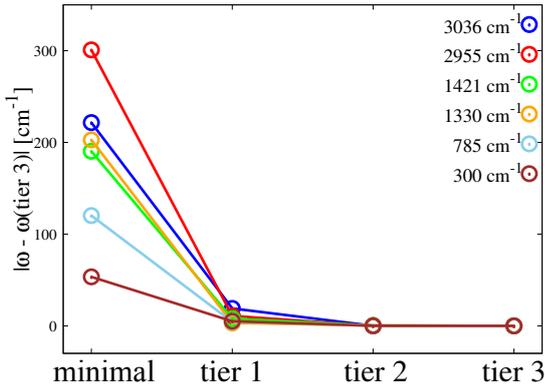}
 \caption{Convergence of the infrared-active vibrational frequencies of ethane with respect to the basis set size~(see 
 text). We use really-tight grid setting with $N_{r,mult}$=2 and $N_{ang,max}$=590. The benchmark values are calculated using ``tier 3''.}
  \label{fig:c2h6_convergence_basis_Patrick}
\end{figure}

\section{Validation and Results}
\label{sec:results}
To validate our implementation we have specifically investigated the
convergence of vibrational frequencies with respect to the numerical
parameters used in the calculation in Sec.~\ref{section:vibration}.
Furthermore, a systematic validation of the implementation by comparing
to vibrational frequencies obtained from finite-differences is presented
in Sec.~\ref{section:ValFD_Cluster}; these tests are extended
to periodic systems in Sec.~\ref{sec:phonon}. All benchmark data is available in
the NoMaD Repository (https://repository.nomad-coe.eu) via 
http://dx.doi.org/10.17172/NOMAD/2017.02.19-1. Eventually, the 
computational performance of the implementation is discussed in Sec.~\ref{Scaling}.

\subsection{Convergence with respect to Numerical Parameters}
\label{section:vibration}
First, we analyze the convergence behavior of our DFPT implementation with respect
to the numerical parameters used in the calculation,~i.e.,~the basis set size used
in the expansion~(Eq.~\ref{eq:expansion}) of the Kohn-Sham states in numerical,
atom centered orbitals and the (radial and angular) grids used for
the numerical integration. Exemplarily, we discuss these effects using the
six infrared active frequencies of ethane~(C$_2$H$_6$), which in all cases
are computed using a local approximation for exchange and correlation~(LDA 
parametrization of Perdew and Zunger ~\cite{Perdew/Zunger:1981} for the correlation 
energy density of the homogeneous electron gas based on the data of Ceperley and 
Alder~\cite{Ceperley/Alder:1980}). In all cases, the DFPT calculations were performed 
for the respective equilibrium geometry,~i.e.,~the structure obtained by relaxation~(maximum force~$<10^{-4}$~eV/$\mbox{\AA}$) using the exact same computational settings. Due to the fact that the exact same
formalism is used for both for finite systems and periodic materials, the presented
convergence studies are also valid for both cases.

Fig.~\ref{fig:c2h6_convergence_basis_Patrick} shows the absolute change in these
vibrational frequencies if the basis set size is increased.  Here, a minimal basis~(half a basis function per electron
in the spin-unpolarized case) includes the orbitals that would be occupied orbitals in a free atom following the Aufbau principle. 
Additional sets of basis functions are added in ``tier 1'', ``tier 2'',... calculations, see Ref.~\cite{Blum2009} for more details.
The vibrational frequencies converge quickly with the basis set size. Already at a ``tier 1'' level we
get qualitatively correct results with a maximal absolute/relative error of $18$~cm$^{-1}$/$0.6$~\%.
Fully quantitatively converged calculations are achieved with the ``tier 2'' basis set. 

\begin{figure}
\centering
 \includegraphics[width=0.9\columnwidth]{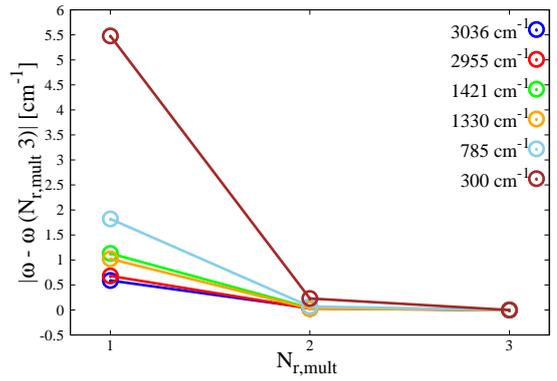}
 \caption{
 Convergence of the infrared-active vibrational frequencies of ethane with respect to the radial grid density,
 as controlled by the parameter~$N_{r,mult}$~(see text). 
 We use a ``tier 2'' basis set and  $N_{ang,max}$=590 here. The benchmark values are calculated using $N_{r,mult}$=3.}
 \label{fig:c2h6_convergence_Nr_Patrick}
\end{figure}

\begin{figure}
 \includegraphics[width=0.9\columnwidth]{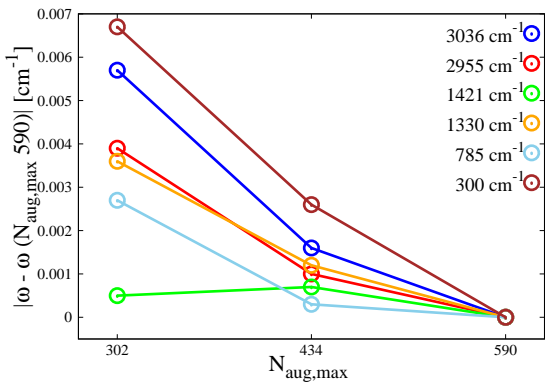}
 \caption{
 Convergence of the infrared-active vibrational frequencies of ethane with respect to the angular integration grid,
 as controlled by the parameter~$N_{ang,max}$~(see text).
 We use a ``tier 2'' basis set and $N_{r,mult}$=2 here. The benchmark values are calculated using $N_{ang,max}$=590.}
 \label{fig:c2h6_convergence_aug_Patrick}
\end{figure}

Atom-centered grids are used for the numerical integrations in {\it FHI-aims}~\cite{Blum2009}: 
Radially, each atom-centered grid consists of $N_r$ spherical integration shells, the outermost of which
lies at a distance $r_{outer}$ from the nucleus. The shell density can be controlled by means of the radial multiplier $N_{r,mult}$. For example, $N_{r,mult}$=2 results in a total of $2N_{r} +1$ radial integration shells. On these
shells, angular integration points are distributed in such a way that spherical harmonics up to a certain order are integrated exactly by the use of Lebedev grids as proposed by Delley~\cite{Delley-aug}. Here, we characterize the angular integration grids by the maximum number of angular integration points~$N_{ang,max}$ used in the
calculation. Fig.~\ref{fig:c2h6_convergence_Nr_Patrick} and  Fig.~\ref{fig:c2h6_convergence_aug_Patrick} show our convergence tests with respect to $N_{r,mult}$ and $N_{ang,max}$,
respectively. In both cases, we find that the computed vibrational frequencies depend only weakly on the chosen
integration grids: For $N_{r,mult}$, even the most sparse radial integrations grids yields qualitative and almost
quantitatively correct frequencies, since the maximum absolute and relative errors are $5.5$~cm$^{-1}$ and 
$1.8$~\%, respectively. Quantitatively converged results are achieved at the $N_{r,mult}=2$ level with absolute 
and relative errors of $0.2$~cm$^{-1}$ and $0.08$~\%. As Fig.~\ref{fig:c2h6_convergence_aug_Patrick} shows, the
vibrational frequencies are virtually unaffected by the angular integration grids; the maximum absolute 
error is always smaller than $0.01$~cm$^{-1}$.

\subsection{Validation against Finite-Differences}
\label{section:ValFD_Cluster}
\begin{figure}
\centering
\includegraphics[width=0.9\columnwidth]{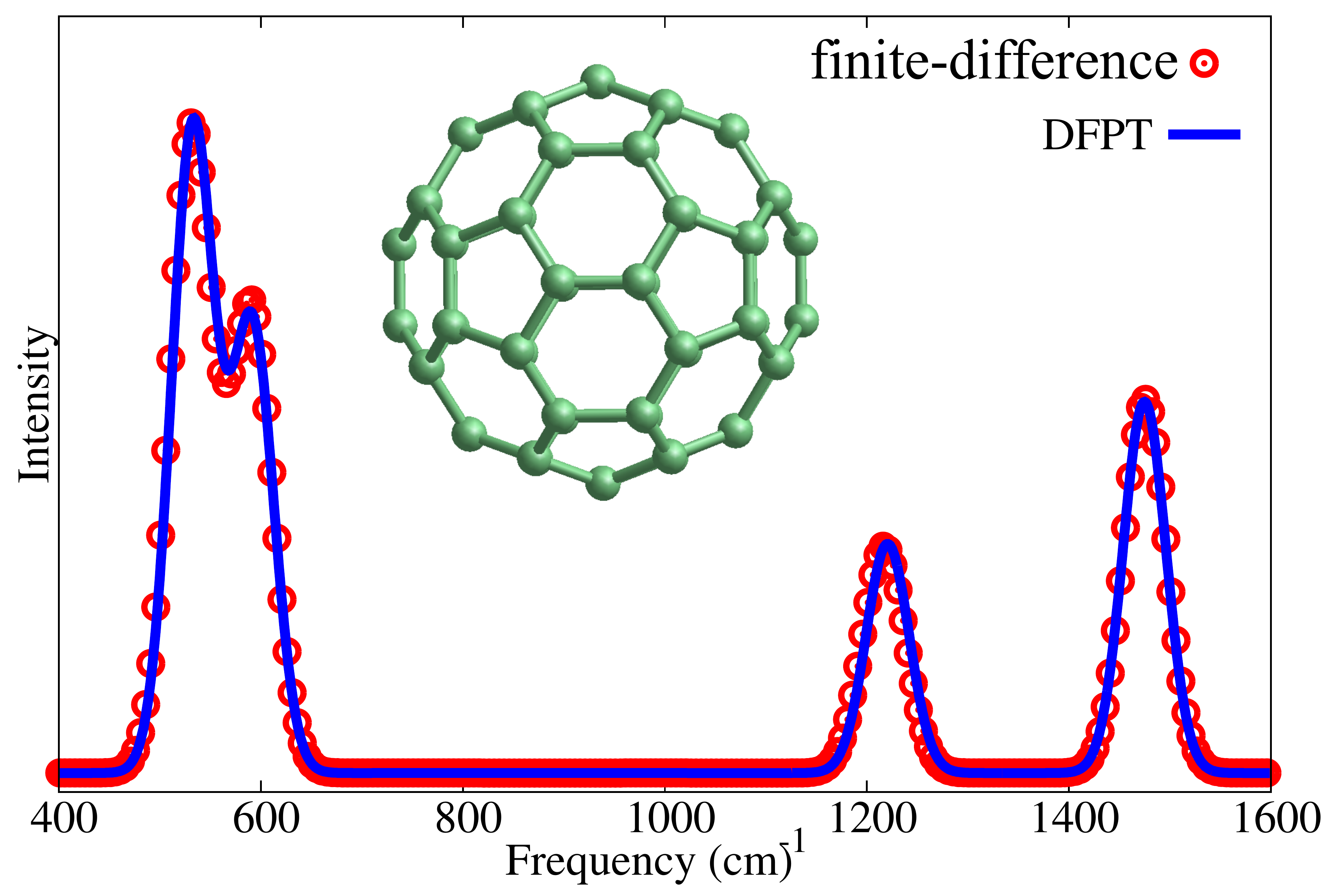}
\begin{tabular}{ c c | c c}
\hline \hline
 Frequency~(cm$^{-1}$) &  & Intensity (a.u.) &  \\
 fd     &DFPT & fd    & DFPT \\\hline 
  530.8      & 532.8 & 0.607 &  0.604\\
  591.3      & 591.6 &  0.435  & 0.417\\
 1217.0     & 1220.5&0.211  &  0.214  \\
 1475.5     &  1474.9& 0.349&   0.347 \\
\hline \hline
\end{tabular}
\caption{IR spectrum for the C$_{60}$~molecule computed at the LDA level of theory using tight grid settings, a ``tier 1'' basis set, and a Gaussian broadening of 30~cm$^{-1}$. The finite-difference (fd) and the DFPT result lie almost on top of each other, as the exact values listed in the table below substantiate.}
\label{fig:C60-IR}
\end{figure}

To validate our DFPT implementation, we have compared the obtained vibrational frequencies
to finite-difference calculations, in which the Hessian was obtained
via a first order finite-difference expression for the forces and dipole moments~(see below)
using an atomic displacement of~0.0025~{\AA}. Exemplarily, we discuss the performance of
our implementation  using the infrared~(IR) spectrum of the C$_{60}$ molecule.
The IR intensity 
\begin{align}
\label{eq:IRdef}
 I^{IR}_{\lambda}\sim \left| \sum_{I}\dfrac{\partial \vec{\mu}}{\partial \mathbf{R}_{I}} \dfrac{\mathbf{e}_{\lambda I}}{\sqrt{M_I}} \right|^2 = \left| \sum_{I} \dfrac{\mathbf{e}_{\lambda I}}{\sqrt{M_I}}\int n^{(1)}(\mathbf{r}) \,\mathbf{r}\,d\mathbf{r}\, \right|^2 \;, 
\end{align}
for a given vibrational eigenmode~$\mathbf{e}_{\lambda}$ can be computed both with finite-differences and DFPT 
by inspecting the changes induced in the dipole moment~$\vec{\mu}=\int n(\vec{r})\,\mathbf{r}\,d\mathbf{r}$ by the displacements associated with the vibrational mode~$\lambda$. As Fig.~\ref{fig:C60-IR} illustrates, both 
the IR frequencies and intensities agree very well between the finite-difference approach and our DFPT implementation.

To validate our DFPT implementation in a more systematic way, we have also compared the vibrational frequencies of 32 selected molecules with finite-difference calculations, utilizing the exact same first order finite-difference formalism used for the C$_{60}$~molecule. All calculations were performed at the LDA level of theory using fully converged numerical parameters~\footnote{So called ``tier 2'' basis sets and ``really tight'' defaults were used for the numerical settings. Additionally, we increased the order of the multipole expansion to~$l=12$ and the radial integration grid to $N_{r,mult}=4$ for all systems except LiF, NaCl, and P$_2$. An atomic displacement of~0.013~{\AA} was used in the 
finite-difference calculations.} for the equilibrium geometry determined by relaxation~(maximum force~$<10^{-4}$~eV/$\mbox{\AA}$).
A detailed list of results for these calculations is given in the~\ref{sec:32-mol-new}. For the sake of readability, 
we here only discuss the difference between the vibrational frequencies obtained via DFPT and via finite-differences, which we quantify by the mean absolute error (MAE), the maximum absolute  error (MaxAE), the mean absolute percentage error (MAPE) and the maximum absolute percentage error (MaxAPE) for each molecule. These statistical data is succinctly summarized in Tab.~\ref{tab:mol-new}: Overall, we find an excellent agreement between our DFPT implementation and the finite-difference results~(average MaxAE of 1.40~cm$^{-1}$ and average
MaxAPE of~$0.16$\%). Please note that the largest occurring absolute error~(10.13~cm$^{-1}$ in P$_2$) and the largest occurring relative error~($1.46$\% in H$_2$O$_2$) still correspond to relatively moderate relative and absolute errors~($1.26$\% and $5.73$~cm$^{-1}$, respectively). 
The occurrence of these deviations are in part caused by numerical errors,~e.g.,~the ones arising due to the moving integration grid~\cite{Blum2009} and due to the finite multipole expansion~\cite{Blum2009}~(The multipole term in force constants calculation~Eq.(\ref{eq:force_constants}) has been omitted). Such errors affect these two approaches~(finite difference and DFPT) differently. To a large extent, this is mitigated in these benchmark calculations by choosing highly-accurate settings. Still, the finite-difference reference calculations themselves exhibit a certain uncertainty, since they can be sensitive to the atomic displacement chosen for evaluating the numeric derivatives. For instance, this is the case for the P$_2$ molecule, which exhibits the largest absolute error in Tab.~\ref{tab:mol-new}.
For this reason, we have also compared our DFPT calculations with benchmark results~(Gaussian code, aug-cc-pVTZ basis set) reported in the {\it ``NIST Computational Chemistry Comparison and Benchmark Database''}~\cite{cccbdb}. For the 15 dimers contained both in Tab.~\ref{tab:mol-new} and in this database, the mean absolute percentage errors is only 0.5\%.

\begin{table}
\centering
\begin{tabular}{c| c c c c }
\hline \hline
    & MAE     & MaxAE  &   MAPE  &  MaxAPE \\
    & (cm$^{-1}$) & (cm$^{-1}$) & (\%)    & (\%) \\
\hline 
Cl$_2$ & 0.15 & 0.15 & 0.03 & 0.03 \\
ClF & 0.63 & 0.63 & 0.08 & 0.08 \\
CO & 1.42 & 1.42 & 0.07 & 0.07 \\
CS & 0.61 & 0.61 & 0.05 & 0.05 \\
F$_2$ & 0.50 & 0.50 & 0.05 & 0.05 \\
H$_2$& 2.33 & 2.33 & 0.06 & 0.06 \\
HCl & 1.22 & 1.22 & 0.04 & 0.04 \\
HF & 2.80 & 2.80 & 0.07 & 0.07 \\
Li$_2$& 0.40 & 0.40 & 0.12 & 0.12 \\
LiF & 0.32 & 0.32 & 0.03 & 0.03 \\
LiH & 0.18 & 0.18 & 0.01 & 0.01 \\
N$_2$& 1.48 & 1.48 & 0.06 & 0.06 \\
Na$_2$& 0.19 & 0.19 & 0.12 & 0.12 \\
NaCl & 0.64 & 0.64 & 0.17 & 0.17 \\
P$_2$& 10.13 & 10.13 & 1.26 & 1.26 \\
SiO & 0.50 & 0.50 & 0.04 & 0.04 \\
H$_2$O & 1.14 & 1.87 & 0.05 & 0.12 \\
SH$_2$ & 0.29 & 0.59 & 0.02 & 0.05 \\
HCN  & 0.96 & 1.40 & 0.05 & 0.04 \\
CO$_2$& 0.97 & 1.66 & 0.06 & 0.07 \\
SO$_2$ & 0.41 & 0.50 & 0.05 & 0.10 \\
C$_2$H$_2$& 0.82 & 1.47 & 0.05 & 0.04 \\
H$_2$CO& 0.47 & 0.98 & 0.03 & 0.05 \\
H$_2$O$_2$ & 1.27 & 5.73 & 0.26 & 1.46 \\
NH$_3$& 0.47 & 0.75 & 0.03 & 0.02 \\
PH$_3$& 0.18 & 0.32 & 0.01 & 0.03 \\
CH$_3$Cl& 0.35 & 0.77 & 0.02 & 0.03 \\
SiH$_4$ & 0.19 & 0.24 & 0.02 & 0.03 \\
CH$_4$ & 0.35 & 0.65 & 0.02 & 0.05 \\
N$_2$H$_4$& 0.54 & 1.05 & 0.04 & 0.15 \\
C$_2$H$_4$& 0.70 & 2.88 & 0.07 & 0.31 \\
Si$_2$H$_6$ & 0.18 & 0.62 & 0.05 & 0.45 \\
\hline
Average & 1.02  & 1.40 & 0.09 & 0.16  \\
\hline \hline
\end{tabular}
\caption{
Mean absolute error (MAE), maximum absolute error (MaxAE),  mean absolute percentage error (MAPE) and max absolute percentage error (MaxAPE) for the difference between the vibrational frequencies obtained via DFPT and via finite-differences using an atomic displacement of~0.013~{\AA} for a set of 32 molecules. All calculations are performed at the LDA level of theory with fully converged numerical settings and relaxed geometries~(see text and respective footnote). }
\label{tab:mol-new}
\end{table}

\subsection{Extended Systems: Phonons}
\label{sec:phonon}
\begin{figure}
\includegraphics[width=0.95\columnwidth]{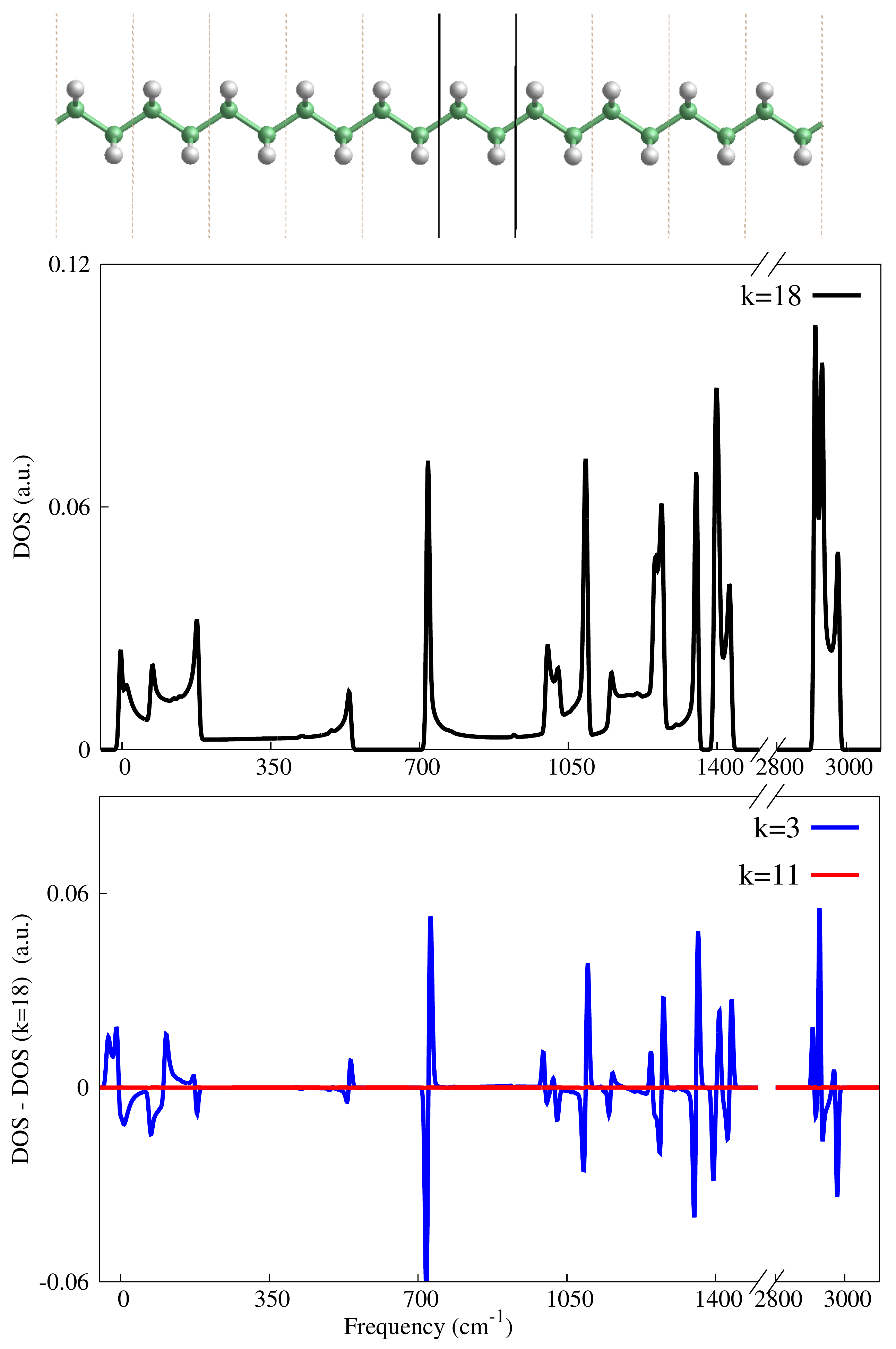}
 \caption{Convergence of the phonon density of states of polyethylene with respect to the number of $\mathbf{k}$-points
 utilized in the primitive Brillouin zone for DFPT calculations of the C$_2$H$_4$ chain. The top panel shows the density of states
 for 18~$\vec{k}$-points and the bottom panel shows the difference with respect to this converged reference. A Gaussian broadening of 5 cm$^{-1}$ and 200 $\mathbf{q}$ points was used in the computation of~$g(\omega)$. }
 \label{fig:convergence_PBC_k}
\end{figure}

\begin{figure}
\includegraphics[width=0.95\columnwidth]{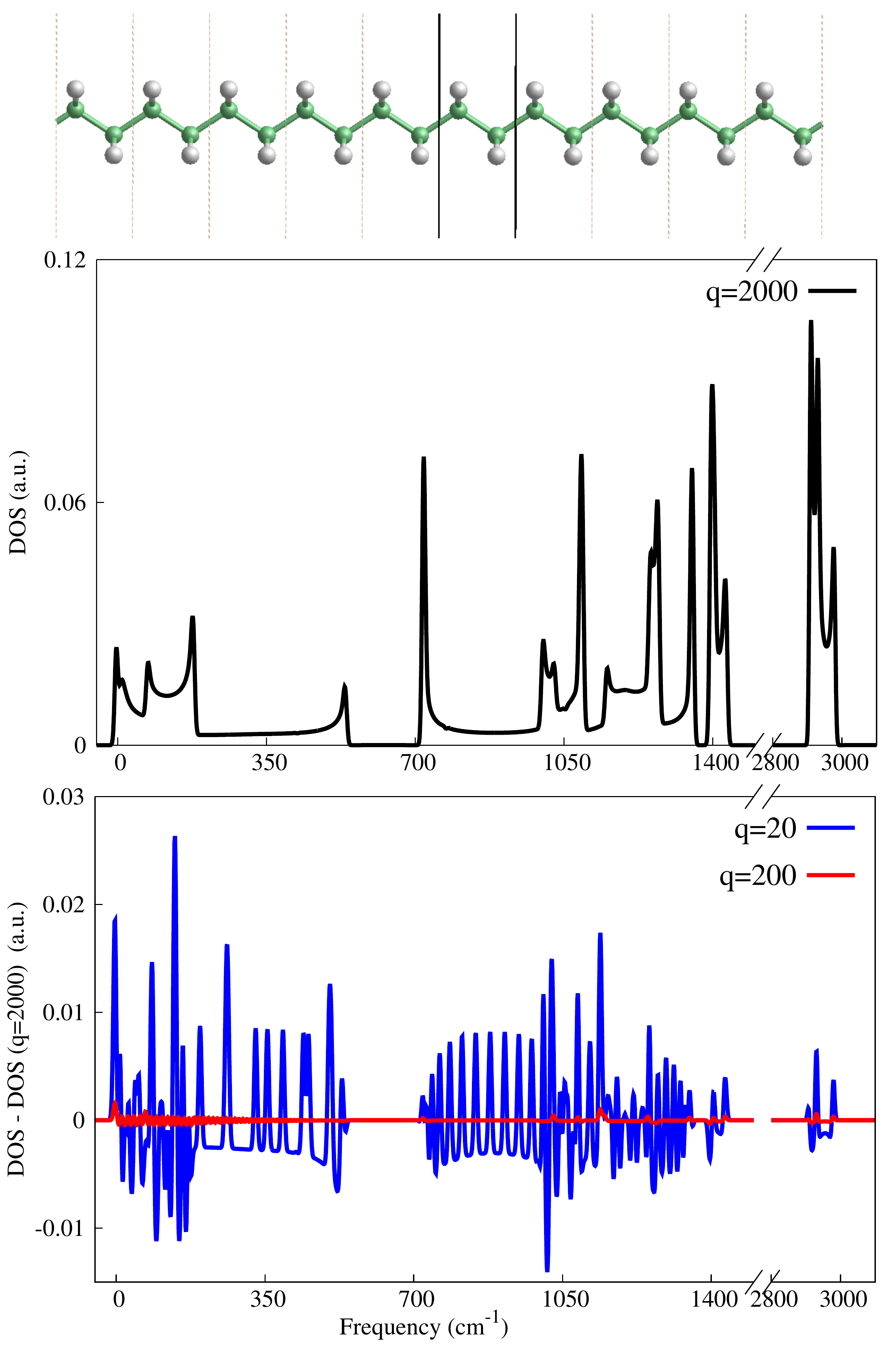}
 \caption{Same as Fig.~\ref{fig:convergence_PBC_k}, but for the convergence with respect to the number
 of $\mathbf{q}$-points in the primitive Brillouin zone. A Gaussian broadening of 5 cm$^{-1}$ is used. } 
 \label{fig:convergence_PBC_q}
\end{figure}

To showcase the ability of our implementation to treat finite systems and periodic
solids on the same footing, we compare the vibrational frequencies of various polyethylene chains~H(C$_2$H$_4$)$_n$H 
with different lengths~($n$ from 1 to 8) to the respective periodic, infinite chain of~C$_2$H$_4$.
In the latter case, we compute the vibrational/phonon density of states~(DOS) 
\begin{equation}
g(\omega)= \dfrac{1}{N_q}\sum_{\mathbf{q}}\sum_{\lambda}
\delta[\omega-\omega_{\lambda}(\mathbf{q})] \;,
\end{equation}
whereby a normalized Gaussian function with a width~$\sigma$ of 5~cm$^{-1}$
is used to approximate the Delta-distribution~$\delta[\omega-\omega_{\lambda}(\mathbf{q})]$. 
It should be noted that the phonon DOS of an infinite 
C$_2$H$_4$~chain is not zero at the $\Gamma$-point, because it is a one-dimensional system~\cite{Huang}.
All calculations have been performed for relaxed equilibrium geometries~(maximum force~$<10^{-4}$~eV/$\mbox{\AA}$) with fully converged numerical parameters,~i.e.,~using the aforementioned 
really-tight integration grids and ``tier 2'' basis sets. For the periodic chain, a reciprocal-space
grid of $11 \times 1 \times 1$ electronic $\vec{k}$-points and a grid of $200 \times 1 \times 1$ vibrational $\vec{q}$-points (in the primitive Brillouin zone) has been utilized to converge the density of states~$g(\omega)$, as substantiated in Fig.~\ref{fig:convergence_PBC_k} 
and Fig.~\ref{fig:convergence_PBC_q}. Whereas the convergence with respect to electronic $\mathbf{k}$-points is reasonably fast, a large amount of vibrational $\mathbf{q}$-points is required to sample the Brillouin zone, especially 
for the relatively moderate broadening~$\sigma$ of 5~cm$^{-1}$. In this context, it is important to realize that
the actual number of $\vec{q}$-points used is not at all computationally critical in our implementation: As discussed
in Sec.~\ref{HARMPHON}, our implementation involves determining all non-vanishing force-constants in real-space; the
respective $\vec{q}$-dependent properties can then be determined exactly by a simple Fourier transform with minimal
numerical effort. For instance, using $\vec{q}=2000$ only requires $\sim 1$~s more computational time than the $\vec{q}=20$~case. 

\begin{figure}
\includegraphics[width=1.0\columnwidth]{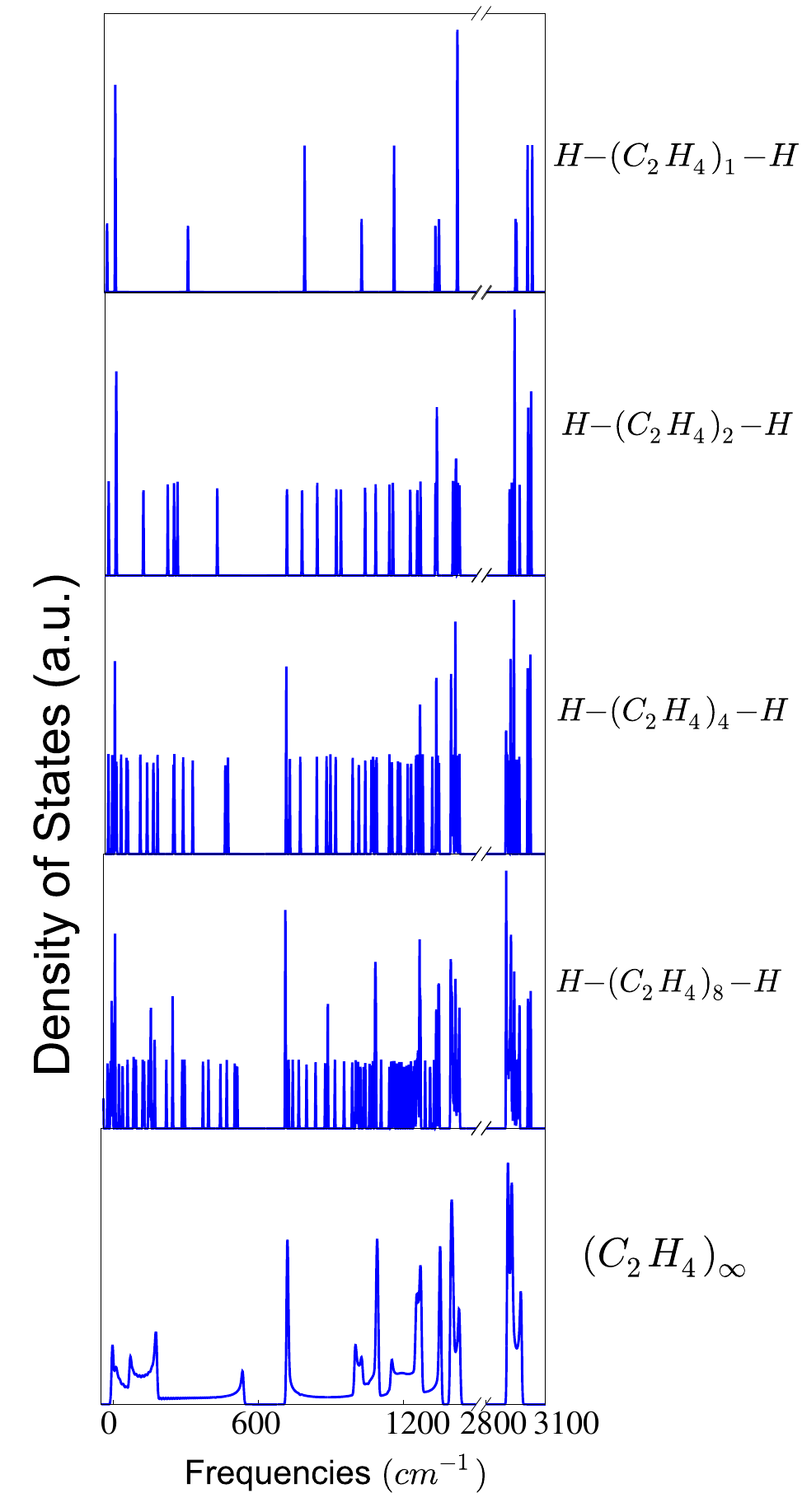}
\caption{Vibrational frequencies for increasingly longer H(C$_2$H$_4$)$_n$H chains compared to the
vibrational density of states~$g(\omega)$ of an infinite C$_2$H$_4$ chain. All calculations were performed 
using the LDA functional and with converged numerical parameters~(see text). Already for a length of~$n$=8, 
the vibrational frequencies of the isolated chain start to resemble the density of states~$g(\omega)$
of the infinitely long polyethylene chain (bottom panel).}
\label{fig:c2h4_dos}
\end{figure}

The outcome of these investigations is summarized in Fig.~\ref{fig:c2h4_dos}, in which the vibrational density of states~($\sigma$=1 cm$^{-1}$) for the isolated H(C$_2$H$_4$)$_n$H~chains with variable length~($n$ from 1 to 8) is compared to the
vibrational density of states~($\sigma$=5 cm$^{-1}$) of the extended, infinitely long  polyethylene (C$_2$H$_4$) chain.
With increasing length~$n$, the vibrational frequencies of the isolated chain start to resemble the density of states~$g(\omega)$ of the infinitely long polyethylene chain. Still, some features, e.g., the low frequency modes
that stem from long-wavelength phonons can only be correctly captured in the periodic DFPT calculation. Please note 
that the differences between the vibrational density of state of the H(C$_2$H$_4$)$_8$H~molecule~(50 atoms) and the 
C$_2$H$_4$~chain~(66 atoms in the DFPT supercell) are to a large extend not caused by the additional force-constants 
accounted for in the periodic case. Rather, the differences stem from the fact that the molecular vibrational density
of states effectively corresponds to a reciprocal-space sampling of~$\vec{q}\approx 8$, which --as Fig.~\ref{fig:convergence_PBC_q} shows-- is not sufficient to capture the contributions of long-range wavelengths to the density of states.

\begin{figure}
\includegraphics[width=1.0\columnwidth]{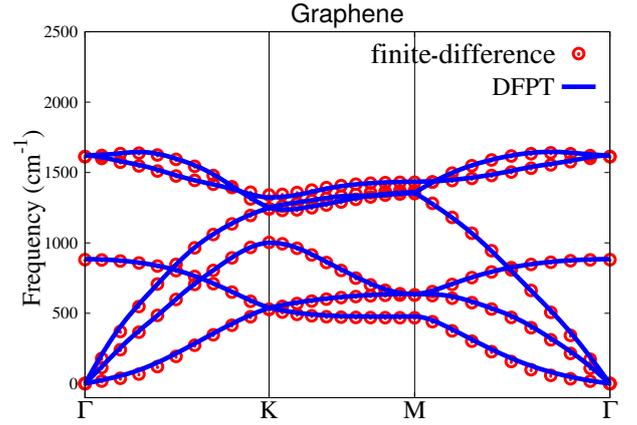}
\caption{Vibrational band structure of graphene computed at the LDA level using
both DFPT~(solid blue line) and finite-difference~(red open circles). All calculations
have been performed using a 11$\times$11$\times$1 $\mathbf{k}$-grid sampling for the 
primitive Brillouin zone, tight settings for the
integration, and a ``tier 1'' basis set.}
\label{fig:graphen}
\end{figure}

Eventually, we have also validated our real-space implementation against finite-difference calculations performed using
{\it phonopy}~\cite{Togo-2008,phonopy} for two realistic periodic systems. As a two-dimensional example, we use graphene, the vibrational properties of which have been controversially debated in the literature~\cite{Maultzsch2004,Piscanec2004}, especially
regarding the role of long-ranged interactions that are not treatable in real-space. As discussed in Sec.~\ref{sec:Ves1} already, correction terms that can account for such interactions are not yet part of the implementation discussed in this work.
To avoid possible artifacts due these effects, we have thus performed finite-difference calculations~(displacement 0.008~$\mbox{\AA}$) in the exact same $11 \times 11 \times 1$~supercell~(242 atoms) that is also inherently used 
in the DFPT calculations itself~(see Fig.~\ref{fig:uc_and_sc}). In both the case of DFPT and
finite-differences, all calculations have been performed for relaxed equilibrium geometries~(maximum force~$<10^{-4}$~eV/$\mbox{\AA}$) with 11$\times$11$\times$1~$\mathbf{k}$-points in the primitive unit cell, 
tight settings, the ``tier 1'' basis set, and the LDA functional. As Fig.~(\ref{fig:graphen}) shows, we find an excellent
agreement between our DPFT implementation and the finite-difference calculations: By using such extended supercells, 
even the parabolic dispersion~\cite{Liu2007} in the lowest acoustic branch and the Kohn anomalies at $\Gamma$~and~$K$ are captured in a qualitatively correct fashion by both our real-space DFPT and the finite-difference
approach, as shown by Maultzsch~{\it et al.}~\cite{Maultzsch2004} before. Our implementation is thus ideally suited
to further investigate to which extent long-range corrections to the perturbation potential will alter these effects.

\begin{figure}
\includegraphics[width=1.0\columnwidth]{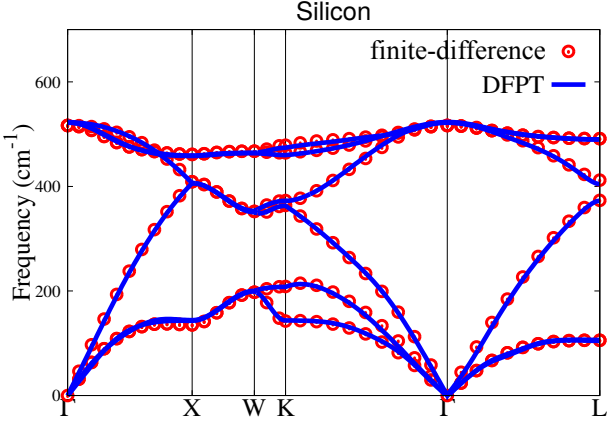}
\caption{Vibrational band structure of silicon in the diamond structure 
computed at the LDA level using both DFPT~(solid blue line) and finite-difference~(red open circles). 
All calculations have been performed using 7$\times$7$\times$7 $\mathbf{k}$~points in the primitive Brillouin zone, 
tight settings for the integration, a ``tier 1'' basis set, and the LDA functional.}
\label{fig:Si-band}
\end{figure}

For a three-dimensional system, we have used silicon in the diamond structure as an example. All calculations
have been performed for relaxed equilibrium geometries~(maximum force~$<10^{-4}$~eV/$\mbox{\AA}$) with 
7$\times$7$\times$7 $\mathbf{k}$~points in the primitive Brillouin zone, tight settings for the integration, 
a ``tier 1'' basis set, and the LDA functional. Finite-difference calculations have been performed again using {\it phonopy}~\cite{Togo-2008,phonopy} with a $5\times 5 \times 5$ supercell of the conventional cubic fcc cell~($1000$ atoms) and a finite displacement of~$0.01$~$\mbox{\AA}$, which yields fully converged vibrational band structures~(error $<1$~cm~$^{-1}$). This was systematically checked by running finite-difference calculations for up to $9\times 9 \times 9$ supercells of the primitive unit cell~($1458$ atoms). As shown in Fig.~(\ref{fig:Si-band}), our DPFT implementation again yields an excellent agreement with the respective finite-difference calculations.

\subsection{Performance and Scaling of the Implementation}
\label{Scaling}
\begin{figure}
 \includegraphics[width=0.9\columnwidth]{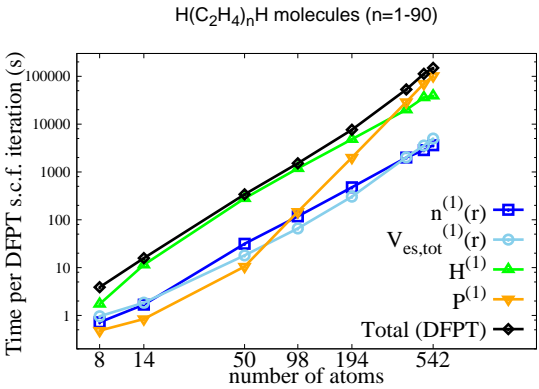}
 \caption{H(C$_2$H$_4$)$_n$H molecules: CPU time of one full DFPT cycle required to compute all perturbations/responses associated with the~$3(6n+2)$~(3 is for three cartesian directions, 6n+2 is the number of atoms.) degrees of freedom on 32 CPU cores~(see text). Following the flowchart in Fig.~\ref{fig:DFPT_flowchart}, also the timings required for the
 computation of the individual response properties~(density n$^{(1)}$, electrostatic potential V$_{es,tot}^{(1)}$,  Hamiltonian matrix H$^{(1)}$, density matrix P$^{(1)}$) are given. Here we use light settings for the integration, a ``tier 1'' basis set, and the LDA functional.}
 \label{fig:scaling_02_c2h4_mol}
\end{figure}
To systematically investigate the performance and scaling of our implementation,
we here show timings for the H(C$_2$H$_4$)$_n$H molecules with 
variable length~$n=1-90$ and the polyethylene chain~C$_2$H$_4$. In the latter case,
we have systematically increased the number of building units in the unit cell
from~(C$_2$H$_4$)$_1$ to (C$_2$H$_4$)$_{12}$. All calculations use a ``tier 1'' basis set,
light settings for the integrations, and the LDA functional. $11 \times 1 \times 1$ 
$\vec{k}$-points were used to sample the primitive Brillouin zone in the periodic case.
We performed all these calculations on a single node featuring two Intel Xeon E5-2698v3 CPUs~(32 cores)
and $4$~Gb of RAM per core. 

For the timings of the finite molecules shown in Fig.~\ref{fig:scaling_02_c2h4_mol}, we find that
the integration of the Hamiltonian response matrix~$H^{(1)}$ determines the computational time for 
small system sizes~i.e.,~for less than 200 atoms. As it is the case for the update of the response
density~$n^{(1)}$, which involves similar numerical operations, we find a scaling of~$O(N^2)$ for
this step~(see Tab.~\ref{tab:scaling}). This is not too surprising, since these operations, which 
scale with~$O(N)$ at the ground-state DFT level~\cite{Blum2009}, need to be performed $3N$~times 
when assessing the Hessian at the DFPT level,~i.e.,~once for each cartesian perturbation of each 
atom. For the exact same reasons, the treatment of electrostatic effects, which
scales as~$\sim O(N^{1.6})$ at the ground-state DFT level~\cite{Blum2009}, scales as~$O(N^{2.4})$ for the
computation of the electrostatic response potential~$V_{es,tot}^{(1)}$. For very large system
sizes~($N\gg 100$), the update of the response density matrix~$P^{(1)}$ becomes dominant,
since it scales as~$O(N^{3.8})$ in this regime. As discussed in Sec.~\ref{sec:Implementation}, 
the computation of~$P^{(1)}$ requires matrix multiplication operations, which traditionally scale~$O(N^3)$,
for each of the $3N$~individual perturbations. To assess very large systems~($N\ll 1000$), it would thus
be beneficial to switch to a more advanced formalism for this computational step~\cite{Ochsenfeld-1997,Liang-2005}.

\begin{figure}
 \centering
 \includegraphics[width=0.9\columnwidth]{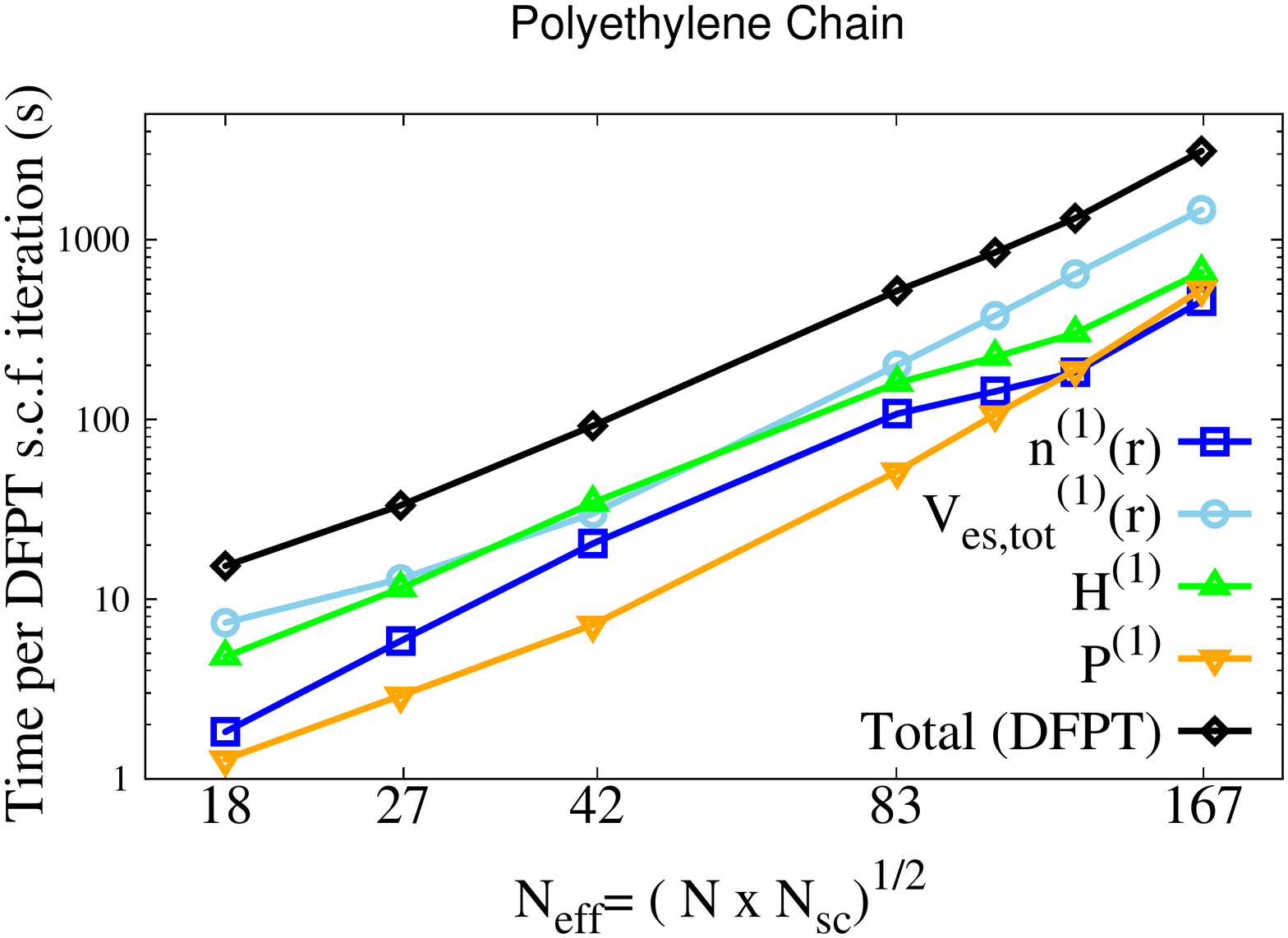}
 \includegraphics[width=0.9\columnwidth]{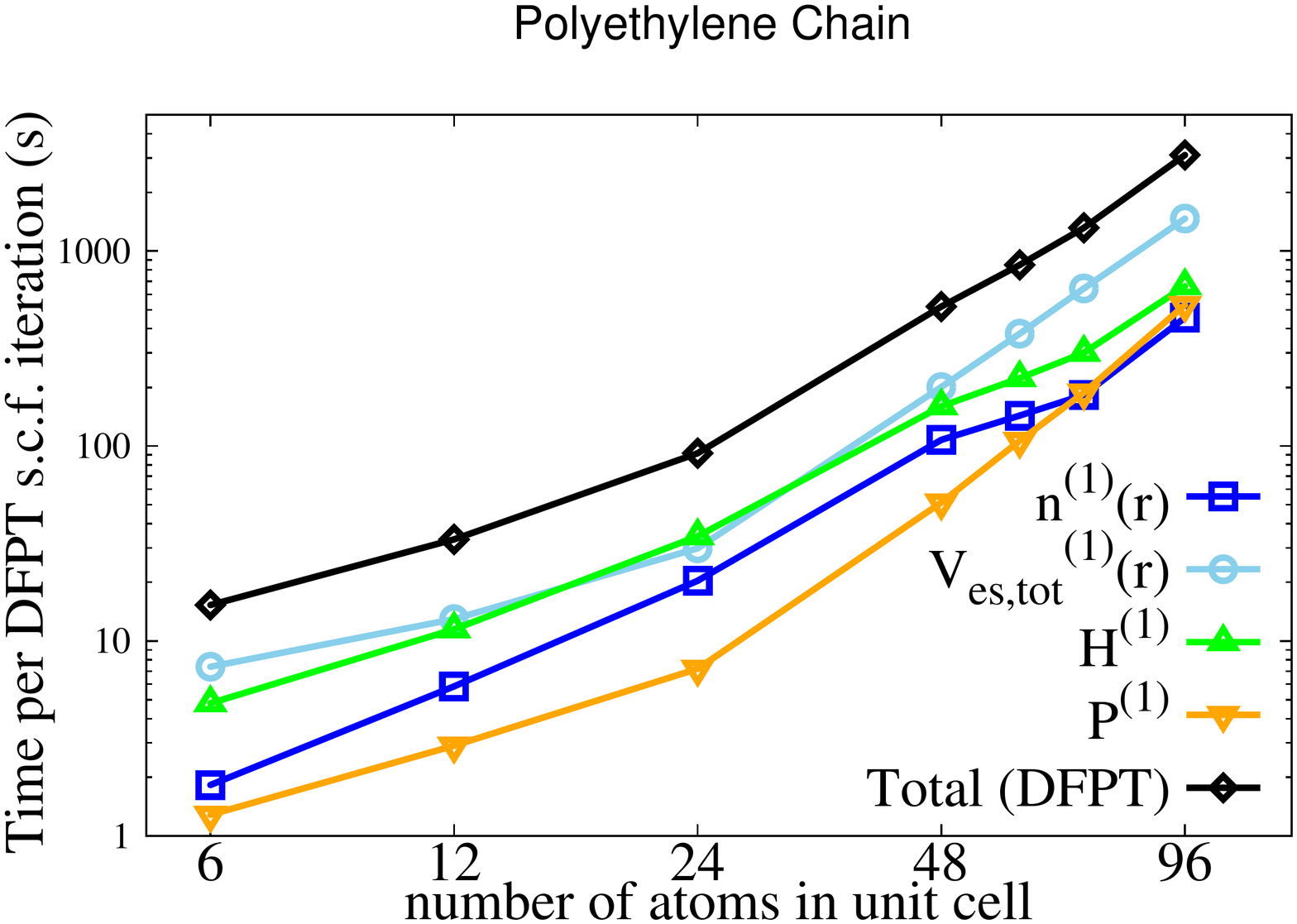}
  \caption{Linear polyethylene (C$_2$H$_4$)$_n$ chain: CPU time per DFPT cycle on 32 CPU cores as a function of the effective number of atoms~$N_{eff}$~(see text) in the upper panel and as function of the number of atoms present in the unit cell~(lower panel). Following the flowchart in Fig.~\ref{fig:DFPT_flowchart}), also the timings required for the
 computation of the individual response properties~(density n$^{(1)}$, electrostatic potential V$_{es,tot}^{(1)}$,  Hamiltonian matrix H$^{(1)}$, density matrix P$^{(1)}$) are given. Here we use light settings for the integration, a ``tier 1'' basis set, and the LDA functional.}
 \label{fig:scaling_03_c2h4_PBC}
\end{figure}

To understand the timings shown in Fig.~\ref{fig:scaling_03_c2h4_PBC} for the periodic linear chain, 
it is important to realize that such periodic calculations do not directly scale with the number of atoms~$N$,
as it was the case in the finite system, in which an $N\times N$ Hessian was computed. Rather, the
calculations are inherently performed in a supercell~(see Fig.~\ref{fig:uc_and_sc}) that features $N_{sc}$~atoms
in total. As discussed in Sec.~\ref{HARMPHON}, only an $N\times N_{sc}$ subsection of the Hessian needs to be 
determined. Accordingly, the scaling is thus best rationalized as function of the effective
number of atoms~$N_{eff}=\sqrt{N\cdot N_{sc}}$, as shown in Fig.~\ref{fig:scaling_03_c2h4_PBC} 
and Tab.~\ref{tab:scaling}. In this representation, the scaling and the respective exponents closely follow
the behaviour discussed for the finite systems already with one exception: Due to the fact that a sparse matrix
formalism is used in the periodic implementation~(see Sec.~\ref{sec:real-space-FC} and Ref.~\cite{sparse_matrix}),
a more favorable scaling for the construction of the density matrix response~$P^{(1)}$ is found.

As also shown in the lower panel of Fig.~\ref{fig:scaling_03_c2h4_PBC} and Tab.~\ref{tab:scaling},
the scaling does however not follow these intuitive expectations if plotted with respect to the number of atoms~$N$ 
present in the primitive unit cell, since $N_{eff}$, $N_{sc}$, and $N$ are not necessarily linearly related. 
For the case of the linear chain, the number of periodic images~$N_{sc}-N$ with atomic orbitals that reach into the unit
cell should be a constant that is independent of the chain length viz. number of atoms~$N$ present in the unit cell.
Accordingly the ratio~$N_{sc}/N$ decreases from a value of $9$ in the primitive C$_2$H$_4$ unit cell~(6 atoms) to a value of $N_{sc}/N=3$, if a (C$_2$H$_4$)$_4$ unit cell with 24 atoms is used. In this regime, in which $N_{eff}$ is approximately proportional to~$\sqrt{N}$, we find a very favourable overall scaling of $O(N^{1.3})$, whereby neither of the involved steps 
scales worse than~$O(N^{1.7})$.

For larger 
system sizes~($N>24$), however, the scaling deteriorates. The reason for this
behaviour is the rather primitive and simple strategy that we have employed in the generation of the DFPT 
supercells to facilitate the treatment of integrals using the minimum image convention, as discussed in Sec.~\ref{PBCSec}.
Effectively, these supercells are constructed using fully intact, translated unit cells -- 
even if a considerable part of the periodic atomic images contained in this translated unit cell do not
overlap with the original unit cell. For the case of the linear chain, the minimal possible ratio~$N_{sc}/N=3$ 
is thus reached in the $N=24$ case and retained for all larger systems~$N>24$. In this limit, $N_{eff}$ becomes
proportional to~$N$, so that we effectively recover the scaling exponents found for~$N_{eff}$ and for finite molecular systems~(cf.~Tab.~\ref{tab:scaling}).

In summary, we find an overall scaling behavior that is always clearly smaller than $O(N^3)$ for the investigated system
sizes both in the molecular and the periodic case. For the periodic case, we find a particularly favorable scaling regime of $O(N^{1.3})$ for small to medium sized unit cells~$N\leqslant24$. As discussed in more detail in the outlook, this regime can be potentially improved and extended to larger unit cell sizes. Please note that the scaling relations discussed above for the
linear chain are qualitatively also found in the case of 2D and 3D materials. Given that the utilized atomic orbitals are spatially confined within a cut-off radius~\cite{Blum2009}, similar relations between $N_{sc}$ and $N$ are effectively found in the case of graphene and silicon. Although the prefactors depend on the shape and dimensionality of the unit cell, 
the relation~$N_{eff} \propto \sqrt{N}$ also approximately holds in these cases. In this context it is very
gratifying to see that even quite extended systems~(molecules with more than 100~atoms and periodic solids with more than
50 atoms in the unit cell) are in principle treatable within the relatively moderate CPU and memory resources offered by a single state-of-the-art workstation. 

Eventually, let us note that a parallelization over cores viz. nodes is already part of the presented implementation, given that the discussed real-space DFPT 
formalism closely follows the strategies used for the parallelization of ground-state DFT calculations in {\it FHI-aims}~\cite{Blum2009,Havu/etal:2009}: 
The parallelization of the operations performed
on the real-space grid closely follows the strategy  described in ~\cite{Havu/etal:2009}; 
For the matrix operation, MPI based ScaLapack routines have been used to achieve a reasonable performance both regarding computational
and memory parallelization.

\begin{figure}
 \centering
 \includegraphics[width=0.9\columnwidth]{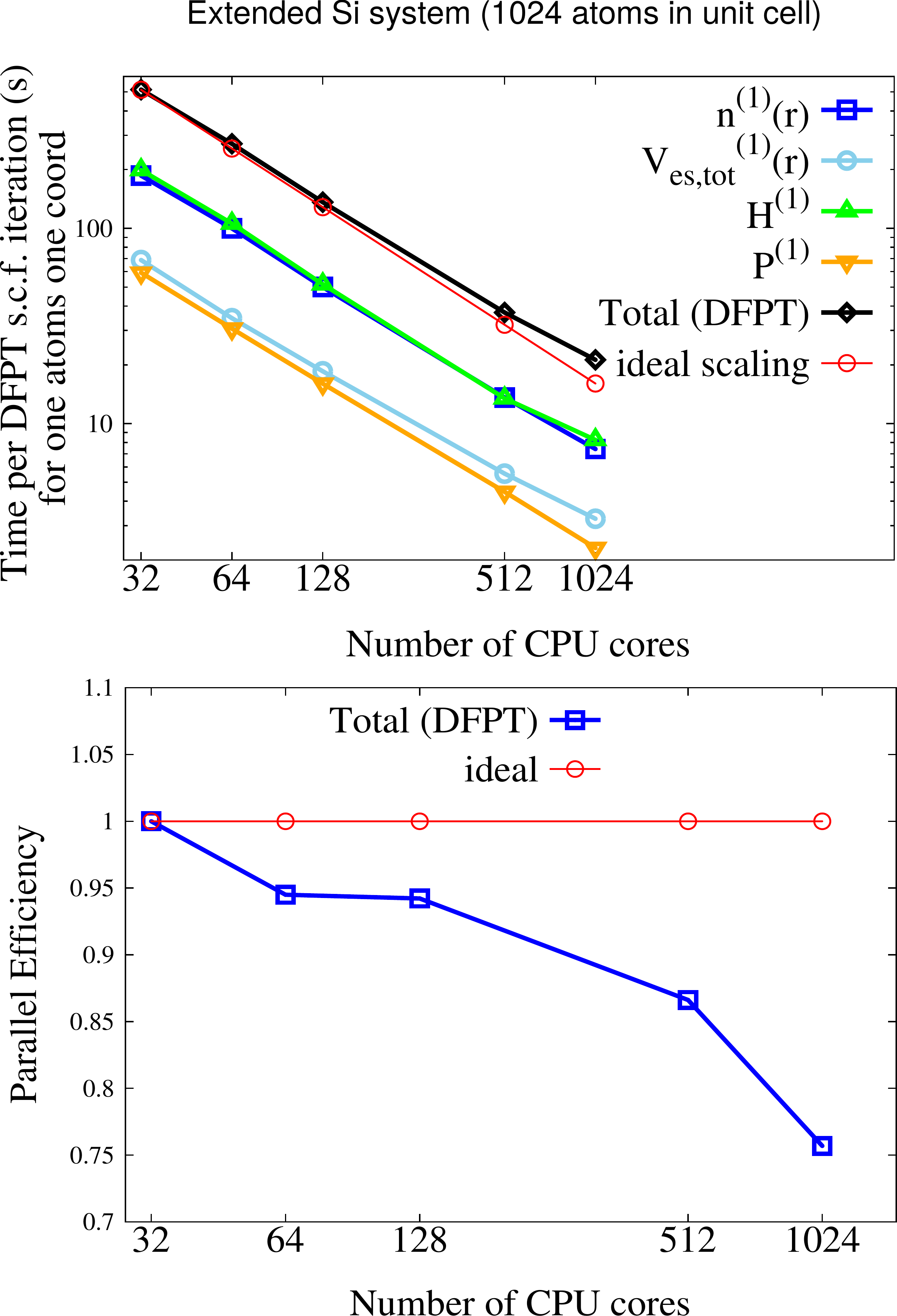}
  \caption{Parallel scalability for a unit cell containing 1024 Si atoms. Here the CPU time per DFPT cycle for the perturbation of one atom in
  one cartesian coordinate is plotted as a function of the number of CPU cores. 
  The timings required for the  computation of the individual response properties~(density n$^{(1)}$, electrostatic potential V$_{es,tot}^{(1)}$,  Hamiltonian matrix H$^{(1)}$, density matrix P$^{(1)}$) are also given. Then red line corresponds to ideal scaling. 
The parallel efficiency is shown in the lower panel.
Here we use light settings for the integration, a ``tier 1'' basis set, and the LDA functional.}
 \label{fig:speedup_04_si_PBC}
\end{figure}

The parallel scalability for a unit cell containing 1024 Si atoms is shown in Fig.\ref{fig:speedup_04_si_PBC}. All calculations use a ``tier 1'' basis set,
light settings for the integrations, and the LDA functional. One $\vec{k}$-point is sufficient to sample the reciprocal space due to the large unit cell. 
Here we give the CPU time required for one single perturbation (one atom and one cartesian coordinate). Clearly, almost ideal scaling is achieved.

\begin{table}
\centering
\begin{tabular}{c |  c | c   c   c }
\hline \hline    
                   & H(C$_2$H$_4$)$_n$H & \multicolumn{3}{c}{C$_2$H$_4$ chain} \\
                   &  $N$    &  $N_{eff}$  &  $N\leqslant24$ & $N>24$  \\\hline 
n$^{(1)}$          & 2.0   & 2.0          & 1.7  & 2.0   \\
V$_{es,tot}^{(1)}$ & 2.4   & 2.4           & 1.0  & 2.8   \\ 
H$^{(1)}$          & 2.0   & 2.2           & 1.4  & 2.0   \\ 
P$^{(1)}$          & 3.8   & 2.7           & 1.2  & 3.3  \\\hline 
Total              & 2.6   & 2.4           & 1.3  & 2.5  \\\hline \hline 
\end{tabular}
\caption{Fitted CPU time exponents~$\alpha$ for the H(C$_2$H$_4$)$_n$H molecules (n=8-90) and 
the periodic polyethylene chain C$_2$H$_4$ discussed in the text. The fits were performed using the expression~$t = c N^{\alpha}$ for the CPU time as function of the number of atoms~$N$ viz.~the effective number of atoms~$N_{eff}$.}
\label{tab:scaling}
\end{table}

\section{Conclusion and Outlook}
\label{Conclusions}
In this paper, we have derived and implemented a reformulation of density-functional perturbation theory in real-space
and validated the proposed approach by computing vibrational properties of molecules and solids. In particular, we have
shown that these calculations can be systematically converged with respect to the numerical parameters used in the computation. Also, we have demonstrated that the computed vibrational frequencies are essentially equal to those
obtained from finite-differences -- both for finite molecules and extended, periodic systems. Comparison of our results with vibrational frequencies stemming from different 
codes and implementations is urgently needed,
but would go beyond the scope of this work.

The key idea of the proposed approach relies on the localized nature of the response density in non-polar materials,
which enables the treatment of perturbations directly in real-space. On the one hand, this allows utilizing the 
computationally favorable real-space techniques developed over the last decades,~e.g.,~massively parallel grid
operations that scale $O(N)$~\cite{Blum2009,Havu/etal:2009}. 
On the other hand, the proposed approach allows us to determine the full, non-vanishing response in real-space 
in one DFPT run. In turn, simple and numerically cheap Fourier transforms--without the need of invoking any Fourier interpolation--give 
access to the exact associated response properties in reciprocal-space. We 
have explicitly demonstrated the viability of this approach for lattice dynamics calculations in periodic systems: In that
case, we get fully $\vec{q}$-point converged densities of states and vibrational band structures along arbitrary paths
from one DFPT run in real-space. Conversely, traditional reciprocal-space implementations would in principle 
have required a single DFPT run for each individual value of~$\vec{q}$. In practice, this is often circumvented in
reciprocal-space implementations, since efficient and accurate interpolation schemes for vibrational frequencies
exist~\cite{Parlinski-1997}. For the exact same reasons, finite-difference strategies can yield accurate results
even in very limited supercells~\cite{Togo-2008,phonopy}. However, this is no longer the case if more complex 
response properties such as the electron-phonon coupling~\cite{Giustino2007,Ponce2015} need to be assessed. In that case, reciprocal-space
formalisms either need to sample the Brillouin zone by brute-force~\cite{Ponce2015} or to rely on approximate interpolation
strategies,~e.g.,~using a Wannierization of the interactions in real-space~\cite{Giustino2007}. The approach discussed in
this work allows to overcome these limitations and to consistently assess all these properties using the well-controlled
wavefunction expansion already used in the ground-state DFT and thus potentially lays the foundation for future research directions in this field.

This is further substantiated by the scaling behavior discussed in the previous section. Despite being a proof-of-concept
implementation that has not undergone extensive numerical optimization, we find the code to exhibit quite favorable 
scaling properties and a promising performance that can be even improved further. For instance, the  exploitation of space and point group symmetry would straightforwardly lead to significant savings in computational time, especially for high-symmetry periodic systems. Along these lines, symmetry can also be used to optimize the construction of the supercell used
in the DFPT calculations.
For the sake of simplicity, this procedure so far relies on translated images of the complete and intact unit cell.  For particularly large and/or oblique unit cells this can result in a significant computational overhead, 
since the supercell can contain periodic images of atoms that do not interact with the unit cell at all. Accordingly, optimizing 
the supercell construction procedure can immediately lead to computational savings without loss of accuracy. Following these strategies, linear scaling should be achievable~\cite{Ordejon-1995}
for large system sizes (hundred and more atoms per unit cell). This would facilitate DFPT calculations of vibrational properties and of the electron-phonon coupling for fully converged $\mathbf{q}$-grids in complex systems, such as organic molecules adsorbed on surfaces. For such kind of applications, additional computational savings can be gained in our proposed real-space approach by artificially restricting the calculation to the actual degrees of freedom of interest,~e.g.,~the ones of the absorbed molecule.

The formalism described in this paper could 
also be extended to all type of perturbations,~e.g. homogeneous electric field perturbations, in this case only one perturbation per 
cartesian direction needs to be considered regardless of the system size.

\section{Acknowledgments}
H.S. acknowledges Wanzhen Liang and Xinguo Ren for inspiring discussions.
We further acknowledge Volker Blum for his continued support during this project. The project received funding from the Einstein foundation (project ETERNAL) and the European Union’s Horizon 2020 research and innovation program under grant agreement no. 676580 with The Novel Materials Discovery (NOMAD) Laboratory, a European Center of Excellence. P.R. acknowledges financial support from the Academy of Finland through its Centres of Excellence Program (Project No. 251748 and 284621).

\bibliography{CPC} %
\bibliographystyle{model1a-num-names}
\biboptions{square,numbers,comma,sort&compress}

\appendix

\section{Convergence Behaviour of Forces with Respect to the Degree of Self-Consistency}
\label{sec:eq9}
\begin{figure}
\includegraphics[width=0.95\columnwidth]{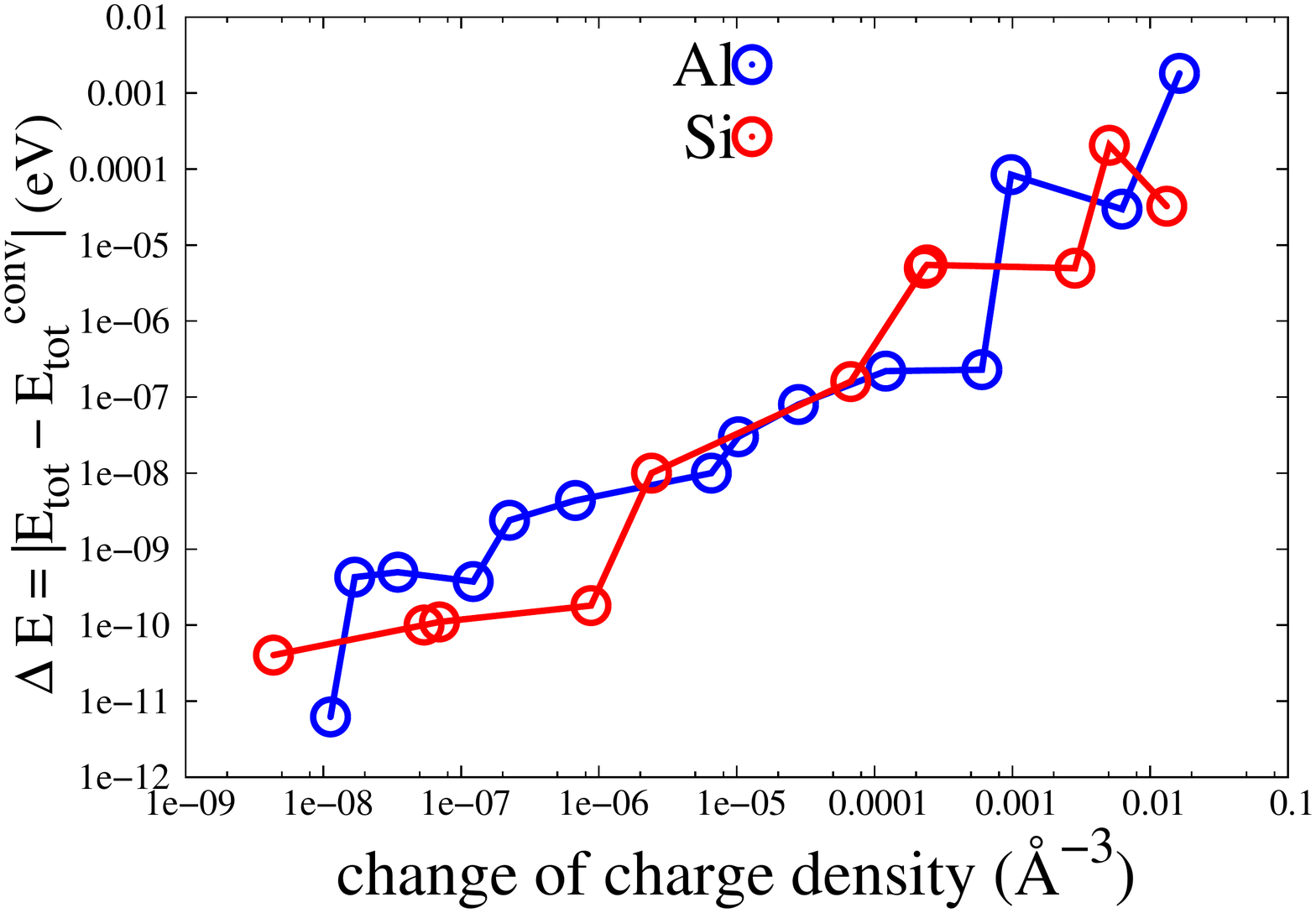}
\includegraphics[width=0.95\columnwidth]{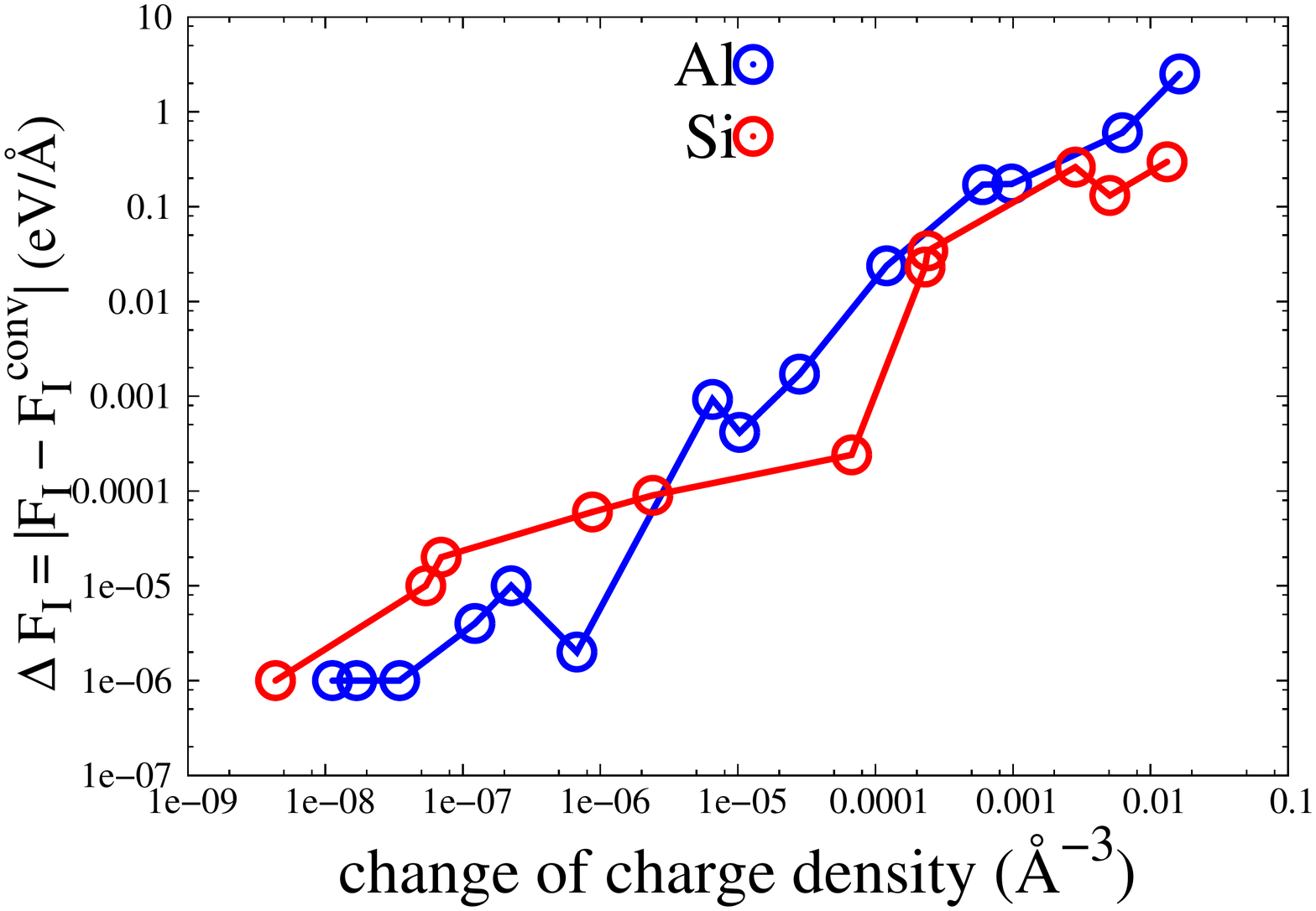}
\caption{ The convergence behaviour of the total energy~$\Delta E = |E_{tot} - E_{tot}^{conv}|$ (top panel)  and the forces~$\Delta \vec{F}_I = |\vec{F}_I - \vec{F}_I^{conv}|$ (bottom panel).}
\label{fig:E_and_F_vs_n}
\end{figure}

To investigate to which extent the last term of  Eq.~(\ref{eq:force_derivative})  really vanishes in practice, we have chosen Si (diamond structure) and Al (fcc) as examples.
In both cases, one atom was displaced by $0.1$~$\mbox{\AA}$, which results in forces on this atom in the order of~$10^{0}$~eV/$\mbox{\AA}$ and~$10^{-1}$~eV/$\mbox{\AA}$, respectively. To investigate what happens in calculations, in which full self-consistency has not yet been reached, we have then run a series of calculations with different break conditions for the self-consistency cycle. We only used the maximally allowed change in charge density as break condition and varied its value between~$10^{-2}$ and~$10^{-8}$~electrons. For the last setting, full self-consistency is achieved: Indeed, the observed change in energy/eigenvalues in the last iteration of such fully converged calculations is $10^{-11}$~eV / $10^{-7}$~eV for Si, and $10^{-12}$~eV / $10^{-7}$~eV for Al. In Fig~\ref{fig:E_and_F_vs_n}, we then show the respective convergence behaviour of the total energy~$\Delta E = |E_{tot} - E_{tot}^{conv}|$  and of the force on the displaced atom~$\Delta {F}_I = |\vec{F}_I - \vec{F}_I^{conv}|$  with respect to these fully converged converged values. As soon as $E_{tot}$ is converged, Eq.~(\ref{eq:force_derivative}) reveals that
\begin{equation}
\Delta \vec{F}_I = -\sum_{\mu i}{\dfrac{\partial E_{tot}}{\partial C_{\mu i}}} \dfrac{\partial C_{\mu i}} {\partial{\mathbf{R}_{I}}} \;. 
\end{equation}
is indeed the error we want to assess.
From Fig.\ref{fig:E_and_F_vs_n} we can see that as the change of charge density approaches zero, the error in the forces starts to vanish $\Delta {F}_I = |\vec{F}_I-\vec{F}^{conv}_I|$ = $10^{-6}$. For the typical self-consistency settings used in {\it FHI-aims} (change in charge density $<10^6$), the error in the force due to the non-fully-achieved self-consistency is thus typically smaller than $1$~meV/$\mbox{\AA}$. In this context, it is however important to note that in relaxation or MD calculations {\it FHI-aims} requires to specify a self-consistency break condition also for the maximum change in the forces, so that in practice these errors are well-controlled.

\section{First Order Density Matrix}
\label{sec:DM1_derivation}
The sum over states in the first order density matrix can be divided into sums over occupied-occupied states, occupied-unoccupied states, and unoccupied-occupied states: 
\begin{align}
P_{\mu m,\nu n}^{(1)}=\sum_{i}^{occ}{f(\epsilon_{i})(C_{\mu m,i}^{(1)}C_{\nu n,i}^{(0)}}+C_{\mu m,i}^{(0)}C_{\nu n,i}^{(1)}) \\
=\sum_{i}\sum_{q}{f(\epsilon_{i}) C_{\mu m, q}^{(0)}U_{q i}^{(1)}C_{\nu n, i}^{(0)}}
+\sum_{i}\sum_{p}{f(\epsilon_{i}) C_{\mu m,i}^{(0)}C_{\nu n,p}^{(0)}U_{p i}^{(1)}}  \nonumber\\ 
=\sum_{i}\sum_{j}{f(\epsilon_{i}) C_{\mu m,j}^{(0)}U_{j i}^{(1)}C_{\nu n, i}^{(0)}}
+\sum_{i}\sum_{a}{f(\epsilon_{i}) C_{\mu m, a}^{(0)}U_{a i}^{(1)}C_{\nu n, i}^{(0)}}  \nonumber\\
+\sum_{i}\sum_{j}{f(\epsilon_{i}) C_{\mu m,i}^{(0)}C_{\nu n,j}^{(0)}U_{j i}^{(1)}}
+\sum_{i}\sum_{a}{f(\epsilon_{i}) C_{\mu m,i}^{(0)}C_{\nu n,a}^{(0)}U_{a i}^{(1)}}  \nonumber\\
=\sum_{i}\sum_{j}{f(\epsilon_{j}) C_{\mu m,i}^{(0)}U_{i j}^{(1)}C_{\nu n, j}^{(0)}}
+\sum_{i}\sum_{a}{f(\epsilon_{i}) C_{\mu m,a}^{(0)}U_{a i}^{(1)}C_{\nu n,i}^{(0)}} \nonumber\\
+\sum_{i}\sum_{j}{f(\epsilon_{i}) C_{\mu m,i}^{(0)}C_{\nu n,j}^{(0)}U_{j i}^{(1)}}
+\sum_{i}\sum_{a}{f(\epsilon_{i}) C_{\mu m,i}^{(0)}C_{\nu n,a}^{(0)}U_{a i}^{(1)}} \nonumber\\
=\sum_{i}\sum_{j}{f(\epsilon_{i})(C_{\mu m,i}^{(0)}C_{\nu n,j}^{(0)}(U_{j i}^{(1)}+U_{i j}^{(1)}))} \nonumber\\
+\sum_{i}\sum_{a}{f(\epsilon_{i})C_{\mu m,a}^{(0)}U_{a i}^{(1)}C_{\nu n, i}^{(0)}}+\sum_{i}\sum_{a}{f(\epsilon_{i}) C_{\mu m,i}^{(0)}U_{a i}^{(1)}C_{\nu n, a}^{(0)}} \;. \nonumber
\end{align}

From Eq.~(\ref{eq:orthonormal}), we have :
\begin{equation}
U^{(1)\dagger}+C^{(0)\dagger}S^{(1)}C^{(0)}+U^{(1)}=0\;.
\end{equation}

So Eq.~(\ref{eq:DM1}) can be simplified as :
\begin{align}
P_{\mu m,\nu n}^{(1)}=\sum_{i}\sum_{j}{f(\epsilon_{i})C_{\mu m, i}^{(0)}C_{\nu n, j}^{(0)}
(-C^{(0)\dagger}S^{(1)}C^{(0)})_{i j}} \\
+\sum_{i}\sum_{a}{f(\epsilon_{i}) C_{\mu m,a}^{(0)}U_{a i}^{(1)}C_{\nu n, i}^{(0)}}+\sum_{i}\sum_{a}{f(\epsilon_{i}) C_{\mu m,i}^{(0)}U_{a i}^{(1)}C_{\nu n, a}^{(0)}}  \nonumber \;.
\end{align}

This means that we only need to calculate the occupied-unoccupied sum for $U_{a i}^{(1)}$ in the CPSCF equation, Eq.~(\ref{eq:cpscf-LCAO}), while the occupied-occupied part is computed from the first order overlap matrix.

\section{First order energy weighted density matrix}
\label{sec:EDM1_derivation}
Similar like the first order density matrix, by using Eq.~(\ref{eq:orthonormal}) and Eq.~(\ref{eq:epsilon1}), we 
can rewrite Eq.~(\ref{eq:EDM1}) into an occupied-occupied part, occupied-unoccupied part, and an unoccupied-occupied part: 

\begin{eqnarray}
W_{\mu m,\nu n}^{(1)} & = &\sum_{i}^{occ}\sum_{j}^{occ}
{f(\epsilon_{i}) C_{\mu m,i}^{(0)}C_{\nu n,j}^{(0)}}\left[ (C^{(0)\dagger}H^{(1)}C^{(0)})_{ij} \right.\\
&&  \left.- (\epsilon_{i}+\epsilon_{j}) (C^{(0)\dagger}S^{(1)}C^{(0)})_{ij}
\right] \nonumber  \\
&& +\sum_{i}^{occ}\sum_{a}^{unocc}{ f(\epsilon_{i}) \epsilon_{i} (C_{\mu m,a} U_{ai}C_{\nu n,i} +C_{\mu m, i} U_{ai}C_{\nu n,a} )  } \;. \nonumber 
\end{eqnarray}

\section{32 molecules' frequencies}
\label{sec:32-mol-new}
Tab.~\ref{tab:dimers-new}, Tab.~\ref{tab:trimers-new}, Tab.~\ref{tab:4-atom-new}, Tab.~\ref{tab:5-atom-new}, Tab.~\ref{tab:6-atom-new},  and Tab.~\ref{tab:8-atom-new} list the vibrational frequencies obtained via DFPT and via finite-differences for systems containing two, three, four, five, six, and eight atoms, respectively. In these comparisons, the atomic displacement is set to ~0.013~{\AA} in finite-difference calculation. We used a ``tier 2'' basis sets and $N_{r,mult}=4$~(except LiF, NaCl and P$_2$, in which $N_{r,mult}=2$ is used) for integration grids and $l$=12 for multipole expansion. All calculations were performed at the LDA level 
for the equilibrium geometry determined by relaxation~(maximum force~$<10^{-4}$~eV/$\mbox{\AA}$).
The statistical data is succinctly summarized in Tab.~\ref{tab:mol-new} in the main text.

\begin{table}
\caption{16 dimers.}
\label{tab:dimers-new}
\begin{tabular}{c| c c c c }
\hline \hline
   &finite-difference   & DFPT  &   ab-err &  rel-err(\%) \\
\hline 

Cl$_2$ & 562.85 & 562.70 & .15 & 0.03 \\
ClF & 805.68 & 805.05 & .63 & 0.08 \\
CO & 2177.77 & 2176.35 & 1.42 & 0.07 \\
CS & 1285.98 & 1285.37 & .61 & 0.05 \\
F$_2$& 1062.79 & 1062.29 & .50 & 0.05 \\
H2 & 4176.79 & 4174.46 & 2.33 & 0.06 \\
HCl & 2881.98 & 2880.76 & 1.22 & 0.04 \\
HF & 3978.39 & 3975.59 & 2.80 & 0.07 \\
Li$_2$& 345.86 & 345.46 & .40 & 0.12 \\
LiF & 930.13 & 929.81 & .32 & 0.03 \\
LiH & 1385.31 & 1385.13 & .18 & 0.01 \\
N2 & 2396.81 & 2395.33 & 1.48 & 0.06 \\
Na$_2$& 164.08 & 163.89 & .19 & 0.12 \\
NaCl & 375.27 & 374.63 & .64 & 0.17 \\
P$_2$ & 804.92 & 794.79 & 10.13 & 1.26 \\
SiO & 1232.65 & 1232.15 & .50 & 0.04 \\
\hline
MAE &        &          & 1.5  &    \\
MAPE&        &          &       & 0.14\% \\
\hline \hline
\end{tabular}
\end{table}

\begin{table}
\caption{5 trimers.}
\label{tab:trimers-new}
\begin{tabular}{c| c c c c }
\hline \hline
   &finite-difference   & DFPT &   ab-err &  rel-err(\%) \\
\hline
H$_2$O & 1544.38 & 1546.25 & 1.87 & 0.12 \\
 & 3712.72 & 3711.88 & .84 & 0.02 \\
 & 3821.62 & 3820.92 & .70 & 0.02 \\
SH$_2$  & 1138.53 & 1139.12 & .59 & 0.05 \\
 & 2623.11 & 2623.01 & .10 & 0.00 \\
 & 2638.81 & 2638.62 & .19 & 0.01 \\
HCN& 718.60 & 718.20 & .40 & 0.06 \\
 & 2152.87 & 2151.78 & 1.09 & 0.05 \\
 & 3338.28 & 3336.88 & 1.40 & 0.04 \\
CO$_2$ & 651.70 & 652.05 & .35 & 0.05 \\
 & 1353.10 & 1352.19 & .91 & 0.07 \\
 & 2415.46 & 2413.80 & 1.66 & 0.07 \\
SO$_2$ & 500.56 & 501.06 & .50 & 0.10 \\
 & 1156.15 & 1155.86 & .29 & 0.03 \\
 & 1359.09 & 1358.64 & .45 & 0.03 \\
\hline
MAE &        &          & 0.76  &    \\
MAPE&        &          &       & 0.05\% \\
\hline \hline
\end{tabular}
\end{table}

\begin{table}
\caption{4-atom molecules.}
\label{tab:4-atom-new}
\begin{tabular}{c| c c c c }
\hline \hline
       &finite-difference &    DFPT&  ab-error  & rel-error\\
\hline
C$_2$H$_2$ & 631.92 & 631.52 & .40 & 0.06 \\
 & 719.65 & 719.22 & .43 & 0.06 \\
 & 719.65 & 719.22 & .43 & 0.06 \\
 & 2023.26 & 2022.38 & .88 & 0.04 \\
 & 3315.34 & 3314.01 & 1.33 & 0.04 \\
 & 3416.76 & 3415.29 & 1.47 & 0.04 \\
H$_2$CO & 1140.29 & 1140.16 & .13 & 0.01 \\
 & 1211.80 & 1212.44 & .64 & 0.05 \\
 & 1458.40 & 1458.69 & .29 & 0.02 \\
 & 1804.60 & 1803.62 & .98 & 0.05 \\
 & 2764.60 & 2764.21 & .39 & 0.01 \\
 & 2814.41 & 2814.05 & .36 & 0.01 \\
H$_2$O$_2$ & 392.08 & 397.81 & 5.73 & 1.46 \\
 & 959.42 & 958.97 & .45 & 0.05 \\
 & 1282.80 & 1282.90 & .10 & 0.01 \\
 & 1388.60 & 1388.65 & .05 & 0.00 \\
 & 3640.98 & 3640.44 & .54 & 0.01 \\
 & 3642.69 & 3641.95 & .74 & 0.02 \\
NH$_3$ & 946.72 & 947.29 & .57 & 0.06 \\
 & 1576.61 & 1577.17 & .56 & 0.04 \\
 & 1577.08 & 1577.18 & .10 & 0.01 \\
 & 3392.19 & 3391.44 & .75 & 0.02 \\
 & 3525.15 & 3524.98 & .17 & 0.00 \\
 & 3525.66 & 3525.00 & .66 & 0.02 \\
PH$_3$& 946.27 & 946.59 & .32 & 0.03 \\
 & 1072.55 & 1072.73 & .18 & 0.02 \\
 & 1072.65 & 1072.74 & .09 & 0.01 \\
 & 2323.49 & 2323.41 & .08 & 0.00 \\
 & 2338.77 & 2338.63 & .14 & 0.01 \\
 & 2338.89 & 2338.64 & .25 & 0.01 \\
\hline
MAE &        &          & 0.64 &    \\
MAPE&        &          &       & 0.07\%\\
\hline \hline
\end{tabular}
\end{table}

\begin{table}
\caption{5-atom molecules.}
\label{tab:5-atom-new}
\begin{tabular}{c| c c c c }
\hline \hline
       &finite-difference &    DFPT&  ab-error  & rel-error\\
\hline
CH$_3$Cl  & 750.93 & 750.79 & .14 & 0.02 \\
 & 981.30 & 981.66 & .36 & 0.04 \\
 & 981.31 & 981.66 & .35 & 0.04 \\
 & 1303.79 & 1303.67 & .12 & 0.01 \\
 & 1398.88 & 1399.40 & .52 & 0.04 \\
 & 1399.34 & 1399.42 & .08 & 0.01 \\
 & 2979.12 & 2978.61 & .51 & 0.02 \\
 & 3079.15 & 3078.82 & .33 & 0.01 \\
 & 3079.60 & 3078.83 & .77 & 0.03 \\
SiH$_4$  & 843.26 & 843.50 & .24 & 0.03 \\
 & 843.26 & 843.50 & .24 & 0.03 \\
 & 843.26 & 843.50 & .24 & 0.03 \\
 & 926.70 & 926.95 & .25 & 0.03 \\
 & 926.70 & 926.95 & .25 & 0.03 \\
 & 2165.07 & 2165.04 & .03 & 0.00 \\
 & 2183.92 & 2183.76 & .16 & 0.01 \\
 & 2183.92 & 2183.76 & .16 & 0.01 \\
 & 2183.92 & 2183.76 & .16 & 0.01 \\
CH$_4$  & 1247.98 & 1248.63 & .65 & 0.05 \\
 & 1247.98 & 1248.63 & .65 & 0.05 \\
 & 1247.98 & 1248.63 & .65 & 0.05 \\
 & 1476.39 & 1476.84 & .45 & 0.03 \\
 & 1476.39 & 1476.84 & .45 & 0.03 \\
 & 2956.13 & 2956.02 & .11 & 0.00 \\
 & 3083.56 & 3083.51 & .05 & 0.00 \\
 & 3083.56 & 3083.51 & .05 & 0.00 \\
 & 3083.56 & 3083.51 & .05 & 0.00 \\
\hline
MAE &        &          & 0.28 &    \\
MAPE&        &          &       & 0.02\%\\
\hline \hline
\end{tabular}
\end{table}

\begin{table}
\caption{6-atom molecules.}
\label{tab:6-atom-new}
\begin{tabular}{c| c c c c }
\hline \hline
     &finite-difference &    DFPT&  ab-error  & rel-error\\
\hline
N$_2$H$_4$ & 489.03 & 489.48 & .45 & 0.09 \\
 & 706.75 & 707.80 & 1.05 & 0.15 \\
 & 869.85 & 870.67 & .82 & 0.09 \\
 & 1140.48 & 1140.22 & .26 & 0.02 \\
 & 1244.10 & 1244.56 & .46 & 0.04 \\
 & 1272.66 & 1273.04 & .38 & 0.03 \\
 & 1600.03 & 1600.02 & .01 & 0.00 \\
 & 1608.97 & 1609.18 & .21 & 0.01 \\
 & 3368.42 & 3367.52 & .90 & 0.03 \\
 & 3371.10 & 3370.59 & .51 & 0.02 \\
 & 3473.44 & 3472.72 & .72 & 0.02 \\
 & 3478.32 & 3477.61 & .71 & 0.02 \\
C$_2$H$_4$ & 795.78 & 796.25 & .47 & 0.06 \\
 & 929.08 & 926.20 & 2.88 & 0.31 \\
 & 940.78 & 939.53 & 1.25 & 0.13 \\
 & 1031.98 & 1029.91 & 2.07 & 0.20 \\
 & 1183.38 & 1183.57 & .19 & 0.02 \\
 & 1322.46 & 1322.65 & .19 & 0.01 \\
 & 1394.15 & 1394.28 & .13 & 0.01 \\
 & 1651.73 & 1651.42 & .31 & 0.02 \\
 & 3040.77 & 3040.49 & .28 & 0.01 \\
 & 3053.43 & 3053.38 & .05 & 0.00 \\
 & 3117.11 & 3116.90 & .21 & 0.01 \\
 & 3144.13 & 3143.72 & .41 & 0.01 \\
\hline
MAE &        &          &  0.59 &    \\
MAPE&        &          &       & 0.05\%\\
\hline \hline
\end{tabular}
\end{table}

\begin{table}
\caption{Si$_2$H$_6$.}
\label{tab:8-atom-new}
\begin{tabular}{c|  c c c c  }
\hline \hline
    &finite-difference &  DFPT&  ab-error  & rel-error\\
\hline
Si$_2$H$_6$& 136.34 & 136.96 & .62 & 0.45 \\
& 336.78 & 336.98 & .20 & 0.06 \\
& 336.81 & 336.99 & .18 & 0.05 \\
& 429.87 & 429.51 & .36 & 0.08 \\
& 592.25 & 592.56 & .31 & 0.05 \\
& 592.41 & 592.57 & .16 & 0.03 \\
& 778.41 & 778.17 & .24 & 0.03 \\
& 845.98 & 846.06 & .08 & 0.01 \\
& 883.04 & 883.20 & .16 & 0.02 \\
& 883.22 & 883.21 & .01 & 0.00 \\
& 896.42 & 896.56 & .14 & 0.02 \\
& 896.54 & 896.56 & .02 & 0.00 \\
& 2145.74 & 2145.56 & .18 & 0.01 \\
& 2149.57 & 2149.48 & .09 & 0.00 \\
& 2159.48 & 2159.42 & .06 & 0.00 \\
& 2159.64 & 2159.43 & .21 & 0.01 \\
& 2168.94 & 2168.92 & .02 & 0.00 \\
& 2169.10 & 2168.93 & .17 & 0.01 \\
\hline
MAE &        &      & 0.18 &    \\
MAPE&        &          &       & 0.05\%\\
\hline \hline
\end{tabular}
\end{table}

\end{document}